\documentclass{JHEP3}
\usepackage{epsfig}

\newcommand{\Zll}{\mbox{$ Z\to \ell^{+} \ell^{-}$}}
\newcommand{\Mll}{\mbox{$M_{\ell\ell}$}}
\newcommand{\Zgammall}{\mbox{$Z/\gamma^{*}\to \ell^{+} \ell^{-}$}}

\newcommand{\Zgam}{\mbox{$Z/\gamma^{*}$}}
\newcommand{\BR}{\mathop{\mathrm BR}\nolimits}
\newcommand{\pT}{\mbox{$p_{{}_{\scriptstyle\mathrm T}}$}}
\newcommand{\pTel}{\mbox{$p_{{}_{\scriptstyle\mathrm T}}^{\,\ell}$}}
\newcommand{\pTmin}{\mbox{$p_{{}_{\scriptstyle\mathrm T}}^{\,\mathrm min}$}}
\newcommand{\etamax}{\mbox{$\eta_{{}_{\,\scriptstyle\mathrm max}}$}}
\newcommand{\etael}{\mbox{$\eta_{{}_{\,\scriptstyle\ell}}$}}
\newcommand{\etal}{{\it et al.}}
\newcommand{\mylink}[1]{\tt{\href{http://#1}{#1}}}
\newcommand{\DNNLO}{\mbox{$\delta_{{\,\scriptstyle\mathrm{NNLO}}}$}}
\newcommand{\Dscale}{\mbox{$\delta_{{\,\scriptstyle\mathrm{scale}}}$}}

\catcode`\@=11
\newcommand{\DOUBLEFIGLAB}[6][ht]{\@dblfloat{figure}[#1]\label{#6}\centerline{%
                \parbox{.45\textwidth}{\centerline{\epsfig{file=#2}}}~~~~
                \parbox{.45\textwidth}{\centerline{\epsfig{file=#3}}}}
                \centerline{\parbox[t]{.45\textwidth}{\caption{#4}}~~~~
                \parbox[t]{.45\textwidth}{\caption{#5}}}\end@dblfloat}
\catcode`\@=12
\title{
Evaluation of the Theoretical Uncertainties in the  $\Zll$  
Cross Sections at the LHC.
}
\author{Nadia E. Adam, Valerie Halyo, Scott A. Yost\\
Department of Physics, \\
Princeton University, \\
Princeton, NJ 08544\\
\vspace{1em}
\email{neadam@princeton.edu}, \email{valerieh@princeton.edu}, \email{syost@princeton.edu}
}
\preprint{Journal reference: JHEP {\bf 05} (2008) 062 }
\date{December 2007}
\abstract{We study the sources of systematic errors in the measurement
of the $\Zll$ cross-sections at the LHC. We consider the systematic errors
 in both the total cross-section
and acceptance for anticipated experimental cuts. We include the best 
available analysis of QCD effects at NNLO in assessing the effect of higher
order corrections and PDF and scale uncertainties on the theoretical 
acceptance. In addition, we evaluate the error due to missing NLO electroweak 
corrections and propose which MC generators and computational schemes should be
implemented to best simulate the events.
}
\begin{document}

%%%%%%%%%%%%%%%%%%%%%%%%%%%%%%%%%%%%%%%%%%%%%%%%%%%%%%%%%%%%%%%%%%%
\section{Introduction}
\label{sec:intro}

A precise measurement of gauge boson production cross-sections for 
$pp$ scattering will be
crucial at the LHC. $W$ and $Z$ bosons will be produced copiously, and 
a careful measurement of their production cross-sections will be important
in testing the Standard Model more rigorously than ever before, and 
uncovering signs of new physics which may appear through radiative corrections.
In addition, these cross-sections have been proposed as a ``standard candle'' 
for measuring the luminosity through a comparison of the measured rates to
the best theoretical calculations of the cross-section.  Investigation of 
this means of measuring luminosity began at the Tevatron and will continue
at the LHC~\cite{lumi,krasny}.

In this paper, we will concentrate on $Z$ production.  
A comparable analysis of systematic uncertainties in $W$ production 
appeared in Ref.~\cite{mangano}. Since that time, both the experimental 
approach to the measurements and the theoretical results needed to calculate
them have both been refined.  NNLO QCD calculations of these processes,
previously available only for the total cross-section~\cite{total} and 
rapidity distribution~\cite{dixon2003}, are now available in differential 
form~\cite{FEWZ}, permitting an analysis of the effect of experimental 
cuts on the pseudorapidity and transverse momentum of the final state leptons.

The high luminosity ($10^{33} - 10^{34}$ cm$^{2}$s$^{-1}$) at the 
LHC insures that systematic errors will play a dominant role in
determining the accuracy of the cross-section. Thus, we present an analysis
of the effect of the theoretical uncertainty in the evaluation of the
acceptance, and propose which among the various available MC generators
and computational schemes should be implemented to best simulate the events. 

This paper is organized as follows. Sec.\ \ref{sec:generators} will give an
overview of the calculation and the computational tools used in the analysis.
The next four sections are each devoted to estimating a class of systematic
errors: electroweak corrections in Sec.\ \ref{sec:EWK}, NNLO QCD in 
Sec.\ \ref{sec:QCD}, QCD scale dependence in Sec.\ \ref{sec:scale}, and parton
distribution function uncertainties in Sec.\ \ref{sec:PDF}. Finally, the 
results are compiled and summarized in Sec.\ \ref{sec:conclusions}.

\section{Theoretical Calculations and MC Generators}
\label{sec:generators}

The dominant production mechanism for $Z$ bosons is the Drell-Yan
process~\cite{DY}, in which a quark and antiquark annihilate to 
form a vector boson, which subsequently decays into a lepton pair. It is 
through this lepton pair $\ell^+\ell^-$ that the production process is observed.
Such pairs may as well be produced via an intermediate
photon $\gamma^{\ast}$, so both cases should be considered together. 
The $Z$ production cross-section may be inferred experimentally from the 
number $N^{\,\rm obs}_{\Zgam}$ of observed events via the relation
\begin{equation}
N^{\,\rm obs}_{\Zgam} = \sigma^{\rm tot}\BR(\Zgam\to\ell^+\ell^-) 
A_{\Zgam}\int{\cal L} dt.
\label{eq:cs1}
\end{equation}
$A_{\Zgam}$ is the acceptance obtained after applying the experimental 
selection criteria. For example, if the cuts require $\pTel > \pTmin$,
$0<\eta^{\ell} <\etamax$, then
\begin{eqnarray}
A_{\Zgam}(\pTmin,\etamax) &=& \frac{1}{\sigma^{\rm tot}\BR(\Zgammall)}
\int_{\pTmin}^{\sqrt{s}/2} \int_{-\etamax}^{\etamax}
d\pT^{\ell^+}\,d\pT^{\ell^-} d\eta_{\ell^+}d\eta_{\ell^-} 
\nonumber\\
& \times & \frac{d^4\sigma(pp\rightarrow\Zgammall)}
{d\pT^{\ell^+}\,d\pT^{\ell^-} d\eta_{\ell^+}d\eta_{\ell^-}}\,.
%\BR(\Zgammall)\,,
\label{eq:acc}
\end{eqnarray}
In practice, further cuts on the invariant mass $\Mll$ of the lepton pair
may be included to prevent the cross-section from being dominated by photons,
 which give a divergent contribution at low energies.
 
Alternatively to the $Z$ production cross-section measurement,
the corrected $Z$ yield
can be used as a standard candle for a luminosity monitor in LHC if one 
calculates the cross-section and solves for  $\int{\cal L} dt $.
The theoretical cross-section may be constructed 
by convoluting a parton-level cross-section 
${\widehat\sigma}_{ab}$ for partons $a$ and $b$ with the parton 
density functions (PDFs) $f_a$, $f_b$ for these partons, 
\begin{equation}
\sigma^{\rm th}(pp\rightarrow\Zgammall) = \sum_{a,b} \int_0^1 
dx_1 dx_2 f_a(x_1)f_b(x_2)\; {\widehat\sigma}_{ab}(x_1,x_2), 
\label{eq:thxs}
\end{equation}
integrating over the momentum fractions $x_1, x_2$,
and applying cuts relevant to the experiment.  Theoretical 
errors come from limitations in the order of the calculation 
of $\sigma_{ab}$, on its completeness
(for example, on whether it includes electroweak corrections or 
$\gamma^*/Z$ interference, and on whether any phase space variables or spins 
have been averaged), and from errors in the PDFs.  \footnote{
The representation eq.\ \ref{eq:cs1} of the cross-section is intended to
illustrate the manner in which the measurement may be used to infer the 
$Z$ production cross-section, and does not imply that a narrow resonance 
approximation is actually used in calculating the cross-section \ref{eq:thxs}.}

Since the final state may include additional partons which form a shower, 
the output from the hard QCD process must be fed to a shower generator 
to generate a realistic final state seen in a detector. This is possible only
if the cross-section is simulated in an event generator. Calculating the
acceptance for all but the simplest cuts will normally require an
event generator as well. 

Thus, when constructing a simulation of an experiment, there is a range of
choices which can be made among the tools currently available. An efficient
calculation requires selecting those adequate to meet the anticipated 
precision requirements, without performing unnecessarily complex calculations.
For example, while NNLO calculations are now available, 
the cross-sections are very complicated, do not always converge well,
 and require substantial time to
calculate. For certain choices of cuts, it may be found that the effect of
the NNLO result can be minimized, or that it can be represented by a simplified
function for the parameters of interest. We will compare several 
possible schemes for calculating the $Z$ production cross-section and 
acceptance, and consider the systematic errors arising for these schemes.

The most basic way to generate events is through one of the showering 
programs, such as PYTHIA~\cite{pythia}, HERWIG~\cite{herwig}, 
ISAJET~\cite{isajet} or SHERPA~\cite{sherpa}. These vary somewhat
in their assumptions and range of effects included, but they all 
start with hard partons at a high energy scale and branch to form partons
at lower scales, which permits a description of hadronization and realistic
events. On their own, these programs typically rely on a leading order 
hard matrix element and include only a leading-log resummation of soft and
collinear radiation in the shower, limiting their value in describing events
with large transverse momentum. In addition, ISAJET lacks color-coherence, 
which is important in predicting the correct distribution of soft
jets~\cite{coherence}.

Fully exclusive NLO QCD calculations are available for 
$W$ and $Z$ boson production~\cite{NLO}.  The MC generator
MC@NLO~\cite{MCNLO} combines a parton-level NLO QCD calculation with
the HERWIG~\cite{herwig} parton shower, thus removing some of the limitations
of a showering program alone. 

Since $\alpha\approx \alpha_s^2$ at LHC energies, NLO electroweak (EWK)
corrections should appear at the same order as NNLO QCD. 
The MC@NLO package is missing EWK corrections, but the most important of these 
corrections under the $Z$ peak is expected to be QED final state radiation~\cite{LESHOUCES}. This 
can be obtained by combining MC@NLO with PHOTOS~\cite{PHOTOS}, an add-on 
program which generates multi-photon emission from events created by the 
host program. Another program, HORACE~\cite{HORACE}, is available which
includes exact ${\cal O}(\alpha)$ EWK corrections together 
with a final state QED parton shower. 

The other available NLO and NNLO calculations are implemented as MC 
integrations, which can calculate a cross-section but do not provide 
unweighted events. Some of these are more differential than others. 
For example, the NNLO rapidity distribution is available in a 
program Vrap~\cite{dixon2003}, but this distribution alone is not 
sufficient to calculate acceptances with cuts on the lepton 
pseudorapidities and transverse 
momenta. A differential version of this NNLO calculation is implemented in 
a program FEWZ~\cite{FEWZ}, but this is not an event generator.  
Another available program
is  ResBos-A~\cite{RESBOS}, which resums soft and collinear
initial state QCD radiation to all orders and includes 
NLO final state QED radiation. Resummation gives this program an
advantage in realistically describing the small $\pT$ regime.

Our analysis is conducted for di-lepton final states. The available 
calculations typically set the lepton masses to zero, so the lepton masses
will be neglected throughout this paper and the choice of final
state lepton has no effect on the calculations. In all results, $\ell$
may be interpreted as either an electron or muon. We have chosen three sets 
of experimental cuts to reflect detector capabilities and to demonstrate 
the impact of physics effects on the acceptances depending
on the selection criteria.

\section{Electroweak Corrections}
\label{sec:EWK}

As noted above, both NNLO QCD corrections and NLO electroweak (EWK) corrections
 are expected to be needed to reach precisions on the order of $1\%$ or 
better in $Z$ boson production.  NLO electroweak\cite{Baur} and 
QCD corrections\cite{NLO} are known 
both for $W$ boson production and $Z$ boson production. However, current
state-of-the-art MC generators do not include both sets of
corrections. The generator MC@NLO~\cite{MCNLO} combines a MC event 
generator with NLO calculations of rates for QCD processes and 
uses the HERWIG event generator for the parton showering, but it does not
include EWK corrections. Final state QED can be added using 
PHOTOS~\cite{PHOTOS,GoloWas}, a process-independent module for 
adding multi-photon emission to events created by a host generator.
However, some ${\cal O}(\alpha)$ EWK corrections are still missing.

To study the error arising from missing ${\cal O}(\alpha)$ EWK
corrections, we used HORACE~\cite{HORACE}, a MC event
generator that includes initial and final-state QED radiation in 
a photon shower approximation and exact ${\cal O}(\alpha)$ EWK
corrections matched to a leading-log QED shower. To determine the
magnitude of the error, we then compared the
results from this generator to a Born-level calculation with final-state
QED corrections added by PHOTOS. 

Specifically, we compared $pp \to Z/\gamma^* \to \ell^{+}\ell^{-}$ events
generated by HORACE with the full {\cal O}($\alpha$) corrections and
parton-showered with HERWIG, to these events generated again by 
HORACE, but without EWK corrections (Born-level), showered with 
HERWIG+PHOTOS.  CTEQ6.5M parton distribution functions~\cite{CTEQ} were
used in the calculations.  The results are shown in
Table~\ref{table:horace_xs_comp} and in Figs.~\ref{fig:horace_mass} 
--~\ref{fig:horace_FSR}.  

For Table~\ref{table:horace_xs_comp}, the 
events are generated in the kinematic region defined by final state
invariant mass  $\Mll > 40$
  GeV/$c^2$, and for each lepton, 
$p_{T}^{\ell} > 5$ GeV/$c$, and $|\etael| < 50.0$.
Two sets of cuts are used: 
\begin{center}
\begin{tabular}{llll}
Loose Cut: & $\Mll > 40$ GeV/$c^2$, & $\pTel > 5$ GeV/$c$, &
  $|\etael| < 50.0,$\\
Tight Cut: & $40 < \Mll < 140$ GeV/$c^2$, & $\pTel > 20$ GeV/$c$, &
  $|\etael| < 2.0.$\\
\end{tabular}
\end{center}

\TABLE[ht]{
\label{table:horace_xs_comp}
\begin{tabular}{|l|c|c|c|c|}
\multicolumn{4}{c}{\bf Photonic and Electroweak Corrections}\\
\hline
 & Born & Born+FSR & ElectroWeak & Difference \\
\hline
$\sigma$ (No PS) & $1984.2 \pm 2.0 $ & $1984.2 \pm 2.0$ & $1995.7 \pm 2.0$
& 0.58 $\pm$ 0.14\% \\ 
$\sigma$ (Loose Cut) & $1984.2 \pm 2.0$ & $1964.6 \pm 2.0$ & $1961.4 \pm
2.0$ & 0.16 $\pm$ 0.14\% \\
$\sigma$(Tight Cut) & $612.5 \pm 1.1$ & $597.6 \pm 1.1$ & $595.3 \pm
1.1$ & 0.38 $\pm$ 0.26\% \\
$A$ (Loose Cut) & $0.9999 \pm 0.0000$ & $0.9901 \pm 0.0001$ & $0.9828 \pm
0.0001$ & 0.74 $\pm$ 0.02\% \\
$A$ (Tight Cut) & $0.3087 \pm 0.0005$ & $0.3012 \pm 0.0005$ & $0.2983 \pm
0.0005$ & 0.96 $\pm$ 0.21\% \\
\hline
\end{tabular}
\caption{Calculation of the $Z/\gamma^* \to \ell^{+}\ell^{-}$
  cross-section $\sigma$ and acceptance $A$ for various
  EWK corrections generated using HORACE 3.1, for $\ell=e$ or $\mu$.  }
}

\FIGURE[ht]{
\label{fig:horace_mass}
\setlength{\unitlength}{1in}
\begin{picture}(6.5,2.5)(0,0)
\put(0,0.2){\epsfig{file=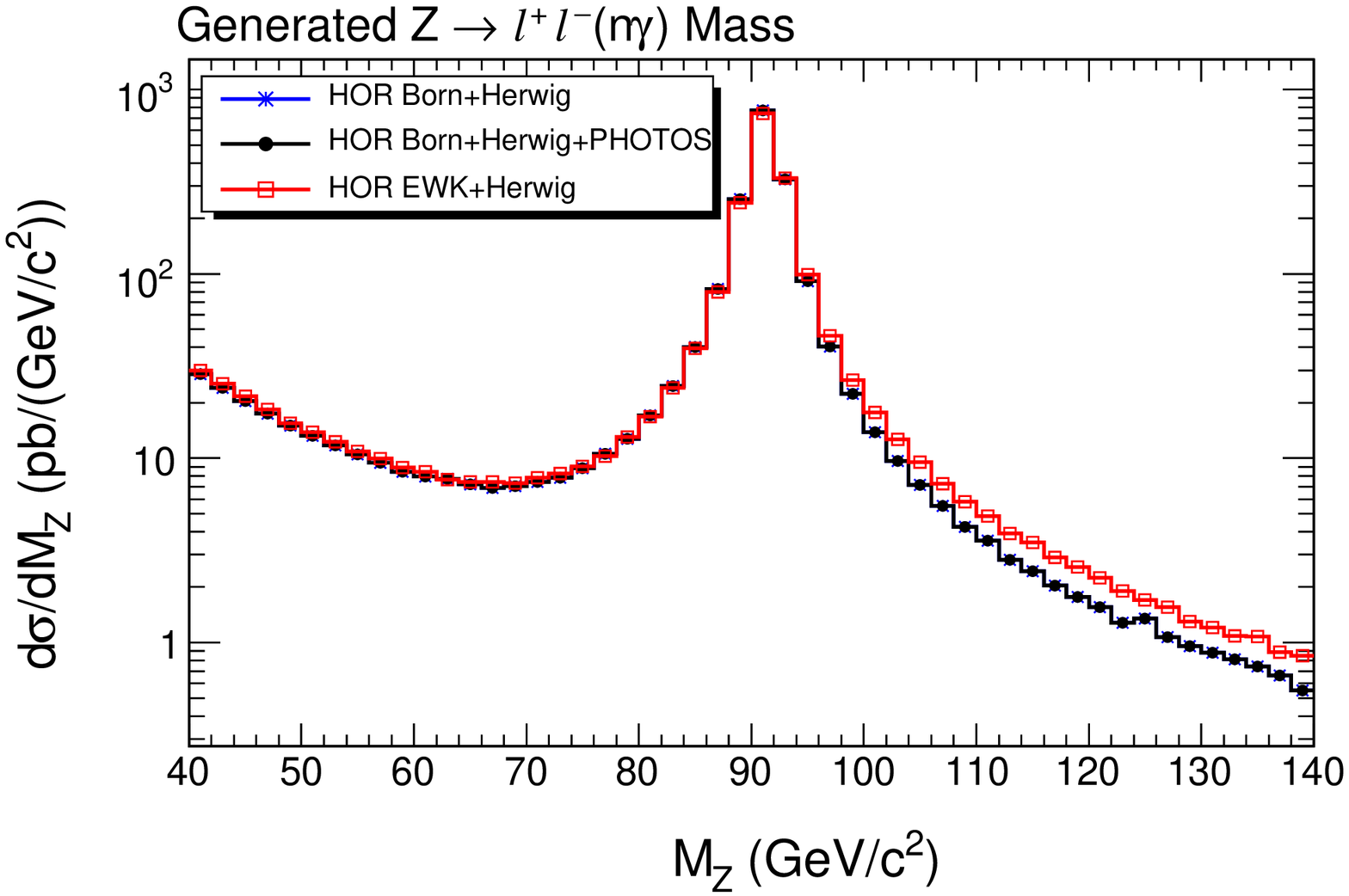,width=3.0in}}
\put(3,0.2){\epsfig{file=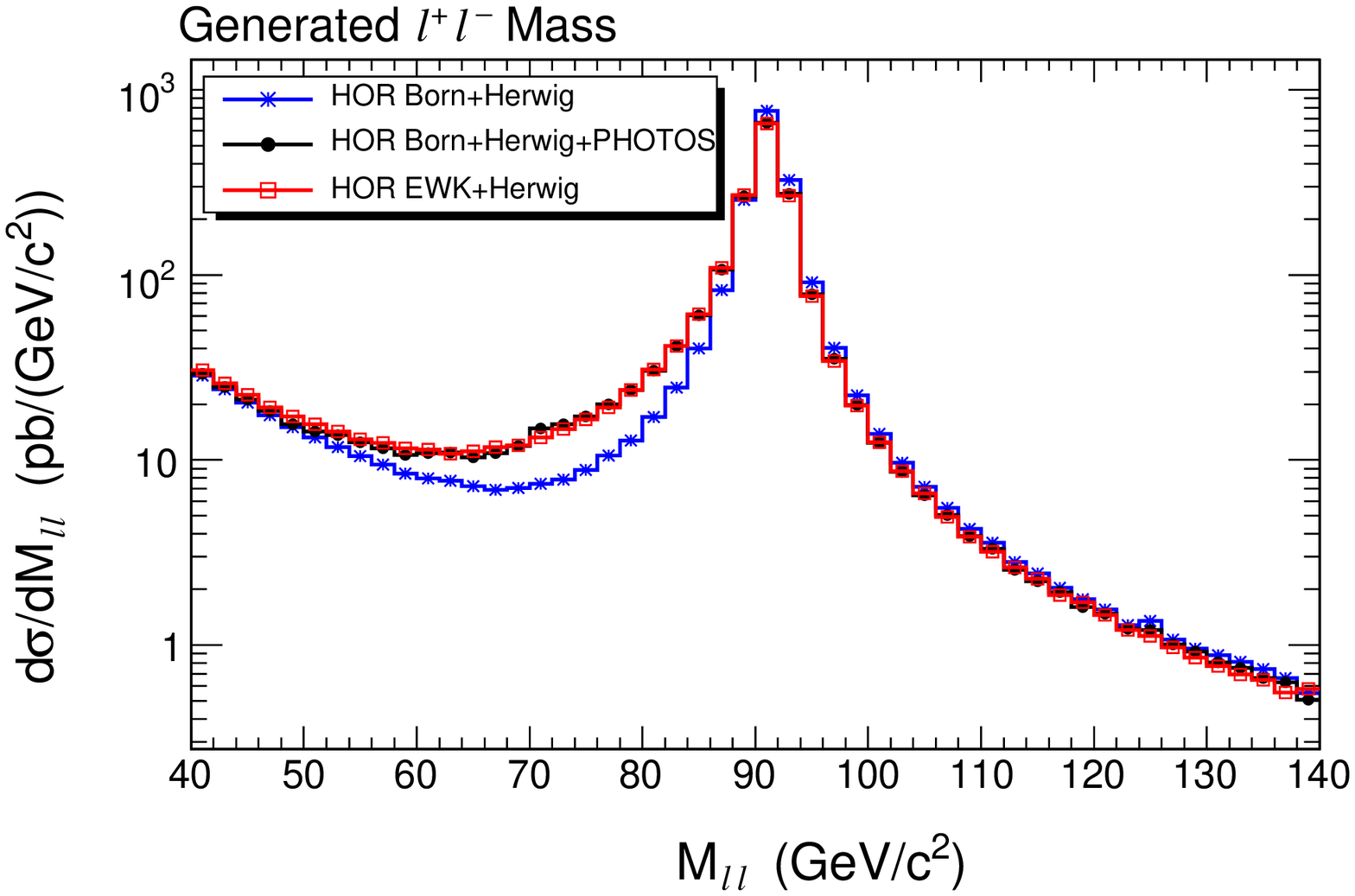,width=3.0in}}
\put(1.5,0){\small (a)}
\put(4.5,0){\small (b)}
\end{picture}
\vspace{-0.5cm}
\caption{ Comparison of the (a) $Z$ boson invariant mass distributions 
and (b) $\ell^+\ell^-$ invariant mass distributions for the
 process $Z/\gamma^* \to \ell^{+}\ell^{-}(n\gamma)$ in
 HORACE 3.1 including ${\cal O}(\alpha)$ EWK corrections 
showered with
 HERWIG (open red squares), HORACE Born-level showered with HERWIG plus
 PHOTOS (black circles), and HORACE Born-level (blue stars).}
}

\DOUBLEFIGLAB[th]{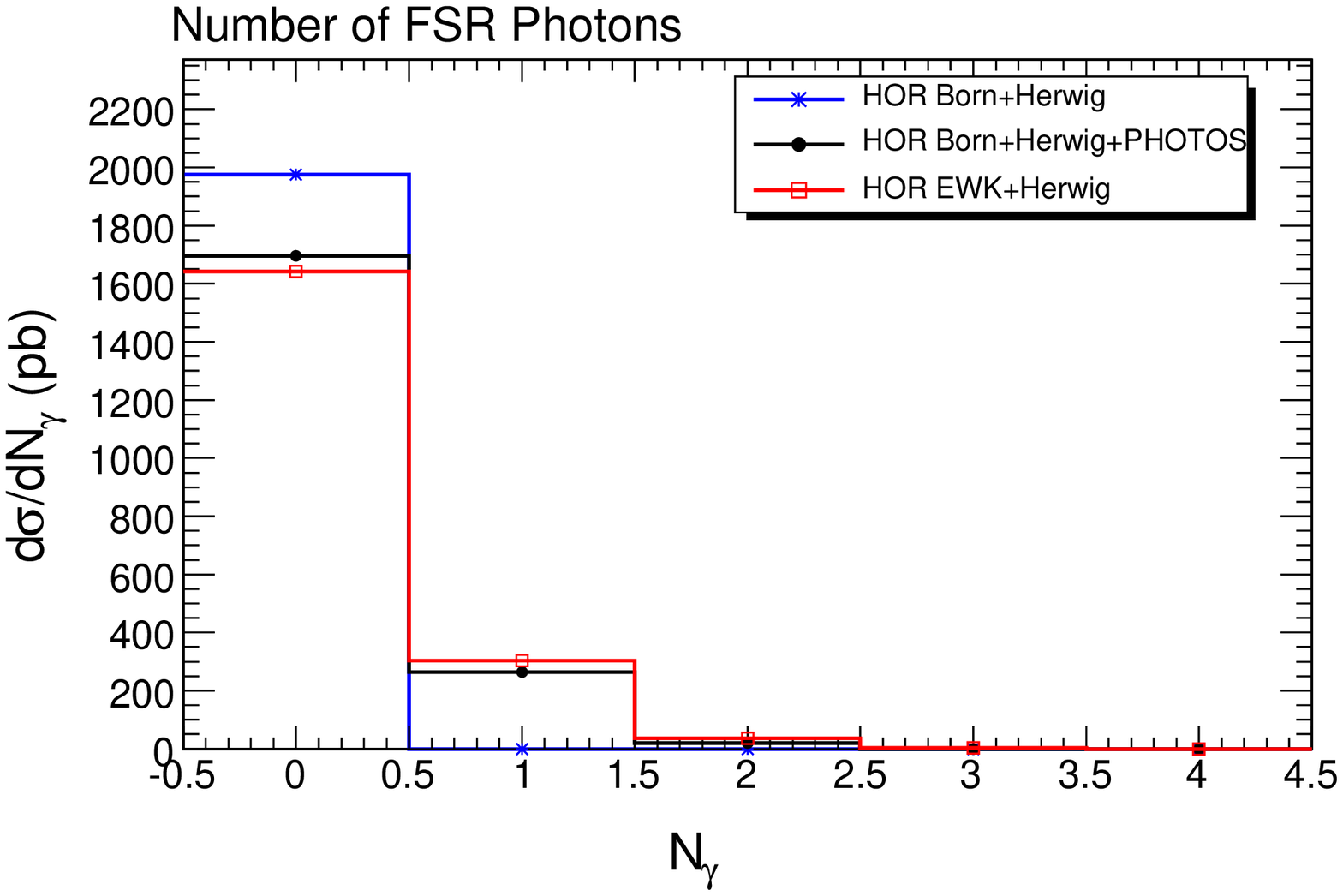,width=3in}{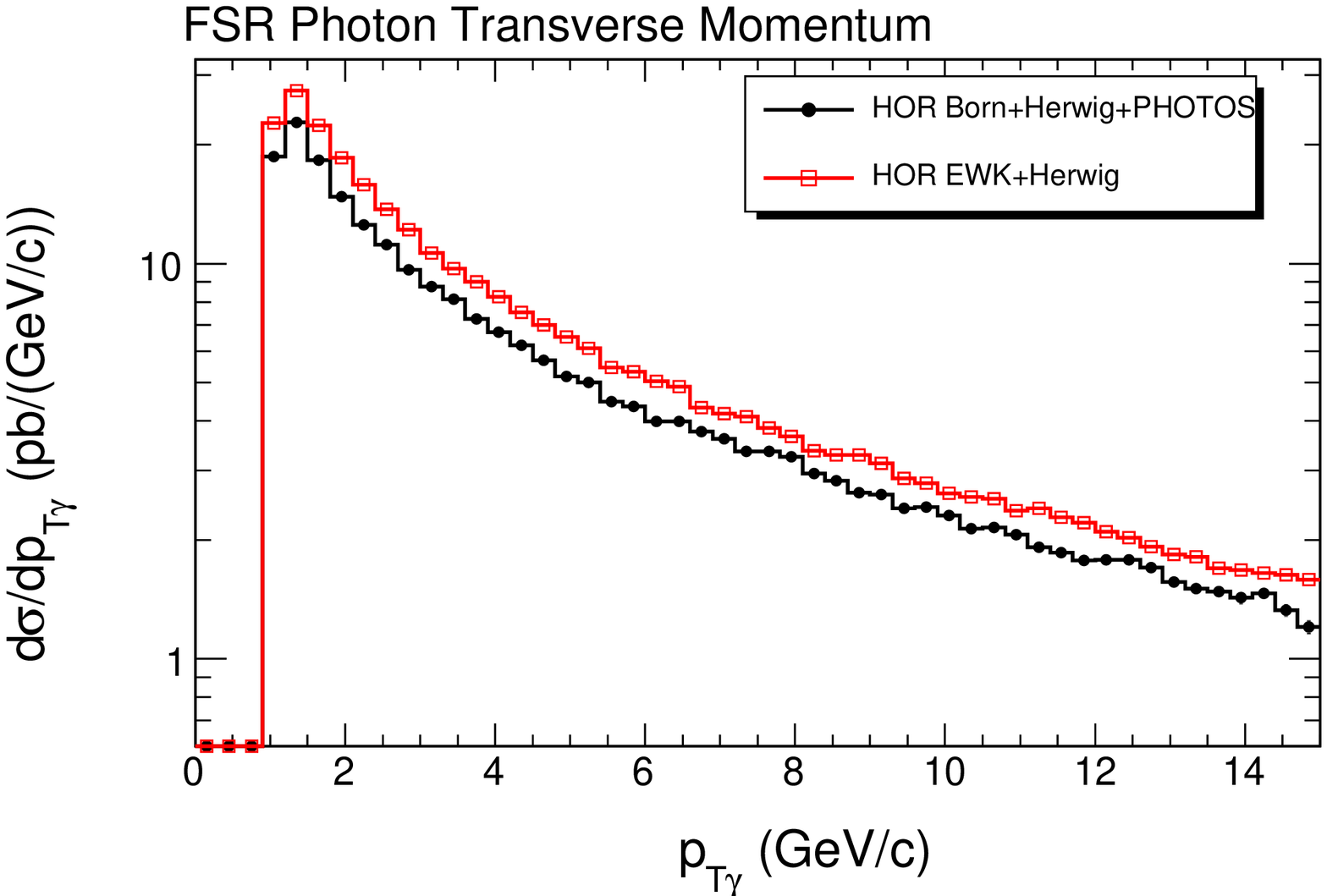,width=3in}{Comparison 
of the number $n$ of final state radiation (FSR) photons
  in $Z/\gamma^* \to \ell^{+}\ell^{-}(n\gamma)$ for
 HORACE 3.1 including ${\cal O}(\alpha)$ EWK corrections
showered with HERWIG (open red squares), HORACE Born-level showered with
HERWIG plus PHOTOS (black circles), and HORACE Born-level (blue stars).}{Comparison of $Z/\gamma^* \to \ell^{+}\ell^{-}(n\gamma)$ final state
  radiation (FSR) transverse momentum distributions for
  HORACE 3.1 including EWK corrections showered with
  HERWIG (open red squares) and HORACE Born-level showered with HERWIG plus
  PHOTOS (black circles).}{fig:horace_FSR}

\FIGURE[ht]{
\setlength{\unitlength}{1in}
\begin{picture}(6.5,2.5)(0,0)
\put(0,0.2){\epsfig{file=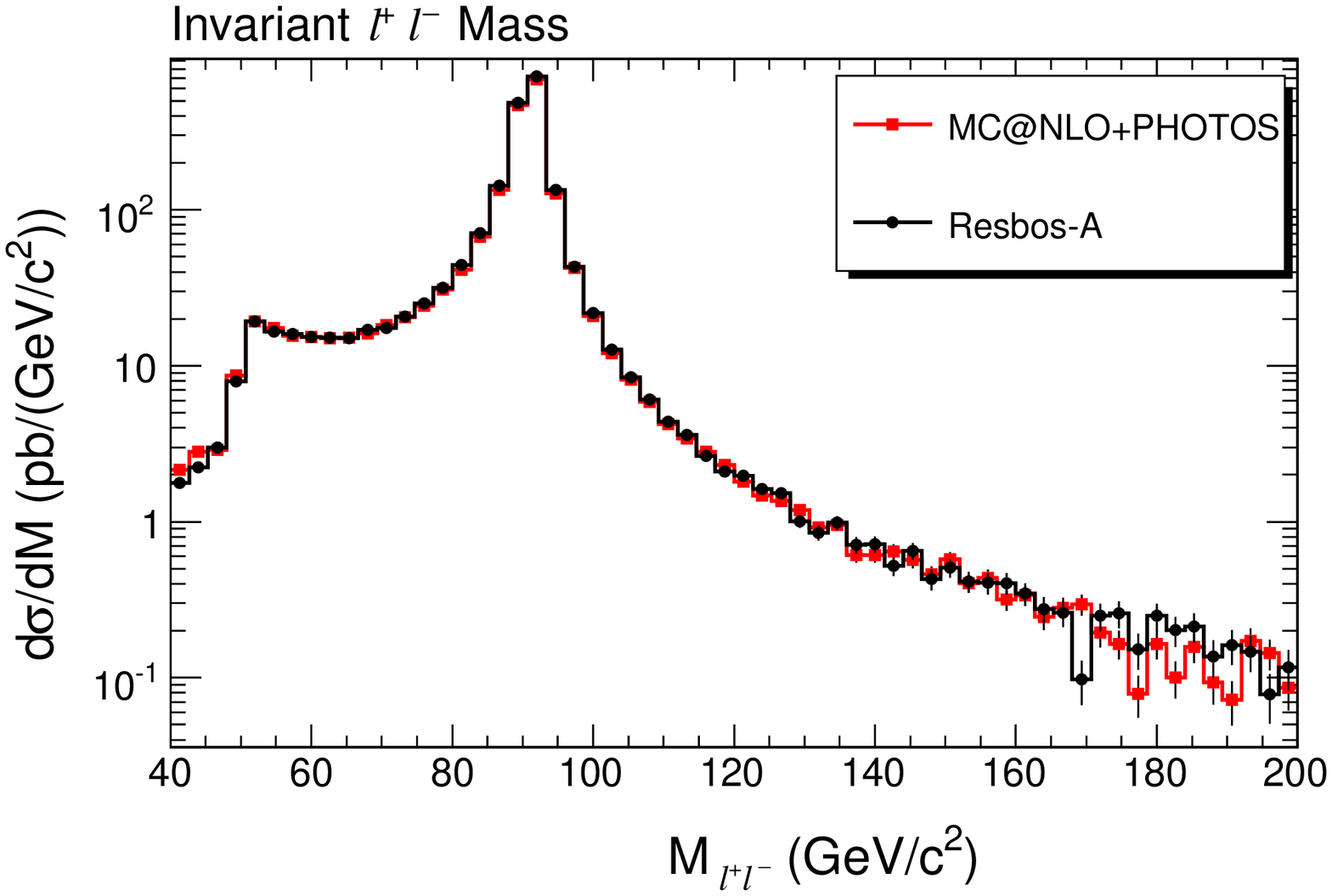,width=3in}}
\put(3,0.2){\epsfig{file=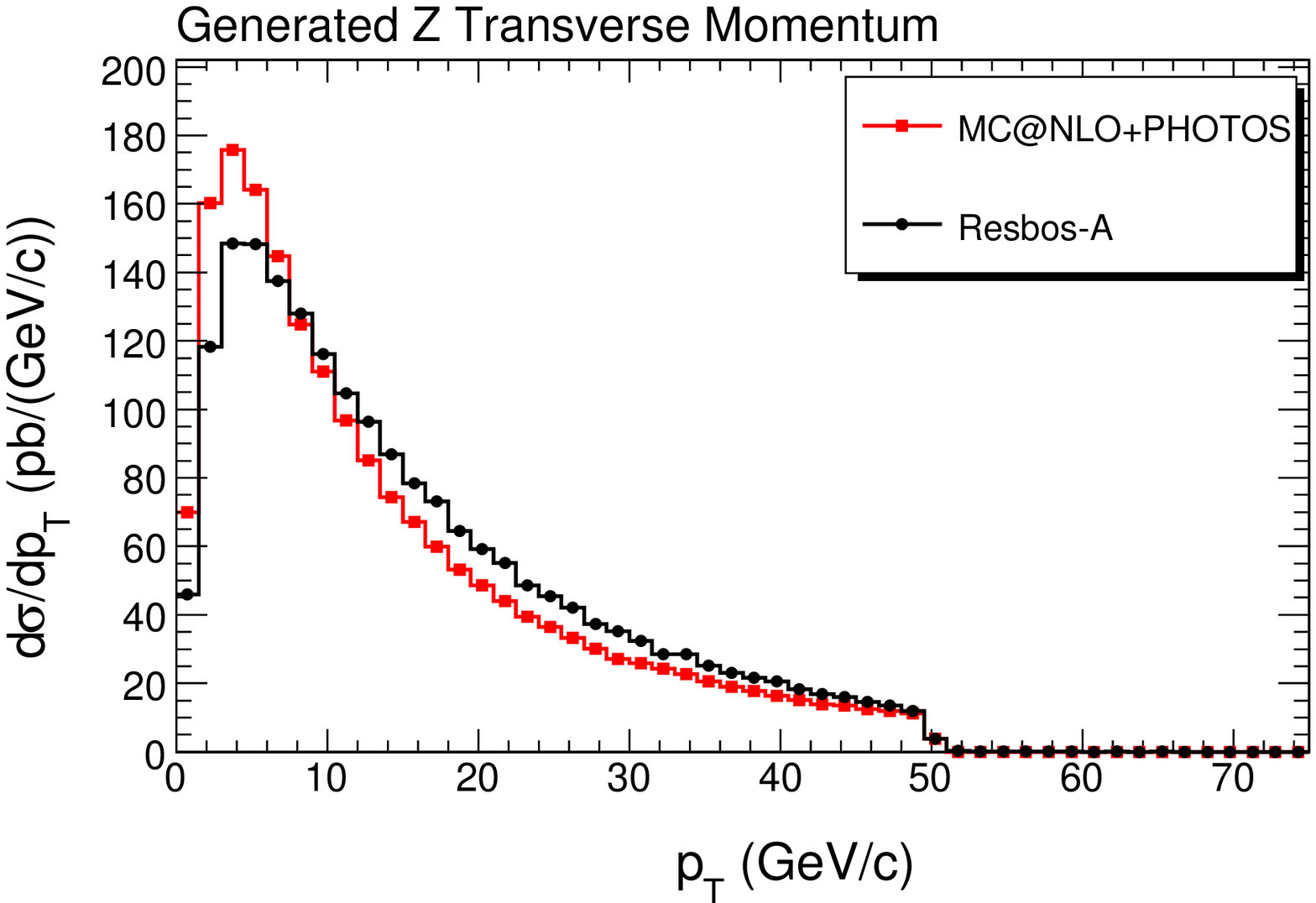,width=3in}}
\put(1.5,0){\small (a)}
\put(4.5,0){\small (b)}
\end{picture}
\vspace{-0.5cm}
\caption{ Comparison of the (a) di-lepton invariant mass and (b) $Z$ transverse momentum
  distributions for the process $Z/\gamma^* \to
  \ell^{+}\ell^{-}(n\gamma)$, in MC@NLO
  with PHOTOS (red squares) and Resbos-A (black circles).}
\label{fig:resbos_mass}
}

The first row in the table shows the total generator-level cross-sections
before QCD parton showering (identified by the label ``No PS''). At
Born level, without the corrections, the ``loose'' cut is essentially 
identical to the generator-level cut. 
The Born+FSR column shows the effect of applying final state 
radiation (FSR) corrections only via PHOTOS. In PHOTOS, FSR 
affects the rates through the cuts only. As noted in Ref.~\cite{GoloWas}, 
the combined effect of the complete first order real and virtual corrections 
not included in the standard version of PHOTOS would increase the total rate by a factor of $1 + 3\alpha/4\pi$, a 0.17\% increase.  The ElectroWeak 
column includes the full HORACE EWK corrections.  In the final column, 
we give the difference between the previous two columns, to compare the 
full EWK correction FSR alone.  The results show agreement within $1\%$ 
between the two schemes. The maximum error in the cross-section is 0.58\%
without added parton shower, and the maximum error in the acceptance is 
0.96\% for the tighter cut.  Therefore, we recommend using MC@NLO 
interfaced with PHOTOS as our primary event generator for measurements 
at the $Z$ peak, until higher precision is required~\cite{LESHOUCES}. 

For completeness, we also compared MC@NLO interfaced with PHOTOS 
with distributions from ResBos-A~\cite{RESBOS}, a MC simulation that
includes final state NLO QED corrections to $W/Z$ boson production and
higher-order logarithmic resummation of soft and collinear QCD radiation.
Fig.~\ref{fig:resbos_mass} shows the distributions
for $\Mll$ and the $\pT$ of the $Z$ boson, respectively.  
The latter exhibits especially well 
the effects of the soft and collinear resummation at low $\pT$ in 
ResBos-A~\cite{RESBOS}.  The only generator-level cut used in these 
comparison plots is the $\Mll > 50$ GeV/$c^2$. Otherwise, the default
parameters were used with no tuning. 

\TABLE[ht]{
\begin{tabular}{|c|c|c|c|}
\hline
  & Invariant Mass (GeV/$c^2$)& Pseudorapidity 
& Transverse Momentum (GeV/$c$)\\
\hline
Cut 1 & $\Mll > 40$ &  $|\etael| < 2.0$ & $\pTel > 20$ \\\hline
Cut 2 &  $\Mll > 40$ &  $1.5< |\etael| < 2.3 $ & $\pTel > 20$ \\ \hline
Cut 3 & $79< \Mll < 104$ &  $|\etael| < 2.0 $ & $\pTel > 20$ \\ \hline
\end{tabular}
\caption{Acceptance regions for final state leptons in NNLO studies}
\label{table:acceptance}
}

\section{NNLO QCD Uncertainties}
\label{sec:QCD}

QCD uncertainties include errors due to missing higher-order
corrections in the hard matrix element, uncertainties in the parton 
distribution functions, and approximations made in the showering
algorithms. In the following, we will evaluate the errors introduced by 
omitting the NNLO corrections by using MC@NLO, and calculate $K$-factors
which can be used to introduce NNLO corrections to the MC@NLO calculation.
We will also examine the effect of uncertainties in the PDFs. For these
studies, we choose three sets of experimental cuts~\ref{table:acceptance} 
to reflect detector capabilities and to demonstrate the impact 
of physics effects on the acceptances depending on the selection criteria.
Here, $\etael$ and $\pTel$ are the pseudorapidity and transverse momentum of the final state leptons.  The different rapidity ranges and invariant di-lepton mass cuts provide useful separation for between 
regions of the $Z$ spectrum which have different sensitivities to some of the sources of uncertainties
and evaluate the impact of mass cut on the theoretical error.

\TABLE[hb]{
\begin{tabular}{|c|c|c|c|c|c|}
\multicolumn{6}{c}{NNLO Cross Sections $\sigma$ (pb)}\\
\hline
Cut & MC@NLO & FEWZ NLO & FEWZ NNLO & $K$-factor & $K\times$ MC@NLO\\
\hline
$\sigma^{\mathrm tot}$& 
$2331\pm 3$&$2358.1\pm 2.3$&$2334.9\pm 4.6$&$0.9902\pm 0.0022$&$2308\pm 6$\\
1 &
$703.6\pm 1.1$&$716.0\pm 0.7$&$726.2\pm 5.5$&$1.0142\pm 0.0077$&$713.6\pm 5.6$\\
2 &
$71.3\pm 0.3$&$74.1\pm 0.07$&$73.39\pm 1.96$&$0.9902\pm 0.0280$&$70.6\pm 1.9$\\
3 &
$623.5\pm 1.0$&$657.2\pm 0.7$&$650.4\pm 4.0$&$0.9897\pm 0.0062$&$617.1\pm 4.0$\\
\hline
\multicolumn{6}{c}{NNLO Acceptances (\%) }\\
\hline
Cut & MC@NLO & FEWZ NLO & FEWZ NNLO & $K$-factor& $K\times$ MC@NLO\\
\hline
1 &
$30.18\pm 0.06$&$30.36\pm 0.04$&$31.10\pm 0.24$&$1.0243\pm 0.0081$&$30.92\pm 0.25$\\
2 &
$3.06\pm 0.01$&$3.143\pm 0.004$&$3.143\pm 0.084$&$1.0000\pm 0.0268$&$3.06\pm 0.08$
\\
3 &
$26.75\pm 0.06$&$27.87\pm 0.04$&$27.86\pm 0.18$&$0.9995\pm 0.0066$&$26.73\pm 0.19$
\\
\hline
\end{tabular}
\caption{Calculation of the $Z/\gamma^* \to \ell^{+}\ell^{-}$ 
($\ell=e$ or $\mu$) cross-section at NLO using MC@NLO, and at NLO and 
NNLO using FEWZ, for the cut region defined in Table \ref{table:acceptance}. 
The acceptances are relative to a total generated cross-section 
$\sigma^{\mathrm tot}$ for $\Mll > 40$ GeV$/c^2$.
}
\label{table:nnlo_calc}
}

We begin by examining the NNLO corrections, using the state-of-the-art
program FEWZ~\cite{FEWZ}, which is differential in the di-lepton invariant mass, 
and the lepton transverse momenta and pseudorapidities. 
The FEWZ program is at NNLO in perturbative 
QCD, and fully differential, giving correct acceptances including spin 
correlations, as well as taking into account finite widths effects and 
$\gamma^*-Z$ interference.  Since we are interested
primarily in studies about the $Z$ peak, we choose the renormalization and
factorization scales to be $\mu_{{}_{\scriptstyle F}} = 
\mu_{{}_{\scriptstyle R}} = M_Z$. Scale dependence
will be discussed in detail the next section.

A comparison of the effects of higher order QCD corrections on the 
cross-section and acceptance 
is presented in Table~\ref{table:nnlo_calc} and Figs~\ref{fig:acc_pt}
--~\ref{fig:acc_m2}. 
Both the NLO and NNLO calculations are done with CTEQ6.5M 
PDFs~\cite{CTEQ}. We comment further on this choice at the end of this section.
All results in the table are calculated at scale $M_Z$. 
In the figures, the NLO results are displayed as a band 
spanning the range of scales from $M_Z/2$ to $2M_Z$. The 
scale dependence of the NNLO result is small enough to be comparable to the
precision of the 

\newpage
\FIGURE[ht]{
\begin{tabular}{cc}
\epsfig{file=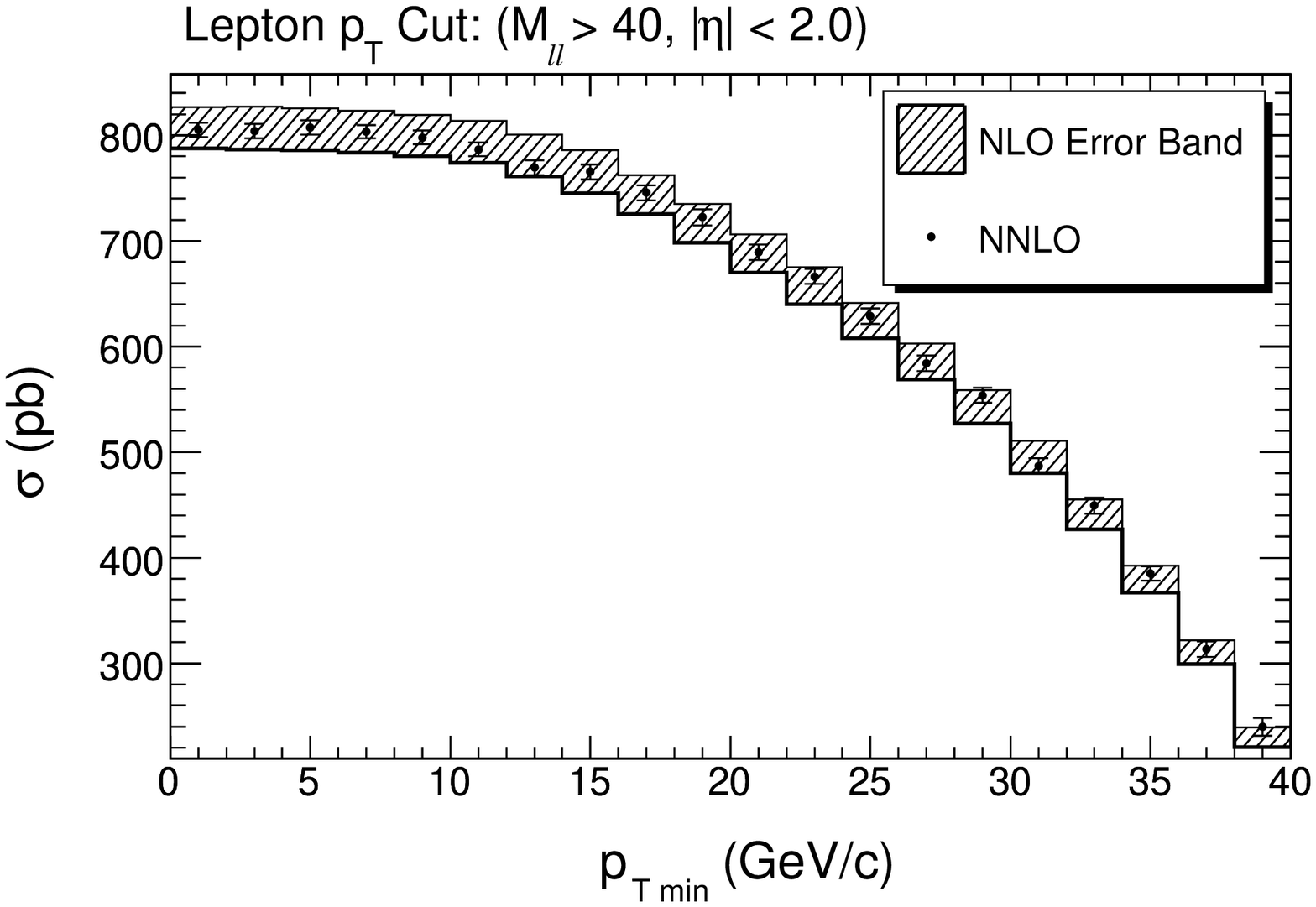,width=7.2cm} &
\epsfig{file=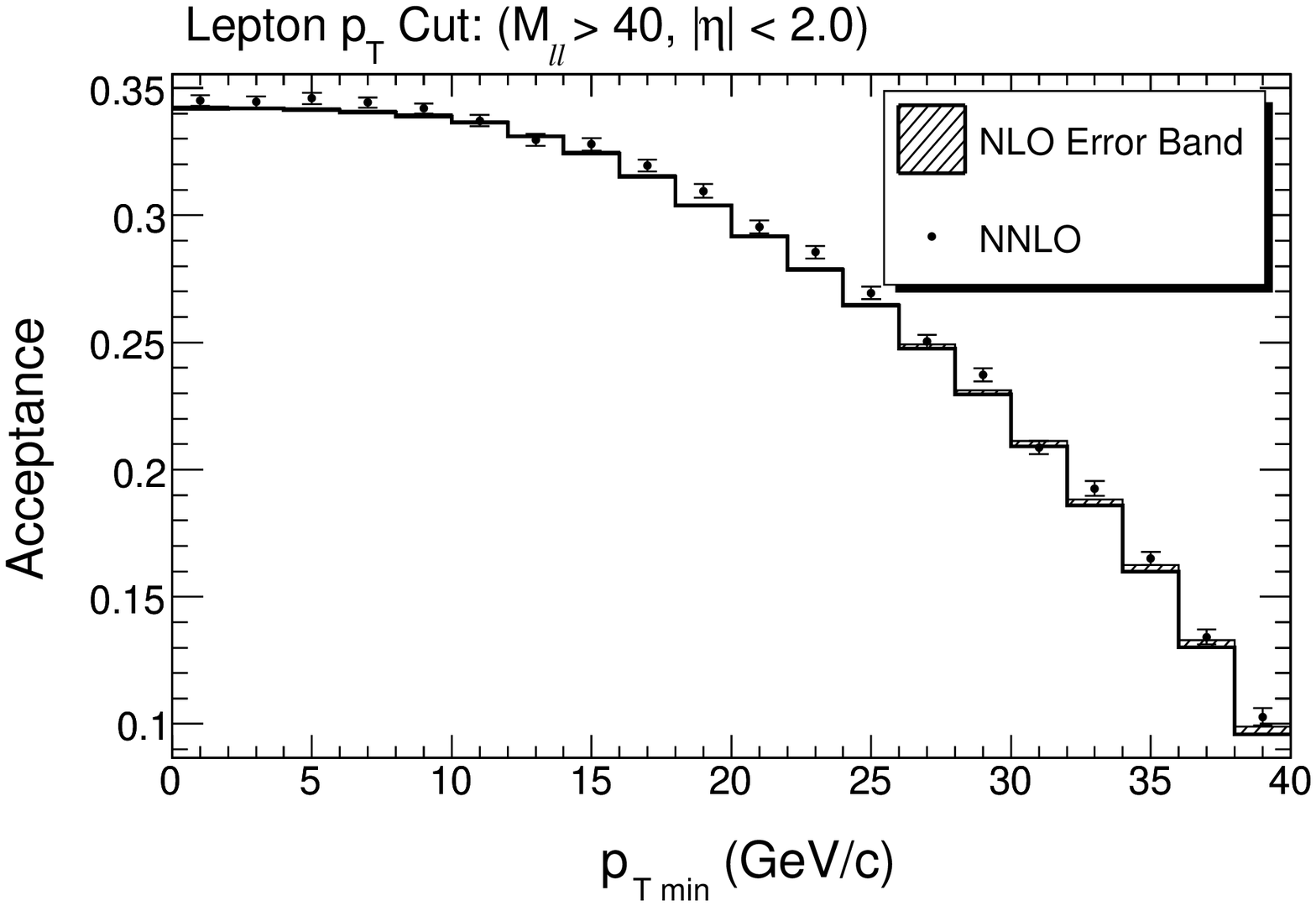,width=7.2cm}\\
(a) & (b) \\
\end{tabular}
\caption{ Cross-section (a) and acceptance (b) versus cut on lepton $\pT$
 at NLO (hashed bands) and NNLO (points), as calculated using FEWZ. }
\label{fig:acc_pt}
}

\FIGURE[ht]{
\begin{tabular}{cc}
\epsfig{file=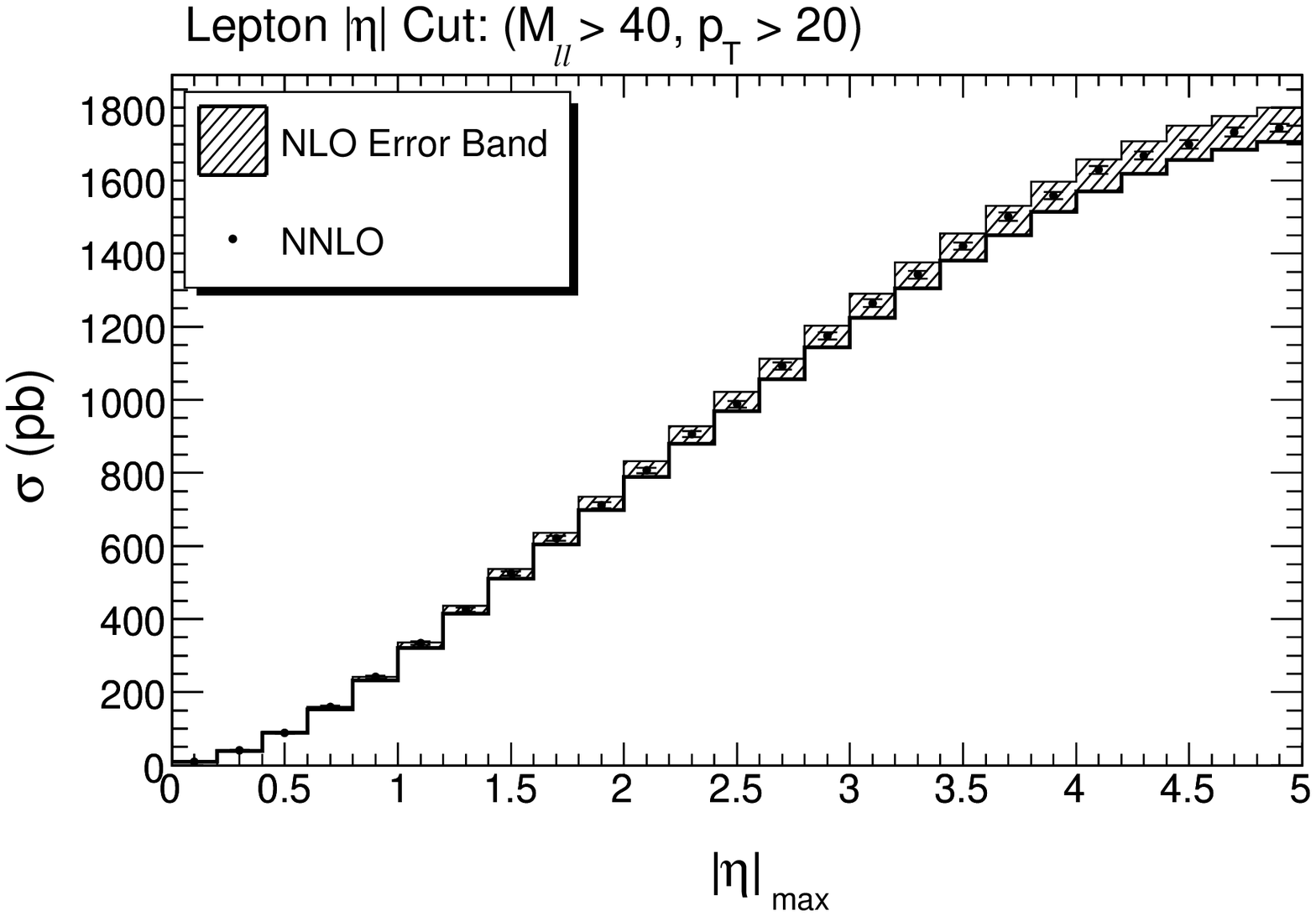,width=7.2cm} &
\epsfig{file=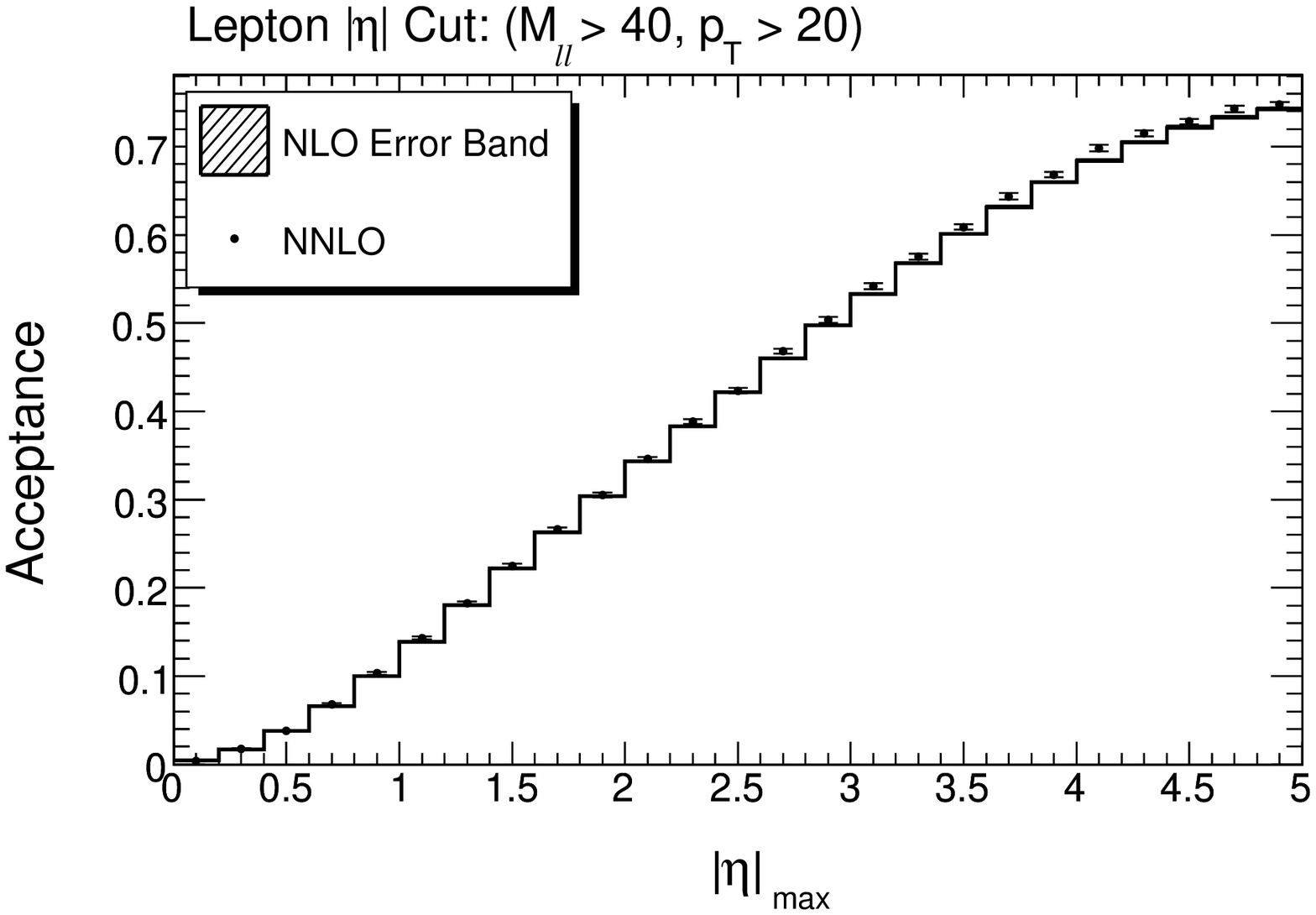,width=7.2cm}\\
(a) & (b) \\
\end{tabular}
\caption{ Cross-section (a) and acceptance (b) versus cut on lepton $|\eta|$
 at NLO (hashed bands) and NNLO (points), as calculated using FEWZ. }
\label{fig:acc_eta}
}

\FIGURE[ht]{
\begin{tabular}{cc}
\epsfig{file=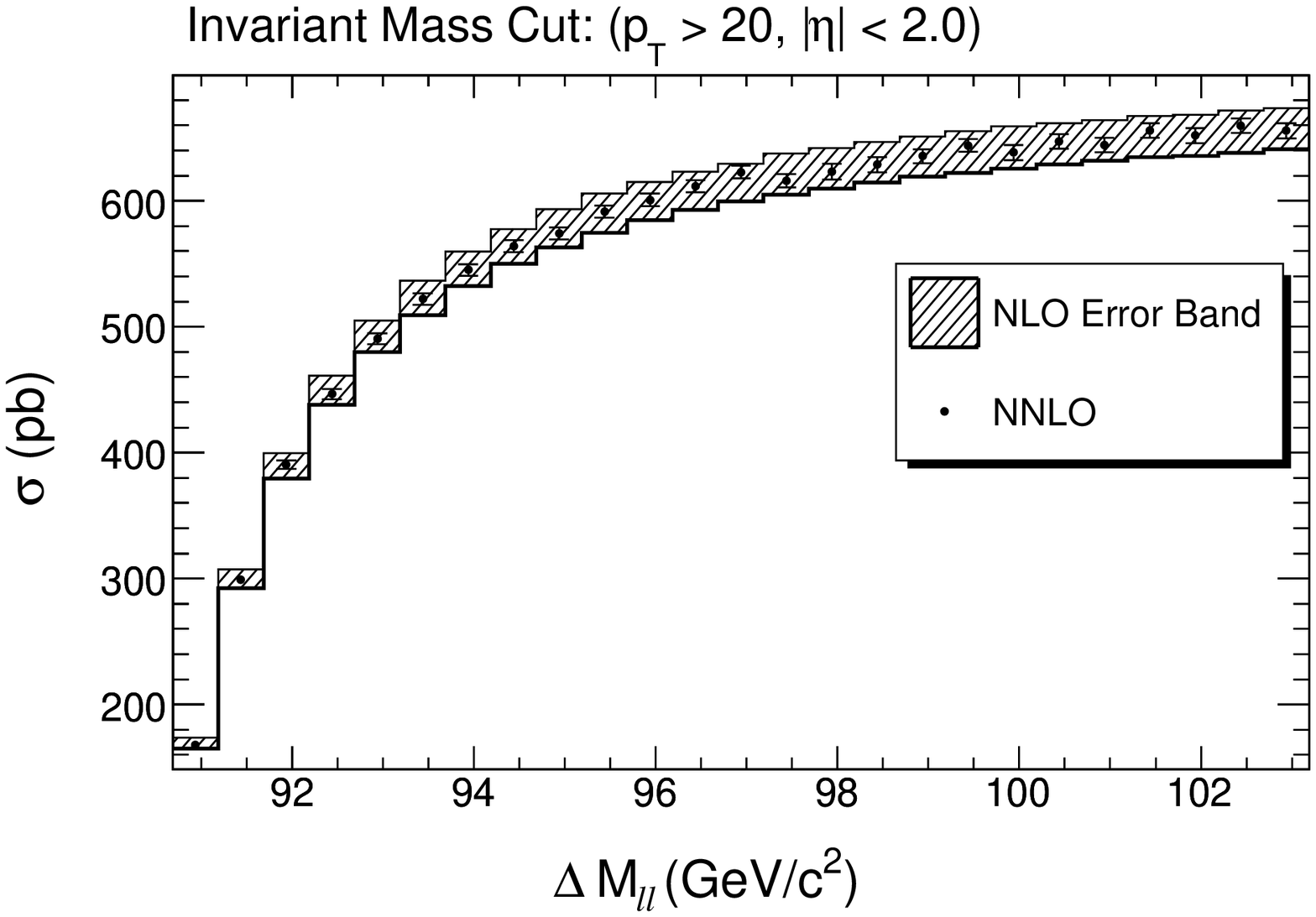,width=7.2cm} &
\epsfig{file=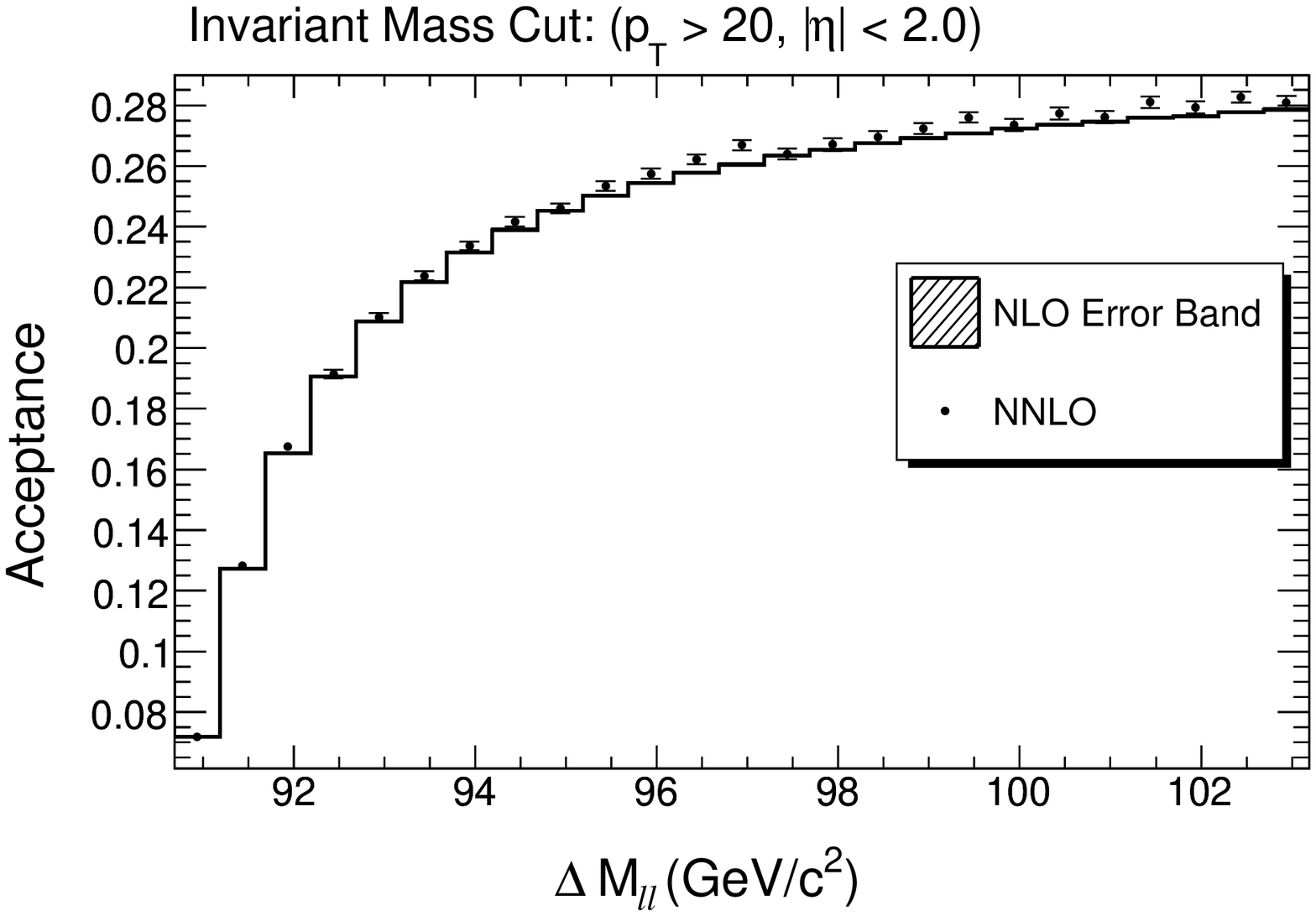,width=7.2cm}\\
(a) & (b) \\
\end{tabular}
\caption{ Cross-section (a) and acceptance (b) versus cut on di-lepton
 invariant mass $\Mll$, at NLO (hashed bands) and NNLO (points), as
 calculated using FEWZ. $\Mll$ is constrained to be in a band of
width $\Delta\Mll$ centered at $M_Z$. }
\label{fig:acc_m2}
}
\clearpage
\noindent
MC evaluation of the integrals, so only the average
of the high and low scales is plotted, with error bars reflecting a 
combination of statistical and scale variation
uncertainties.  Fig.\ \ref{fig:acc_m2} shows the cross-section and acceptance
for an invariant mass band of width $\Delta\Mll$ centered at $M_Z$, to 
study the effect of selecting different cuts about the $Z$ peak. 

Since the NNLO matrix element has not yet been 
interfaced to a shower, we cannot directly compare FEWZ to MC@NLO. The best 
we can do at this time is to use MC@NLO to obtain the NLO showered result, 
and multiply this by a $K$-factor obtained by taking the ratio of the 
NNLO to NLO results derived from FEWZ. This procedure is reasonable
except in threshold regimes where the fixed-order NLO result in FEWZ is 
unreliable.  In fact, the NLO cross-section calculated by FEWZ can 
become negative near the threshold $\pT = M_Z/2$. Similar
methods have been used for calculating NNLO corrections to Higgs 
production~\cite{FEHiP}.  The differences of these $K$-factors 
from unity are shown in Fig.\ \ref{fig:kfactor} 
for both the cross-sections and acceptances, as a
function of cuts on the lepton $\pT$ and $\eta$ as well as on the width 
$\Delta\Mll$ of an invariant mass cut centered at $M_Z$. 
The resulting accepted cross-section is shown in the  $K\times$ MC@NLO
column of Table\ \ref{table:nnlo_calc} and in Fig.~\ref{fig:mcnloxs} as a
function of the same cuts as in Fig.\  \ref{fig:kfactor}.  
The size of $K-1$ is a good indicator of the error due to missing NNLO
if MC@NLO is used without corrections. 

The results in table~\ref{table:nnlo_calc} show $K$-factors corresponding
to an NNLO correction of about $1\%$ for the cross-sections, or 
up to $2.4\%$ for the Cut 1 acceptance. FEWZ's convergence was not as good
for Cut 2, however, leading to an almost $3\%$ technical error in $K$ from the 
evaluation.\footnote{Here, ``technical error'' refers to a 
limitation on the precision arising from the specific calculation 
of the process, rather than from the approximations made in the calculation 
at a theoretical level (NNLO vs. NLO, for example). Technical precision is 
discussed further in the conclusions.}
In the other cases, a $1\%$ evaluation of the NNLO correction
was generally possible within a reasonable run time, which was typically
of order one month on the clusters used in this study. Longer run times do
not appear to improve the convergence significantly. The convergence 
limitations are due to the high dimensionality (11) of the Vegas~\cite{Vegas}
integrals and strong peaking of the integrands.  

We should note that CTEQ6.5M structure functions were used in the above 
calculations because the $K$ factors calculated are intended for rescaling
an MC@NLO calculation using CTEQ PDFs. However, since these PDFs do not include
NNLO corrections, it is useful to note the effect of repeating the calculations
with MRST 2002 NNLO PDFs.  The $K$-factors for the 
cross-section and acceptance for each of the three cuts is shown in 
Table\ \ref{table:PDFNNLO} for both sets of PDFs, for comparison. 
The $K(\sigma)$-factors are ratios of NNLO to NLO cross-sections, and $K(A)$ 
are ratios of NNLO to NLO acceptances relative to a total cross section 
for $\Mll>40$ GeV.  Except for the cross-section $K$ for Cut 1, where 
agreement is to $1.8\%$, the two PDF choices agree to within $1\%$, and 
within the computational uncertainty. 

\FIGURE[ht]{
\begin{tabular}{cc}
Cross-Sections & Acceptances \\
 & \\
\epsfig{file=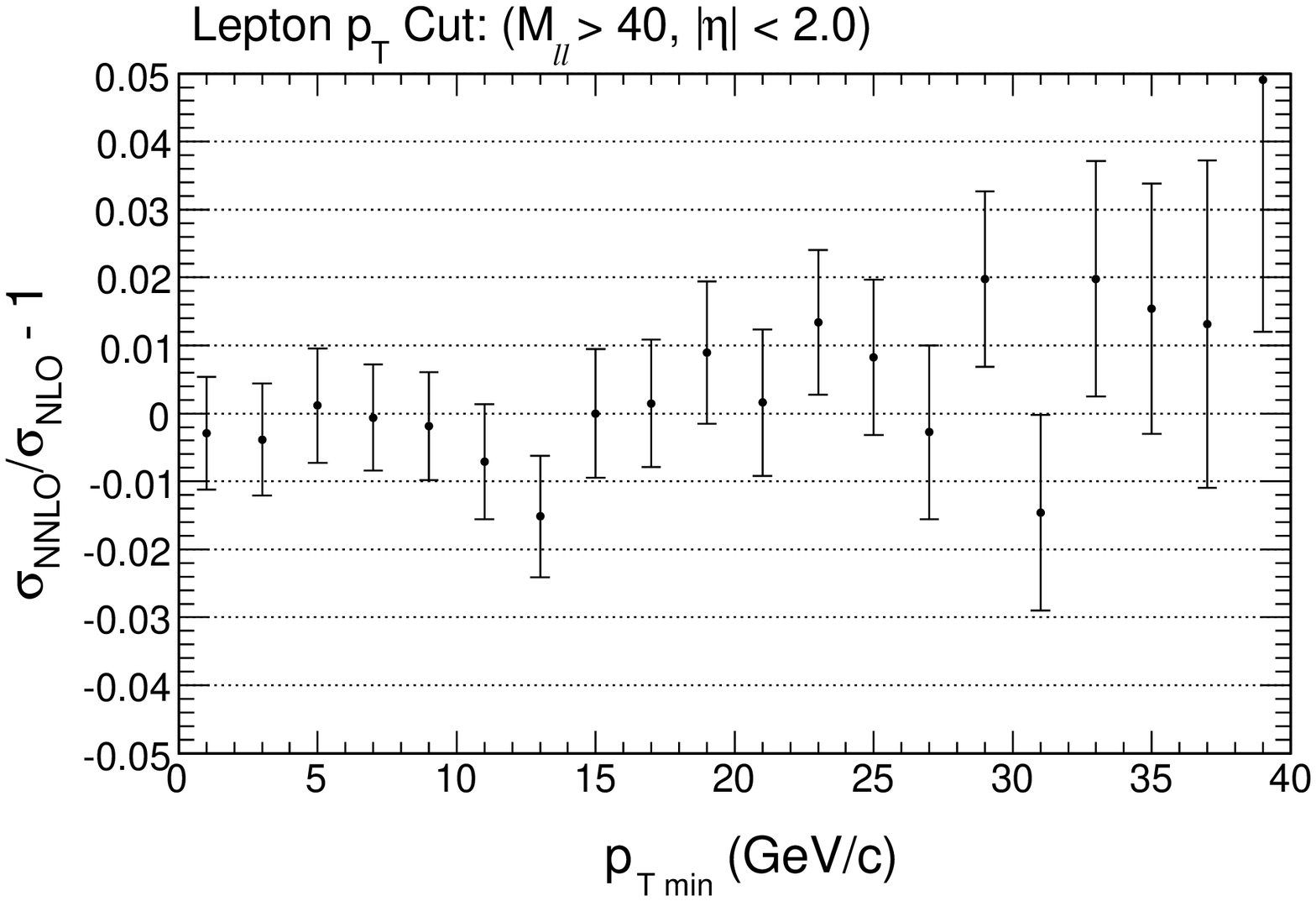,width=7.2cm} &
\epsfig{file=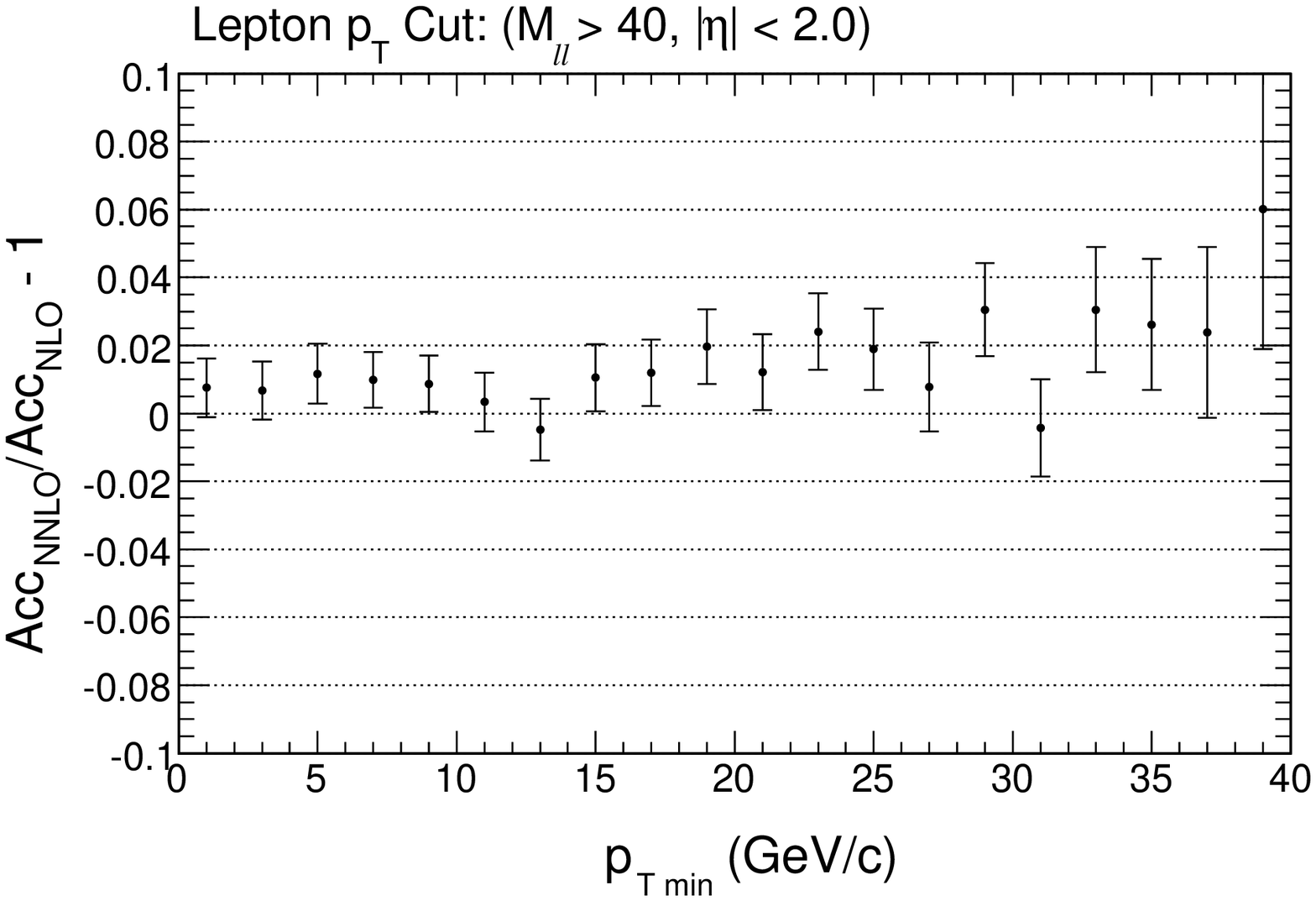,width=7.2cm} \\
\multicolumn{2}{c}{(a)} \\[3em]
\epsfig{file=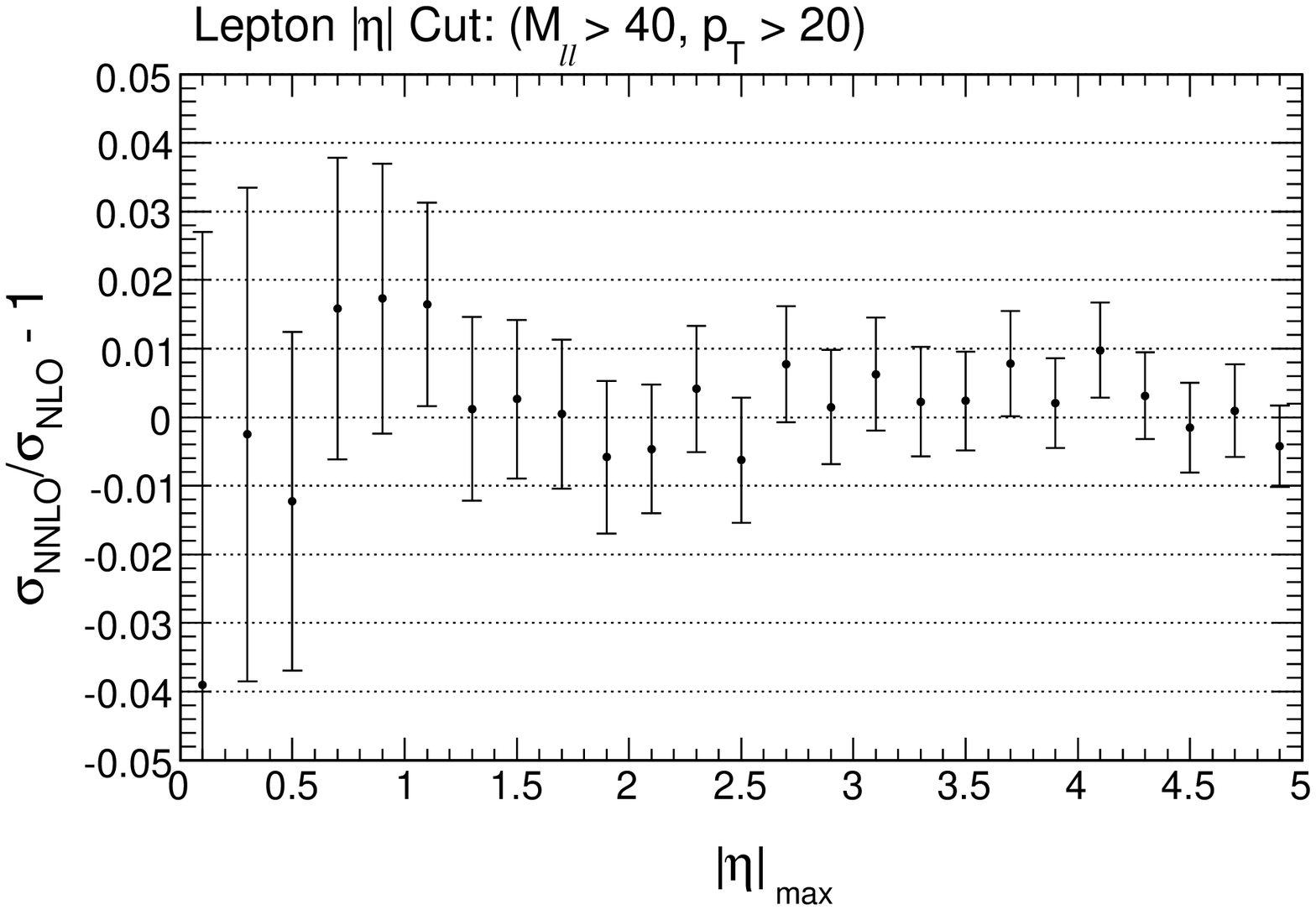,width=7.2cm} &
\epsfig{file=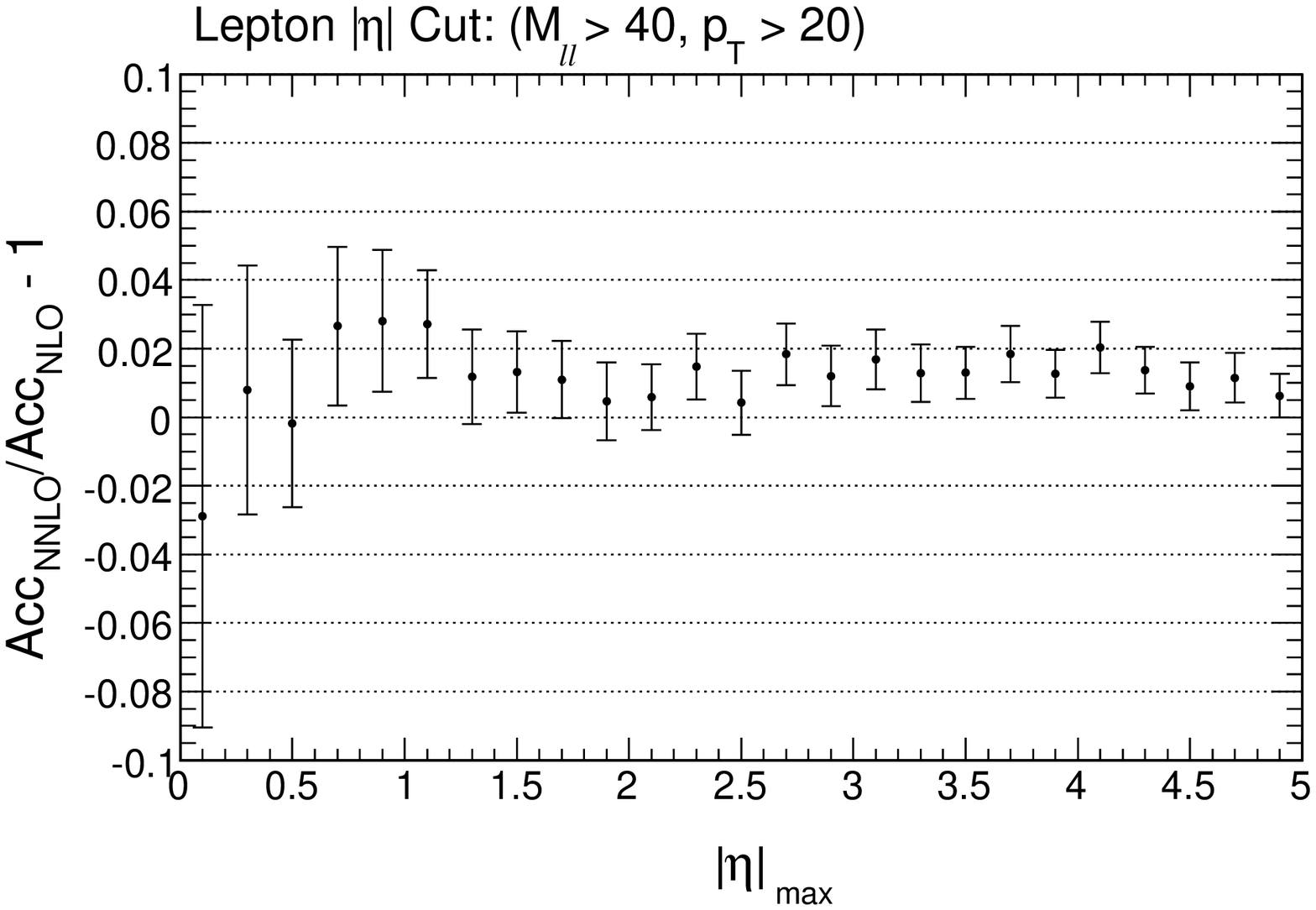,width=7.2cm}\\
\multicolumn{2}{c}{(b)} \\[3em]
\epsfig{file=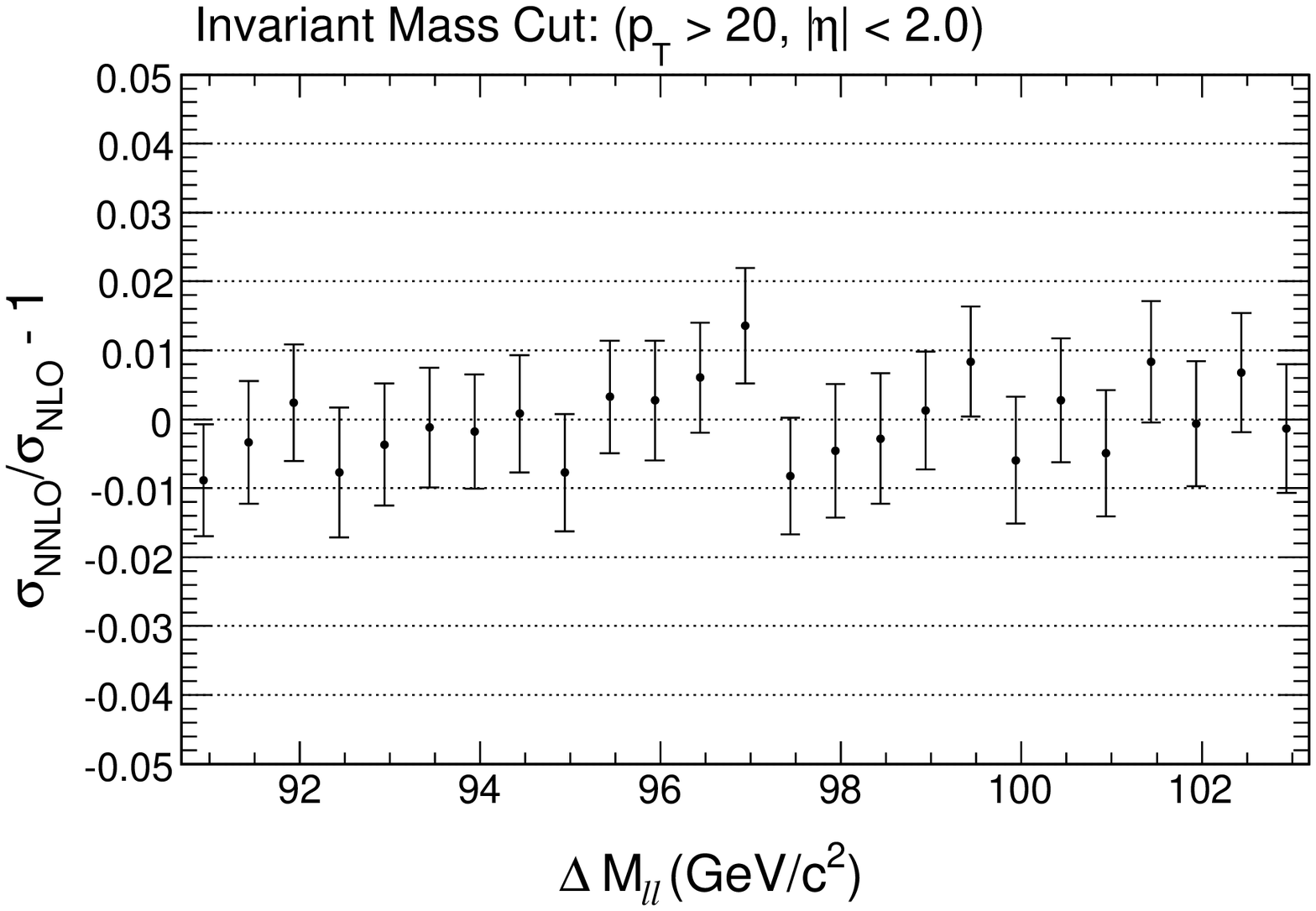,width=7.2cm} &
\epsfig{file=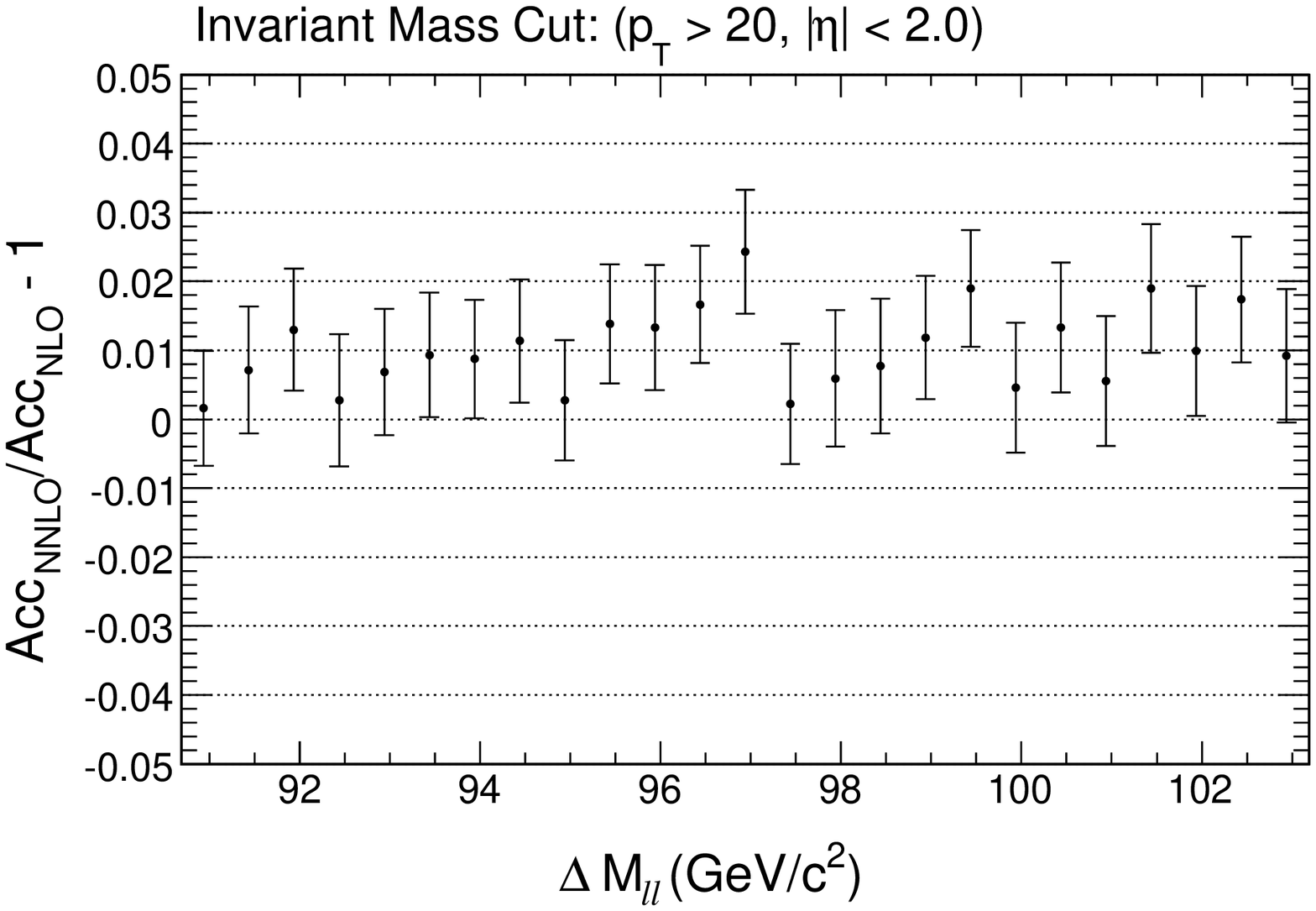,width=7.2cm} \\
\multicolumn{2}{c}{(c)} \\
\end{tabular}
\caption{ Fractional difference in the NNLO and NLO cross-sections (left-hand
side) and acceptances (right-hand side) as a 
function of the lepton (a) $\pT$, (b) $|\eta|$, and (c) $\Mll$ cuts as
in Figs.\  Figs~\ref{fig:acc_pt} --~\ref{fig:acc_m2}.
These differences are the factor $K - 1$ for the cross-section and acceptances,
respectively.}
\label{fig:kfactor}
}
\clearpage 

\FIGURE[ht]{
\begin{tabular}{cc}
\epsfig{file=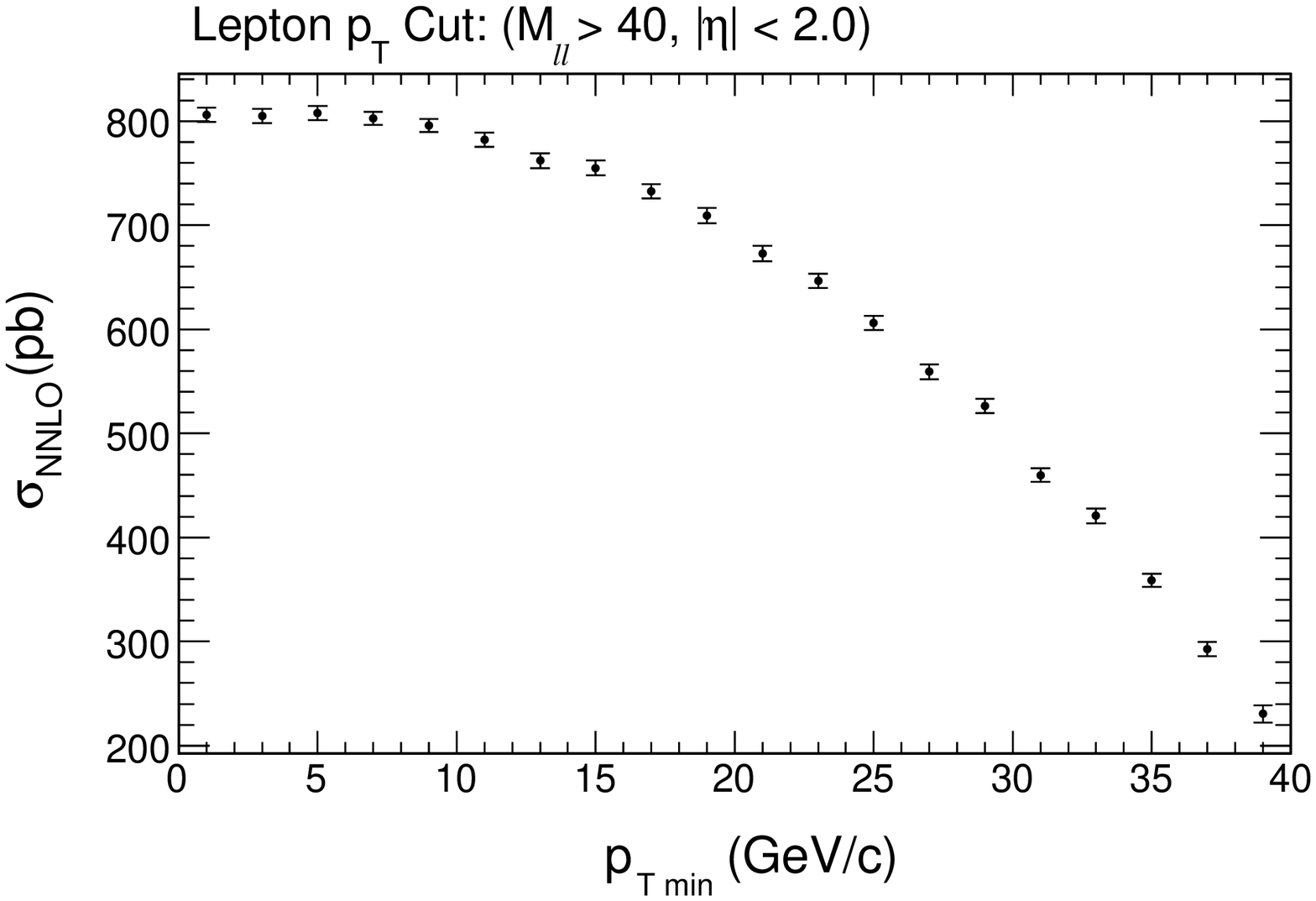,width=7.2cm} &
\epsfig{file=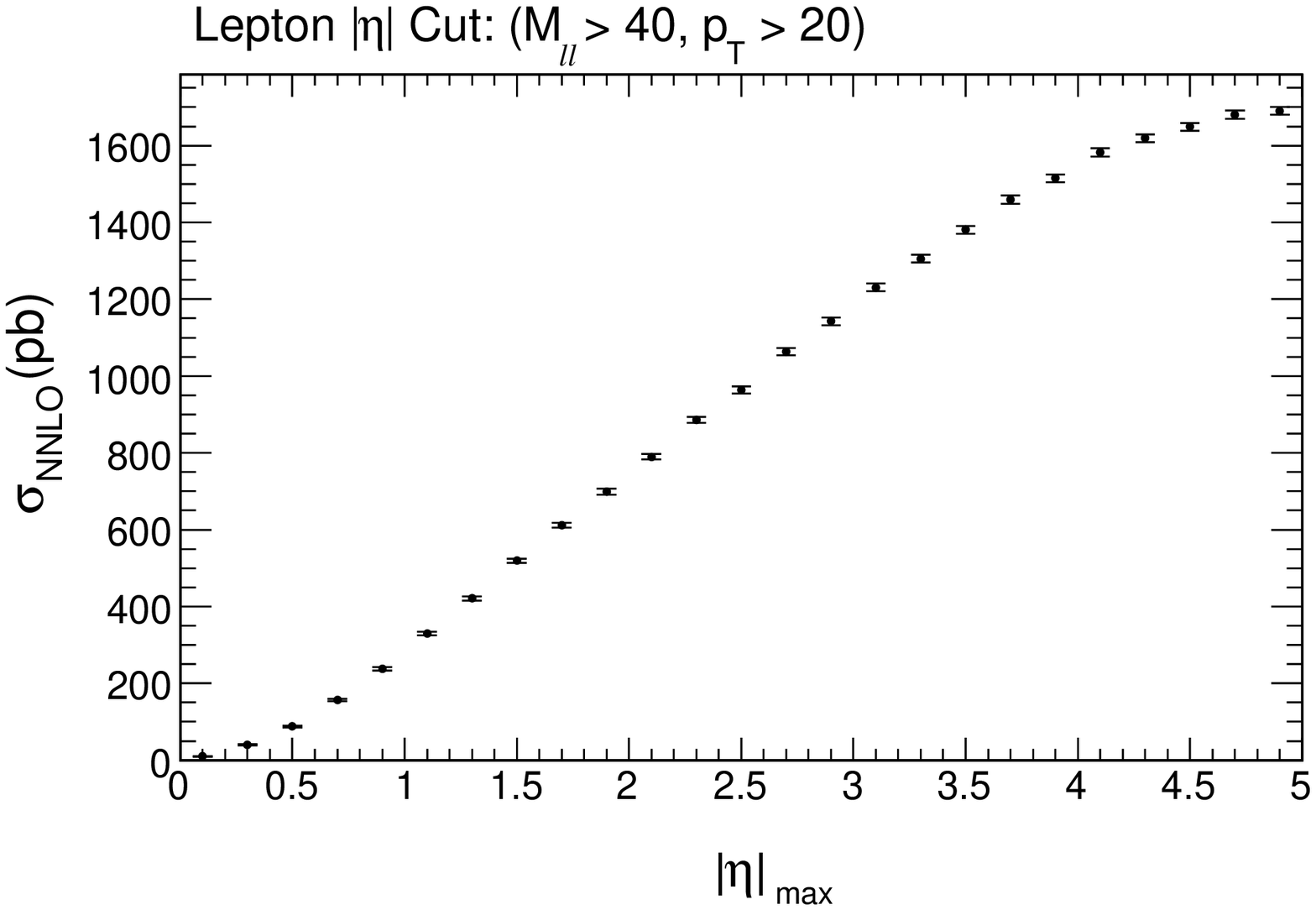,width=7.2cm}\\
~~~~(a) & ~~~~~~(b) \\
\multicolumn{2}{c}{\epsfig{file=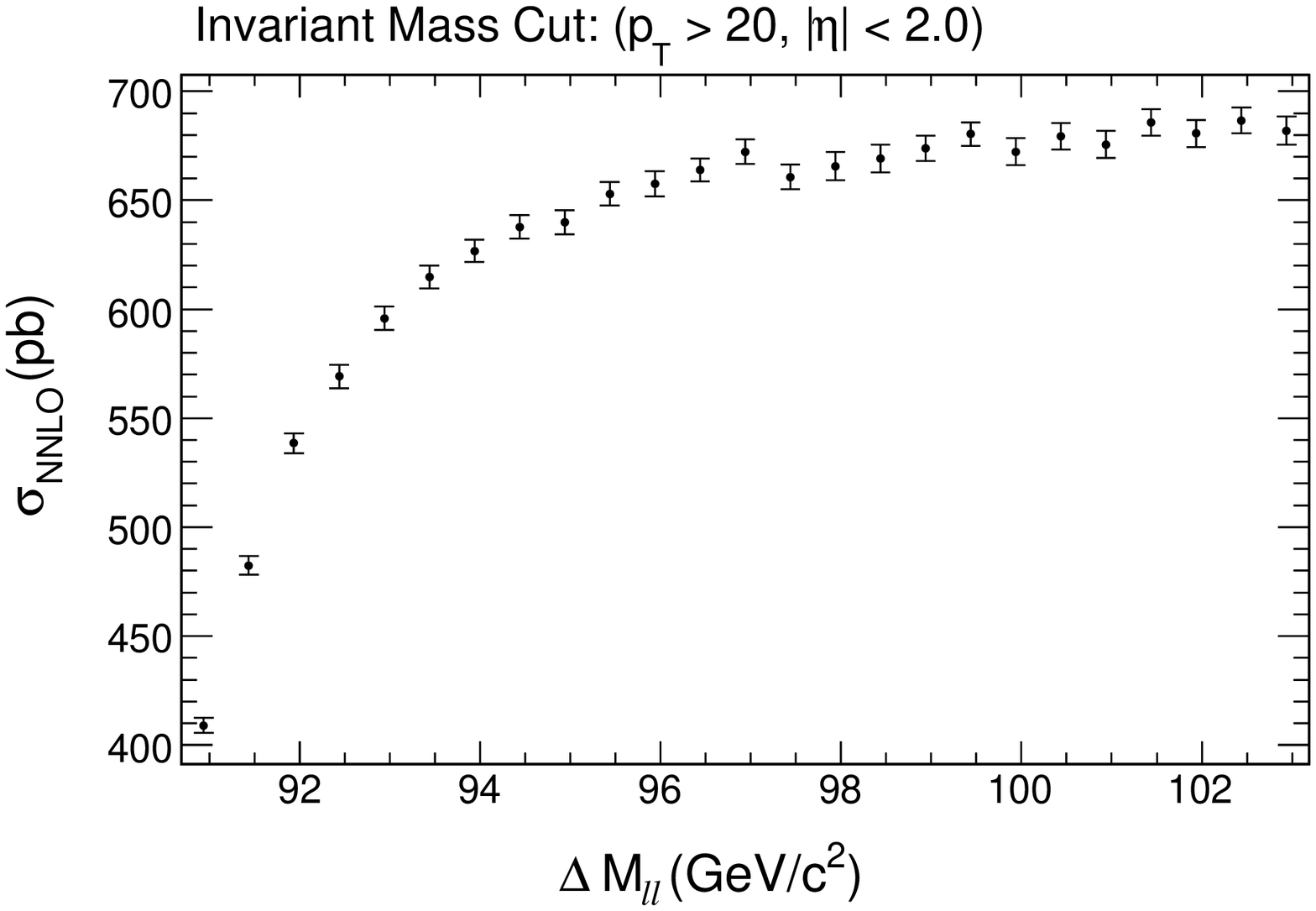,width=7.2cm}}
\\
\multicolumn{2}{c}{~~~~(c)}\\
\end{tabular}
\caption{ Accepted NNLO cross-section estimated from MC@NLO scaled by the $K$-factor versus 
 the lepton (a) $\pT$, (b) $|\eta|$, and (c) $\Mll$ cuts
as in Figs.~\ref{fig:acc_pt} --~\ref{fig:kfactor}.}
\label{fig:mcnloxs}
}

\TABLE[ht]{
\begin{tabular}{|c||c|c||c|c|}
\multicolumn{5}{c}{$K$-factors with CTEQ and MRST PDFs}\\
\hline
Cut & CTEQ $K(\sigma)$ & MRST $K(\sigma)$ & CTEQ $K(A)$ & MRST $K(A)$\\
\hline
1 & $1.0142 \pm 0.0077$ & $0.9963 \pm 0.0109$ & 
    $1.0243 \pm 0.0081$ & $1.0211 \pm 0.0113$ \\
2 & $0.9902 \pm 0.0280$ & $0.9830 \pm 0.0348$ & 
    $1.0000 \pm 0.0268$ & $1.0074 \pm 0.0357$ \\
3 & $0.9897 \pm 0.0062$ & $0.9851 \pm 0.0082$ &
    $0.9995 \pm 0.0066$ & $1.0096 \pm 0.0087$ \\
\hline
\end{tabular}
\caption{Comparison of the $K$ factors for cross-sections and acceptances 
calculated using CTEQ6.5M (NLO) and MRST 2002 (NNLO) PDFs. $K(\sigma)$ is
the $K$-factor for the cross-section and $K(A)$ is the $K$-factor for
the acceptance, as displayed in Table\ \ref{table:nnlo_calc}.}
\label{table:PDFNNLO}
}

\section{Scale Dependence}
\label{sec:scale}

Perturbative QCD calculations at fixed order depend on the factorization 
and renormalization scales introduced in the calculation. Thus, 
the previous calculations have an added uncertainty due to the choice of
certain fictitious scales appearing in the calculation. In a complete, all
order calculation, there would be no dependence on
these scales. However, in a fixed order calculation matched to PDFs, 
a dependence on the factorization scale $\mu_{{}_{\scriptstyle F}}$ 
and renormalization scale $\mu_{{}_{\scriptstyle R}}$ 
appear in the final results.  The effect of scale choice is
significant at NLO. Adding NNLO effects is found to reduce scale dependence 
considerably~\cite{dixon2003,FEWZ}, though it can remain significant near
thresholds where large logarithms render a fixed-order result unreliable. 

As is customary, we will choose the renormalization and factorization
scales to be identical, and investigate the scale dependence by varying them
by a factor of 2 or 1/2 about a central value of 
$\mu_{{}_{\scriptstyle F,R}}=M_Z$, which
is typical of the scales in our acceptance, and was the central 
value chosen in the previous section. 

Table~\ref{table:nnlo_calc} included only the central scale $M_Z$. 
Tables~\ref{table:nnlo_xsection_different_scales1} and 
\ref{table:nnlo_acceptance_different_scales1} and show the total cross
sections and acceptances for lepton production calculated by FEWZ at 
three different renormalization and factorization scales 
$M_Z/2$, $M_Z$, and $2M_Z$.
The acceptances for the final state leptons are as defined in
Table~\ref{table:acceptance}. For a measure
of the size of the scale dependence, the final column of each table shows
the maximum difference between the three values divided by average, with
an error calculated assuming the statistical errors in the three MC runs in
each row are independent. 

We can see that the scale dependence of the cross-sections at NLO
is typically of order $\pm 2.5\%$. The scale dependence of NLO acceptances is
dramatically reduced due to correlations in the scale dependence of the cut and
uncut cross-sections used to compute it. Adding NNLO reduces the scale 
dependence of the cross-sections to $\pm 1.2\%$ or less.  These estimates 
were obtained using the CTEQ6.5M NLO PDFs, but are compatible with the scale 
dependence found using MRST 2002 NNLO PDFs for the same calculations, within 
the limits of the computational errors. A better evaluation 
of the NNLO result may show the scale dependence to be even smaller, since it
is at the same level as the MC precision.  This is particularly true for Cut 2,
for large values of $|\eta|$, where the cross-section is relatively small and
the MC errors relatively large. Improving the convergence of FEWZ would reduce
these errors.

\section{Uncertainties Due to the Parton Distribution Function}
\label{sec:PDF} 

Phenomenological parameterizations of the PDFs are taken from a global fit 
to data.  Therefore, uncertainties on the PDFs arising from diverse 
experimental and theoretical sources will propagate from the global analysis 
into the predictions for the $W/Z$ cross-sections. Figure~\ref{fig:pdf_xs} 
shows the results of the inclusive $Z$ to di-lepton production
cross-section using various  CTEQ~\cite{CTEQ} and MRST~\cite{MRST}  PDFs.
The upward shift of about 7\% (between CTEQ6.1 and 6.5 and MRST2004 and
2006) results from the inclusion of heavy quark effects in the latest
PDF calculations. The acceptance due to the cuts in Table \ref{table:acceptance}
 using each of these PDFs is shown in Fig.~\ref{fig:pdf_acc_mu}.\footnote{
Some theoretical issues which may affect the contribution of the PDFs to the 
NNLO $K$-factor are not included here as we are concerned primarily 
with the error at NLO. See Refs.\ \cite{guzzi} for details. The comparison
at the end of Sect.~\ref{sec:QCD} suggests that such effects are
likely to be small in the $K$ factors calculated here.}

The uncertainties in the PDFs arising from the experimental statistical and 
systematic uncertainties, and the effect on the production cross-section of 
the $Z$ boson, have been studied using the standard methods proposed 
in Refs.~\cite{CTEQ, MRST}.  For the standard set of 
\newpage
\TABLE[ht]{
\begin{tabular}{|l|ccc|c|}
  \multicolumn{5}{c}{Total Cross-Section (in pb), $\Mll > 40$ GeV} \\
\hline
Order& $M_Z/2$ & $M_Z$ & $2 M_Z$ & $\Delta\sigma/{\overline\sigma}$\\
\hline
NLO   & $2297.4 \pm 2.3$ & $2358.1 \pm 2.3$ & $2418.3 \pm 2.3$ &
        $0.0513 \pm 0.0014 $\\
NNLO  & $2320.4 \pm 4.6$ & $2334.9 \pm 4.6$ & $2350.1 \pm 4.6$ &
	$0.0127 \pm 0.0028$ \\
\hline
 \multicolumn{5}{c}{ Cut Region 1} \\
\hline
Order& $M_Z/2$ & $M_Z$ & $2 M_Z$ & $\Delta\sigma/{\overline\sigma}$\\
\hline
NLO   & $698.2 \pm 0.7$ & $716.0 \pm 0.7$ & $735.2 \pm 0.7$ &
        $0.0516 \pm 0.0014 $\\
NNLO  & $718.9 \pm 6.1$ & $726.2 \pm 5.5$ & $719.8 \pm 4.8$ &
	$0.0101 \pm 0.0108 $ \\   
\hline
 \multicolumn{5}{c}{ Cut Region 2} \\
\hline
Order& $M_Z/2$ & $M_Z$ & $2 M_Z$ & $\Delta\sigma/{\overline\sigma}$\\
\hline
NLO   & $72.55 \pm 0.07$ & $74.12 \pm 0.07$ & $75.78 \pm 0.08$ &
        $0.0436 \pm 0.0014 $\\
NNLO  & $73.78 \pm 2.00$ & $73.39 \pm 1.96$ & $75.10 \pm 1.52$ &
	$0.0231 \pm 0.0351 $ \\
\hline
  \multicolumn{5}{c}{ Cut Region 3} \\
\hline
Order& $M_Z/2$ & $M_Z$ & $2 M_Z$ & $\Delta\sigma/{\overline\sigma}$\\
\hline
NLO   & $640.2 \pm 0.6$ & $657.2 \pm 0.7$ & $673.9 \pm 0.7$ &
        $0.0513 \pm 0.0014 $\\
NNLO  & $661.3 \pm 4.7$ & $650.4 \pm 4.0$ & $656.2 \pm 3.8$ &
	$0.0166 \pm 0.0090 $ \\
\hline
\end{tabular}
\caption{Scale dependence of the total and accepted cross-sections (in pb) for 
  $Z$ boson production with $\Mll > 40$ GeV/$c^2$
  calculated by the {FEWZ} program at order NLO and NNLO. 
  The final column is a measure of scale dependence obtained by
  dividing the maximum spread by the average for the three points. 
  \label{table:nnlo_xsection_different_scales1}}
}
\vspace{0.2in}
\TABLE[ht]{
\begin{tabular}{|l|ccc|c|}
  \multicolumn{5}{c}{ Cut Region 1 (\% Accepted)} \\
\hline
Order& $M_Z/2$ & $M_Z$ & $2 M_Z$ & $\Delta A/{\overline A}$\\
\hline
NLO   & $30.39 \pm 0.04$ & $30.36 \pm 0.04$ & $30.40 \pm
 0.04$ & $0.0013 \pm 0.0020 $\\
NNLO  & $30.98 \pm 0.27$ & $31.10 \pm 0.21$ & $30.40 \pm 0.04$ &
	$0.0153 \pm 0.0111 $ \\
\hline
  \multicolumn{5}{c}{ Cut Region 2 (\% Accepted)} \\
\hline
Order& $M_Z/2$ & $M_Z$ & $2 M_Z$ & $\Delta A/{\overline A}$\\
\hline
NLO   & $3.158 \pm 0.031$ & $3.143 \pm 0.030$ & $3.134 \pm 0.033$ &
        $0.0077 \pm 0.0020 $\\
NNLO  & $3.180 \pm 0.086$ & $3.143 \pm 0.084$ & $3.196 \pm 0.065$ &
	$0.0165 \pm 0.0353$ \\
\hline
  \multicolumn{5}{c}{ Cut Region 3 (\% Accepted)} \\
\hline
Order& $M_Z/2$ & $M_Z$ & $2 M_Z$ & $\Delta A/{\overline A}$\\
\hline
NLO   & $27.87 \pm 0.04$ & $27.87 \pm 0.04$ & $27.87 \pm 0.04$ &
        $0.0001 \pm 0.0020 $\\
NNLO  & $28.50 \pm 0.21$ & $27.86 \pm 0.18$ & $27.92 \pm 0.17$ &
	$0.0229 \pm 0.0095 $ \\
\hline
\end{tabular}
\caption{Scale dependence of the acceptances $A$ in the various cut regions for 
  $Z$ boson production with $\Mll > 40$ GeV/$c^2$
  calculated by the {FEWZ} program at order NLO and NNLO. 
  The final column is a measure of scale dependence obtained by
  dividing the maximum spread of the three preceding columns by their average.
 }
  \label{table:nnlo_acceptance_different_scales1}
}

\clearpage 

\FIGURE[ht]{
\epsfig{file=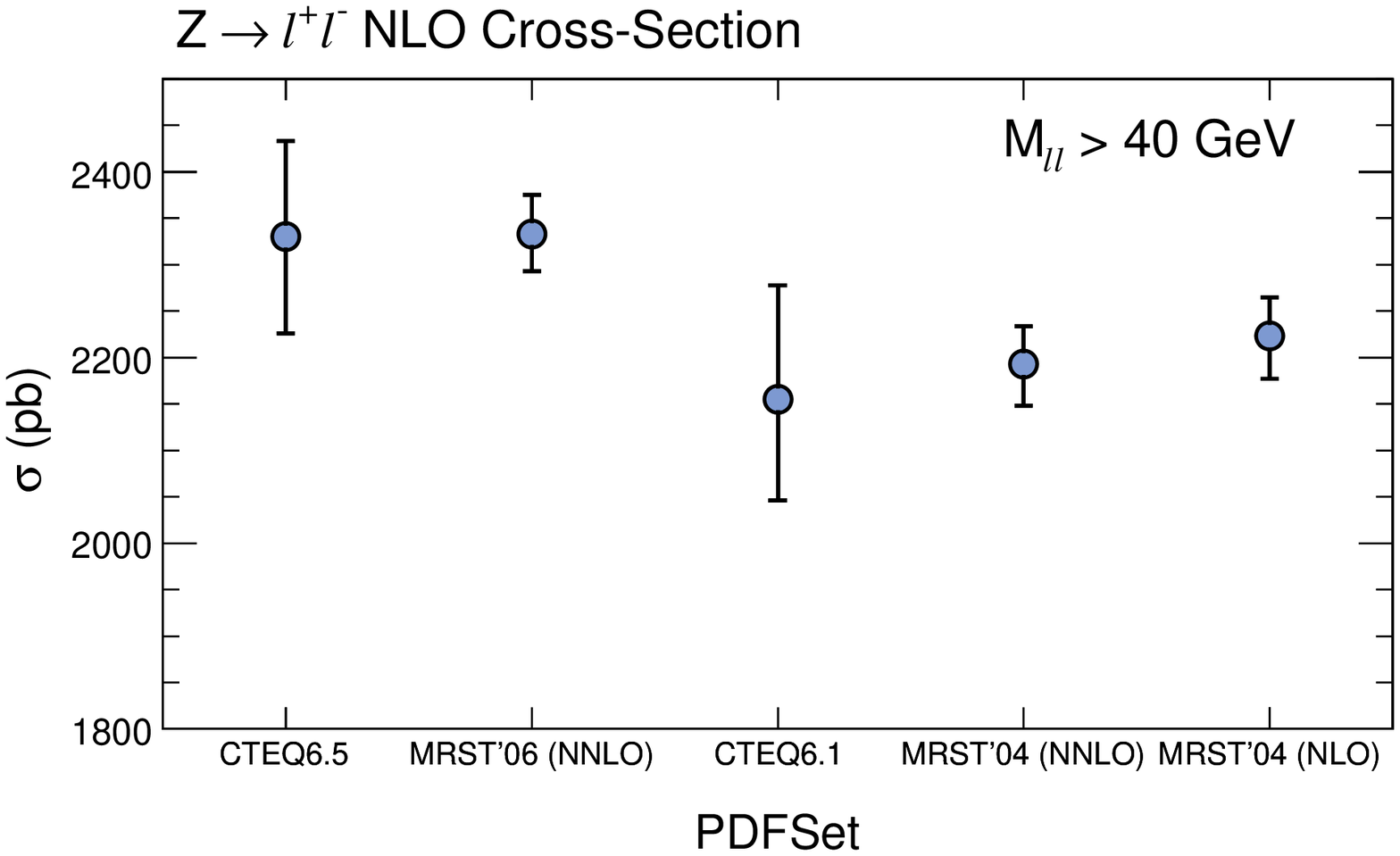,width=9cm}
\caption{ Comparison of $\Zgammall$ cross-sections
  for $\Mll > 40$ GeV/$c^{2}$ for several recent PDF calculations. }
\label{fig:pdf_xs}
}
\noindent
PDFs, corresponding to the minimum in the PDF parameter space, a 
complete set of eigenvector PDF sets, which characterize the region 
nearby the minimum and quantify its error, have been simultaneously 
calculated.  From the minimum set and these ``error'' sets we calculate 
the best estimate and the uncertainty for the 
$Z$ cross-section. We do this using the asymmetric
Hessian error method,~\cite{Asym} where the cross-section results from the 
various eigenvector PDF sets have been combined according to the prescriptions
found in~\cite{CTEQ,Asym}. Fig.~\ref{fig:pdf_xs} list the results for the 
different PDFs and Table~\ref{table:pdfuncertainty} summarizes the results of  
the latest CTEQ and MRST PDF sets. These calculations were done using MC@NLO.
The difference in the uncertainties (approximately a factor of two) between the 
results obtained from the CTEQ and MRST PDF error sets is due to different 
assumptions made by the groups while creating the eigenvector PDF sets.
Note that the first column of Table~\ref{table:nnlo_calc} should agree
with the CTEQ6.5 entries in Table~\ref{table:pdfuncertainty}, and do
within the uncertainties quoted in Table~\ref{table:nnlo_calc}, although
different event samples were used.

\TABLE[ht]{
{\small
\begin{tabular}{|l|ccc|ccc|ccc|}
\hline
 & \multicolumn{3}{c|}{$\Mll > 40$ GeV/$c^2$} & \multicolumn{6}{c|}{
  Cut Region 1} \\
\hline
PDF Set & $\sigma$ (pb) & $\Delta\sigma_+$ & $\Delta\sigma_-$ & $\sigma$ (pb)
& $\Delta\sigma_+$ & $\Delta\sigma_-$ & $A$
& $\Delta A_+$ & $\Delta A_-$  \\
\hline
CTEQ6.5 & 2330 & 103 & 104 & 703.6 & 21.6 & 26.7 & 0.302 & 0.004 & 0.004 \\
MRST2006 & 2333 & 42 & 40 & 712.6 & 13.6 & 16.0 & 0.305 & 0.004 & 0.004 \\
CTEQ6.1 & 2155 & 123 & 109 & 652.1 & 30.2 & 29.5 & 0.303 & 0.007 & 0.005 \\
MRST2004 (NNLO) & 2193 & 41 & 45 & 672.4 & 12.6 & 17.9 & 0.302 & 0.004 &
  0.004 \\
MRST2004 (NLO) & 2223 & 42 & 46 & 662.8 & 12.5 & 17.9 & 0.302 & 0.004 &
  0.004 \\
\hline
 & \multicolumn{3}{c|}{$\Mll > 40$ GeV/$c^2$} & \multicolumn{6}{c|}{
  Cut Region 2} \\
\hline
PDF Set & $\sigma$ (pb) & $\Delta\sigma_+$ & $\Delta\sigma_-$ & $\sigma$ (pb)
& $\Delta\sigma_+$ & $\Delta\sigma_-$ & $A$
& $\Delta A_+$ & $\Delta A_-$  \\
\hline
CTEQ6.5 & 2330 & 103 & 104 & 71.1 & 2.2 & 3.4 & 0.0305 & 0.0001 & 0.0005 \\
MRST2006 & 2333 & 42 & 40 & 72.0 & 0.1 & 2.4 & 0.0308 & 0.0000 & 0.0010 \\
CTEQ6.1 & 2155 & 123 & 109 & 66.3 & 3.0 & 4.1 & 0.0307 & 0.0006 & 0.0014 \\
MRST2004 (NNLO) & 2193 & 41 & 45 & 69.0 & 0.3 & 2.1 & 0.0310 & 0.0002 &
  0.0007 \\
MRST2004 (NLO) & 2223 & 42 & 46 & 68.7 & 0.3 & 2.0 & 0.0313 & 0.0002 &
  0.0007 \\
\hline
 & \multicolumn{3}{c|}{$\Mll > 40$ GeV/$c^2$} & \multicolumn{6}{c|}{
  Cut Region 3} \\
\hline
PDF Set & $\sigma$ (pb) & $\Delta\sigma_+$ & $\Delta\sigma_-$ & $\sigma$ (pb)
& $\Delta\sigma_+$ & $\Delta\sigma_-$ & $A$
& $\Delta A_+$ & $\Delta A_-$  \\
\hline
CTEQ6.5 & 2330 & 103 & 104 & 623.5 & 20.8 & 22.0 & 0.268 & 0.005 & 0.003 \\
MRST2006 & 2333 & 42 & 40 & 634.1 & 10.4 & 15.5 & 0.272 & 0.003 & 0.005 \\
CTEQ6.1 & 2155 & 123 & 109 & 578.9 & 26.6 & 26.8 & 0.269 & 0.002 & 0.007 \\
MRST2004 (NNLO) & 2193 & 41 & 45 & 598.0 & 12.3 & 14.5 & 0.269 & 0.004 &
  0.003 \\
MRST2004 (NLO) & 2223 & 42 & 46 & 588.5 & 12.1 & 14.3 & 0.268 & 0.004 &
  0.003 \\
\hline
\end{tabular}
}
\caption{ Cross-sections $\sigma$, and acceptances $A$, with asymmetric
  Hessian uncertainties as calculated 
  using several recent PDF sets for the three cut regions defined in
  Table~\ref{table:acceptance}. }
\label{table:pdfuncertainty}
}

Finally we study the sensitivity of the kinematic acceptance calculations
to the uncertainties affecting the PDF sets.
Figs.~\ref{fig:pdf_xs_vs_pt_mu} and~\ref{fig:pdf_xs_vs_eta_mu} show the
systematic error on the production
cross-sections as a function of the $|\eta|$ cut and minimum
lepton $\pT$ for variations on the three types of cuts in 
Table\ \ref{table:acceptance}.  The fractional uncertainties, shown in
in the same figures, demonstrate that
the relative uncertainty in the cross-section is very flat as a function
of the kinematic cuts, until the region of extreme cuts and low
statistics in the MC are reached. The corresponding uncertainty on the
acceptance as a function of the kinematic cuts is shown in
Figs~\ref{fig:pdf_acc_vs_pt_mu} and~\ref{fig:pdf_acc_vs_eta_mu}. 
These show a similar dependence to the cross-section uncertainties,
though the fractional errors are smaller.

\section{Conclusions}
\label{sec:conclusions}

To evaluate the overall contribution from theoretical uncertainties to
both the cross-section and acceptance calculations for the decay mode
$Z/\gamma^* \to \ell^{+}\ell^{-}$ ($\ell = e$ or $\mu$) 
at the LHC (and for $\Mll > 40$
GeV/$c^2$) we add the uncertainties from each of the sources considered
in the preceding sections.  We compile the errors assuming that 
the calculation is done with MC@NLO at scale 
$\mu_{{}_{\scriptstyle F}} = \mu_{{}_{\scriptstyle R}} = M_Z$ and interfaced
to PHOTOS to add final state QED radiation. The missing electroweak contribution
may then be inferred from HORACE as in Sec.\ \ref{sec:EWK}. For these
errors we take those resulting from the tight cut set in
Table~\ref{table:horace_xs_comp}; these cuts are considered the most
representative of likely analysis cuts for the LHC experiments. 

QCD uncertainties may be divided into two main classes. If the NNLO
$K$-factor is set to 1, there
is a missing NNLO contribution $\DNNLO = K-1$. Since $K$ has residual NNLO
scale dependence, we must also take this into account and write 
$K = 1 + \DNNLO \pm \Dscale$. The factor $\DNNLO$ can be inferred from 
Table\ \ref{table:nnlo_calc} and $\Dscale$ can
be inferred from half the scale variation of the NNLO entries in Table\ % 
\ref{table:nnlo_xsection_different_scales1}.  
For example, for the total cross-section, we can infer that
$\DNNLO = -(0.98 \pm 0.22)\%$, while $\Dscale = (0.64 \pm 0.16)\%$. Both errors 
in the MC@NLO result are of order 1\%. The 
original NLO scale dependence estimate of 5.13\% is no longer relevant because
$K$ is now known, even though it may be set to 1 in a 
particular calculation. Similar estimates can be made for the two sources of
error in each of the cuts. 

Both classes of QCD errors are also associated with a ``technical
precision'' due to limitations of the computing tools used to evaluate them. 
Significant improvements in the NNLO precision could be obtained if a
program with faster convergence were available.  We therefore include an 
``error on the error'' for the QCD errors, and propagate these through in
the usual fashion to derive a final accuracy for the total QCD
uncertainty estimate. This sets a limitation on how much the NLO calculation
can realistically be improved using currently available NNLO results. 
 
These contributions to QCD errors are summarized in 
Table\ \ref{table:QCD_uncertainty} for the total cross-section and the 
three cuts of Table\ \ref{table:acceptance}. Results are shown both for 
the three cut cross-sections and their ratio for the total cross-section
with $\Mll>40$GeV/$c^2$, and the errors are assumed to be uncorrelated.

If the $K$-factor had not been calculated at NNLO, the error of the 
NLO cross-sections could have been roughly estimated from half the width of the 
scale-dependence band, or half the NLO results for 
$\Delta\sigma/{\overline\sigma}$ in Table\ %
\ref{table:nnlo_xsection_different_scales1}, giving uncertainties of 
$2.2 - 2.6\%$.  The errors calculated from the $K$ factors are in within 
the limits these expectations, up to the technical precision of the 
calculation. A similar error NLO estimate for the error in the 
acceptance based on Table\ \ref{table:nnlo_acceptance_different_scales1} 
would predict at most $0.33\%$ missing NNLO.  
The actual missing NNLO in Cut 1 was found to be larger
than this, but the other two cuts have very small missing NNLO, to within
the technical precision of the calculation. 

\TABLE[ht]{
\begin{tabular}{|l|c|c|c|c|}
\multicolumn{5}{c}{QCD Uncertainties (\%)} \\
\hline
\multicolumn{5}{|c|}{Cross-Section $\Delta\sigma$} \\
\hline
Uncertainty & $\sigma^{\mathrm tot}$  & Cut 1 & Cut 2 & Cut 3 \\ \hline
Missing NNLO & $-0.98 \pm 0.22$  & $1.42 \pm 0.77$   & $-0.98 \pm 2.80$
& $-1.33 \pm 0.62$     \\
Scale Dependence & $0.64 \pm 0.16$ & $0.51 \pm 0.54$   & $1.16 \pm 1.76$
 & $0.83 \pm 0.46$   \\
\hline
Total  & $1.17 \pm 0.20$ & $1.51 \pm 0.75$  & $1.52 \pm 2.25$
& $1.57 \pm 0.58$     \\
\hline
\multicolumn{5}{|c|}{Error in Acceptance ($\Delta A$)} \\
\hline
Uncertainty &  $-$ & Cut 1 & Cut 2 & Cut 3 \\
\hline
Missing NNLO & $-$ & $2.43 \pm 0.81$ & $0.00 \pm 2.68$ & $-0.05 \pm 0.66$ \\
Scale Dependence & $-$ & $0.77 \pm 0.56$ & $0.83 \pm 1.77$ & $1.15 \pm 0.48$ \\
\hline
Total  & $-$ & $2.55 \pm 0.79$ & $0.83 \pm 1.77$ & $2.55 \pm 0.79$  \\
\hline
\end{tabular}
\caption{Summary of QCD uncertainties $\Delta\sigma$ in the 
cross-sections and $\Delta A$ in the acceptances relative to the 
un-cut cross-section $\sigma^{\mathrm tot}$ 
for $\Mll > 40$ GeV/$c^2$. The three cuts are described in 
Table\ \ref{table:acceptance}. Missing NNLO is shown with a sign, 
because it has been calculated.}
\label{table:QCD_uncertainty}
}

The final contribution to the total error considered here is the uncertainty 
from the PDFs.  This may be extracted from the results of Sec.\ \ref{sec:PDF}
by taking the errors from the CTEQ6.5 results for Cut 1 (see
Table~\ref{table:acceptance}). The errors are asymmetric, so we take the
largest of the two (up or down) uncertainties as the total fractional
error for the PDF calculation. We choose the first cut set, since it is
the most representative of likely analysis cuts at the LHC
experiments. CTEQ errors, rather than the MRST errors, are used
because they give a more conservative estimate. The difference between the
results obtained by the latest CTEQ and MRST PDFs is less than the 
maximum error quoted for CTEQ for all three cut regions. 

The errors are added in quadrature, assuming
no correlations, and the results are given in
Table~\ref{table:total_uncertainty}. It should be remarked that there is
no concensus on the best way to combine the errors in 
Table~\ref{table:total_uncertainty}, so this total must be considered
an estimate. The QCD error is taken for Cut 1 for
the same reasons as given above. In addition, as we have discussed above,
we propagate the ``error on the error'' for each of the contributions in
order to have some reasonable estimate of the accuracy of the quoted
total theoretical uncertainty. The exception to this is the PDF error,
which can be considered as an upper limit on the uncertainty. 
We conclude that the event generator MC@NLO interfaced to PHOTOS
should be sufficient to guarantee an overall theoretical uncertainty on the
$Z$ production due to higher order calculation, PDFs, and renormalization scale
at the level of $4\%$ for the total cross-section and at approximately $3\%$
for the acceptance. 

\TABLE[h]{
\begin{tabular}{|l|c|c|}
\multicolumn{3}{c}{Total Theoretical Uncertainty (\%)} \\
\hline
Uncertainty & Cross-Section $\Delta\sigma$ & Acceptance $\Delta A$ \\
\hline
Missing {\cal O}($\alpha$) EWK   & $0.38 \pm 0.26$ & $0.96 \pm 0.21$ \\
Total QCD Uncertainty & $1.51 \pm 0.75$ & $2.55 \pm 0.79$ \\
PDF Uncertainty   & 3.79 & 1.32 \\
\hline
Total Uncertainty & $4.1 \pm 0.3$ & $3.0 \pm 0.7$   \\
\hline
\end{tabular} 
\caption{ Total theoretical uncertainty on the $Z$ production
  cross-section $\Delta\sigma$, and acceptances $\Delta A$.} 
\label{table:total_uncertainty}
}

$Z$ production will provide a valuable tool for studying QCD, measuring 
precision electroweak physics, and monitoring the luminosity. 
As the luminosity increases, the large statistics will
permit a further improvement in the systematic uncertainties due to the PDFs.
Adding complete {\cal O}($\alpha$) EWK corrections to the event generator
would eliminate most of the EWK uncertainty, and incorporating NNLO QCD
corrections would substantially reduce the QCD uncertainties. 

Reaching a combined precision of 1\%, as desired in the
later stages of analysis at high integrated luminosity, will require new tools.
In addition to improved PDFs, an event generator combining NNLO QCD with
complete {\cal O}($\alpha$) EWK corrections will be needed, with exponentiation
in appropriate regimes, and adequate convergence properties to technically
reach the required precision.\cite{QEDxQCD} Measurement strategies
have been proposed that may mitigate some of the effects of systematic errors
on the precision of the $Z$ measurements.\cite{krasny} A combination of 
improved calculations and improved meaurements will be needed to 
permit the desired precision to be reached as the integrated luminosity 
increases to a point where it is needed.

\acknowledgments{ 
This work was supported in part by US DOE grant DE-FG02-91ER40671.
We thank Frank Petriello, Zbigniew W{\c a}s, Carlo M. Carloni Calame, 
C.-P. Yuan, and B.F.L. Ward for helpful correspondence, and
FNAL for computer resources.  S.Y. thanks the Princeton Department of Physics 
for hospitality, and Seigfred Yost for lifelong support. 
}

\FIGURE[ht]{
\begin{tabular}{cc}
\epsfig{file=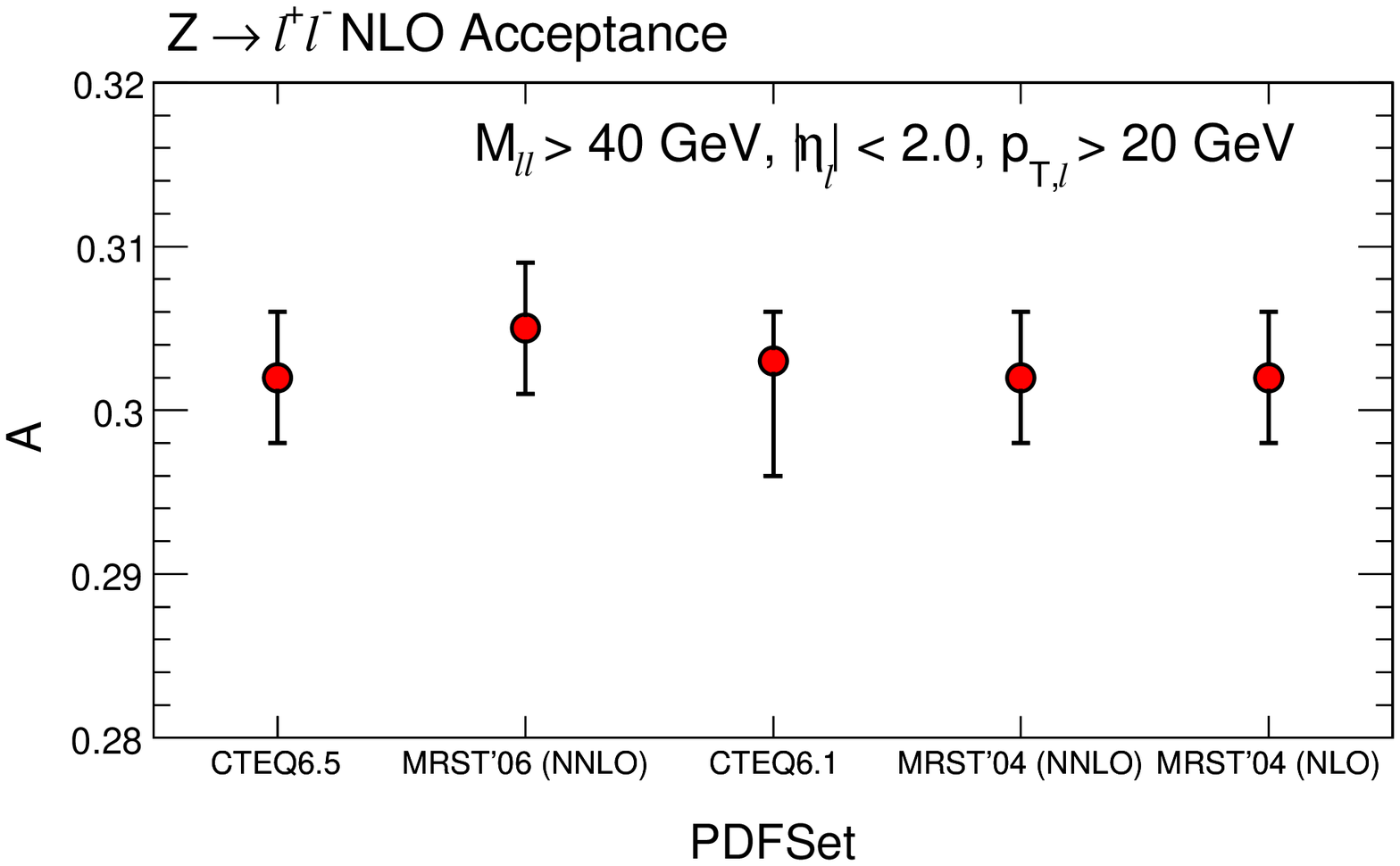,width=7cm} &
\epsfig{file=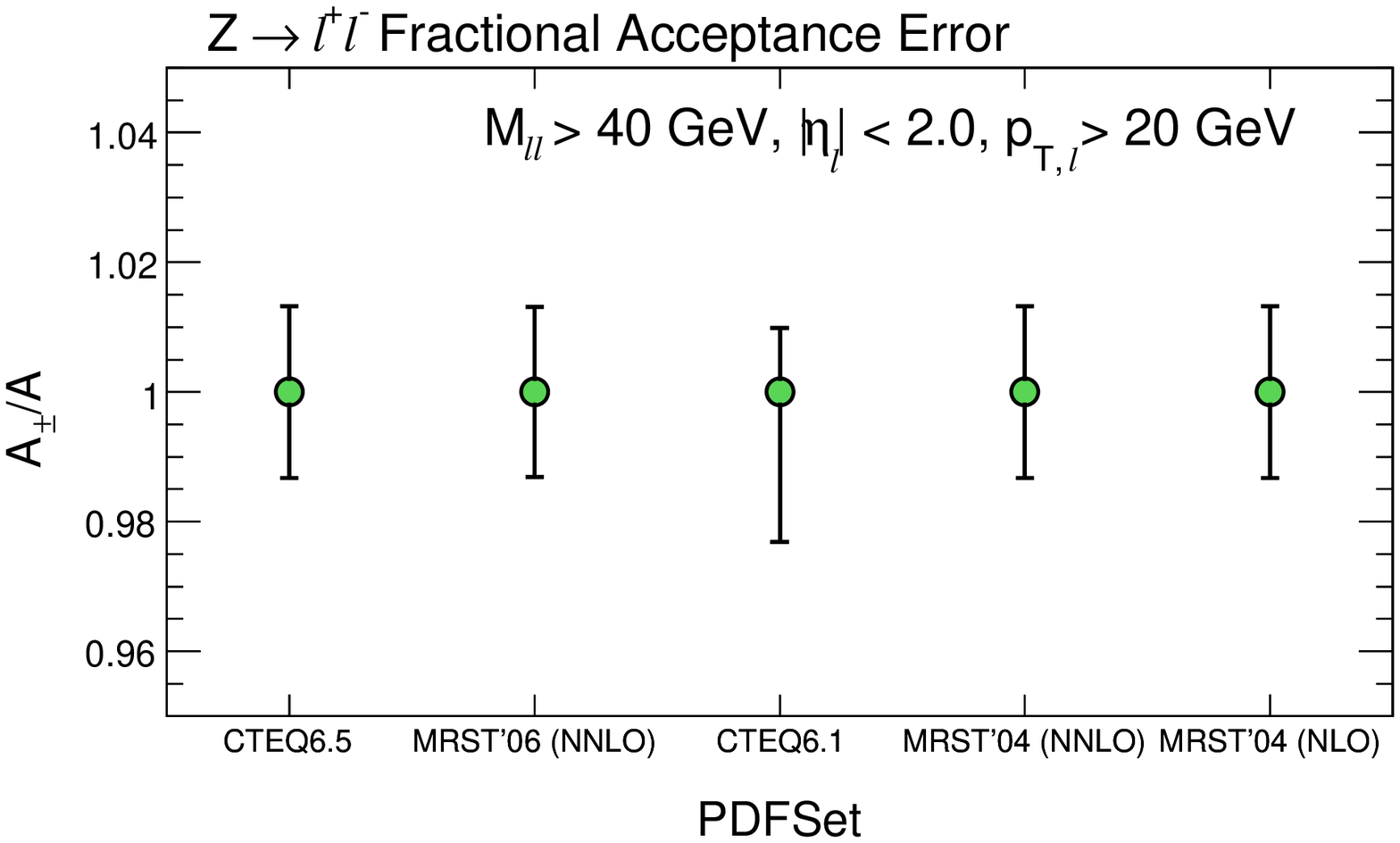,width=7cm}\\
\multicolumn{2}{c}{(a)} \\[3em]
\epsfig{file=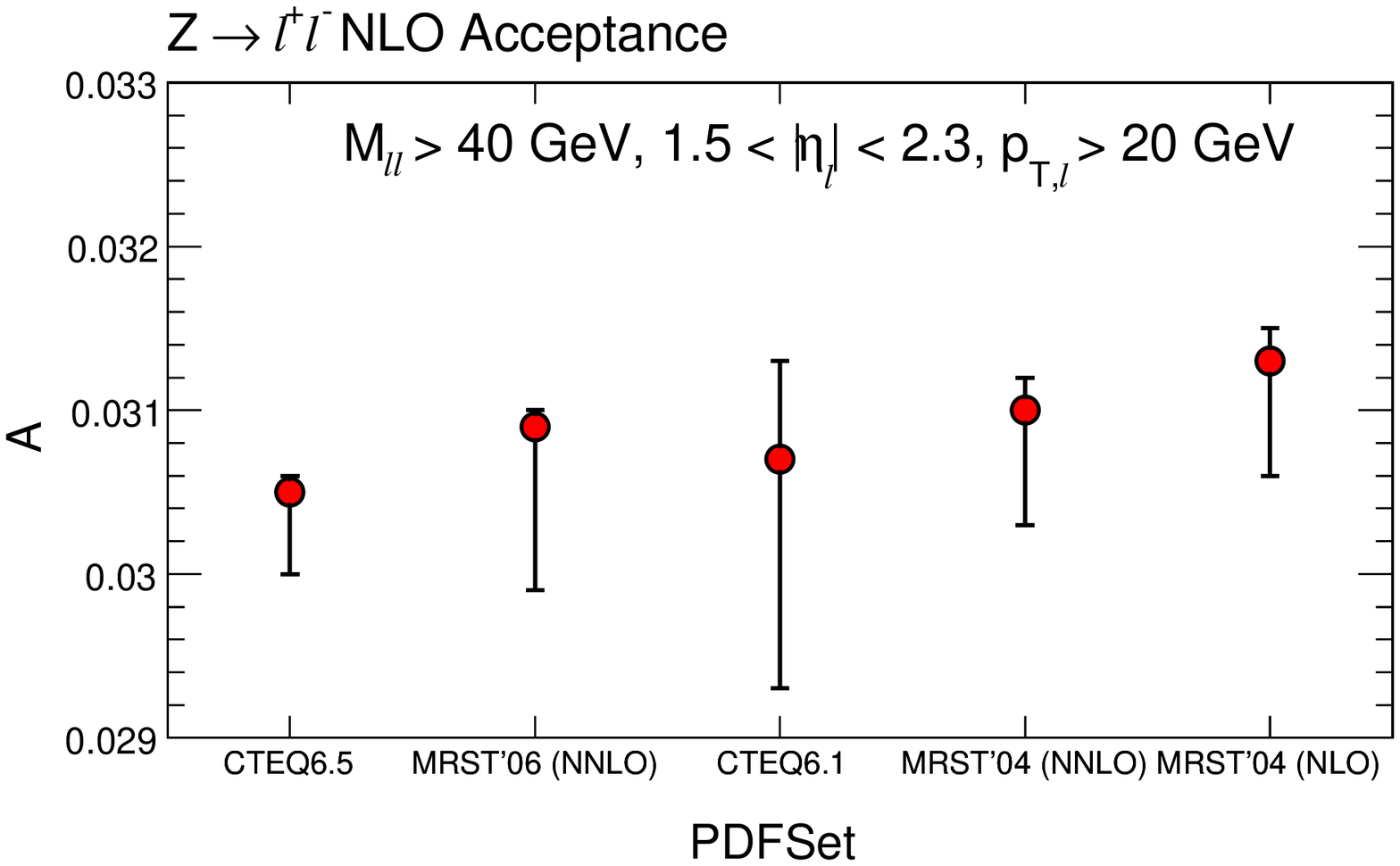,width=7cm} &
\epsfig{file=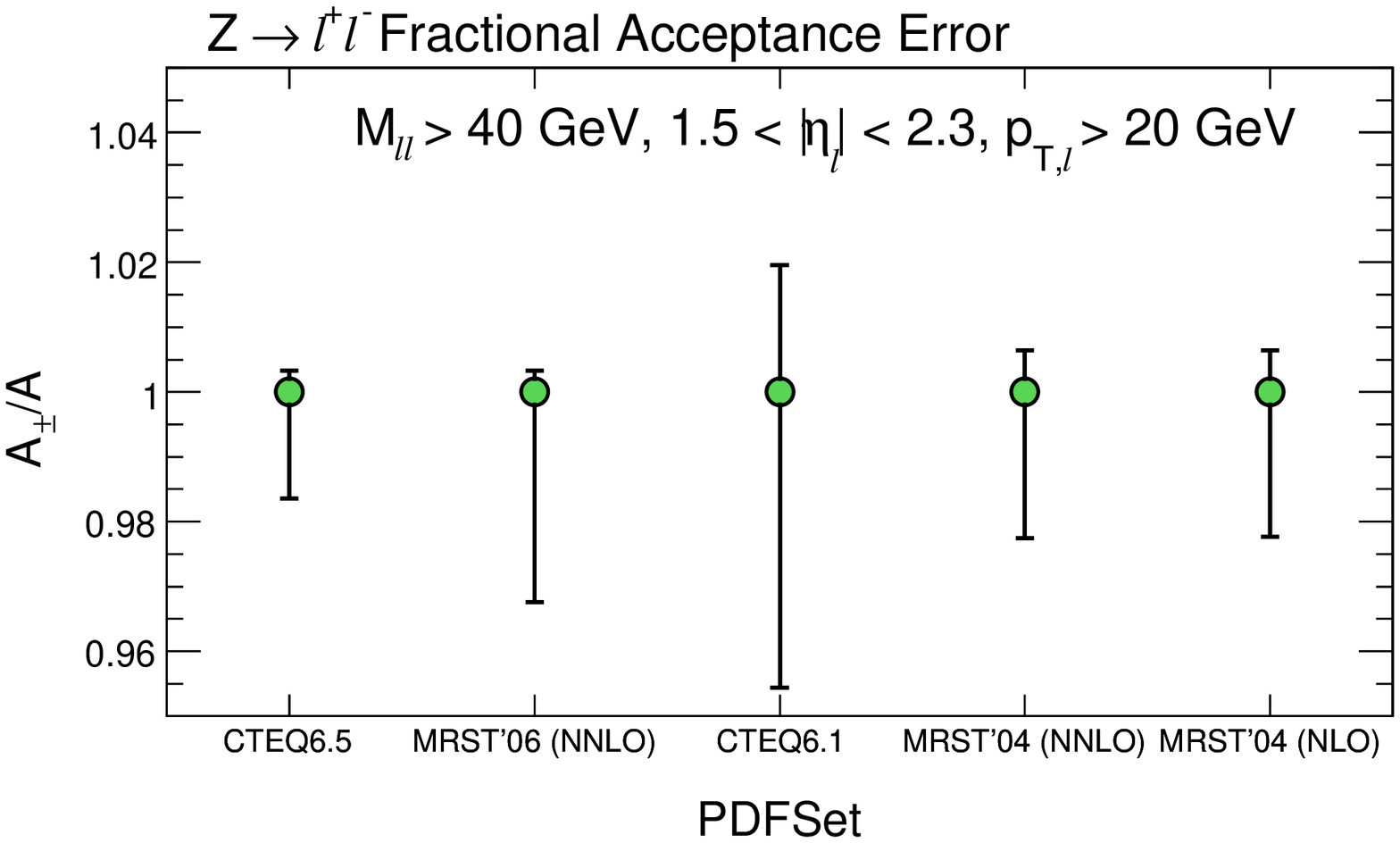,width=7cm}\\
\multicolumn{2}{c}{(b)} \\[3em]
\epsfig{file=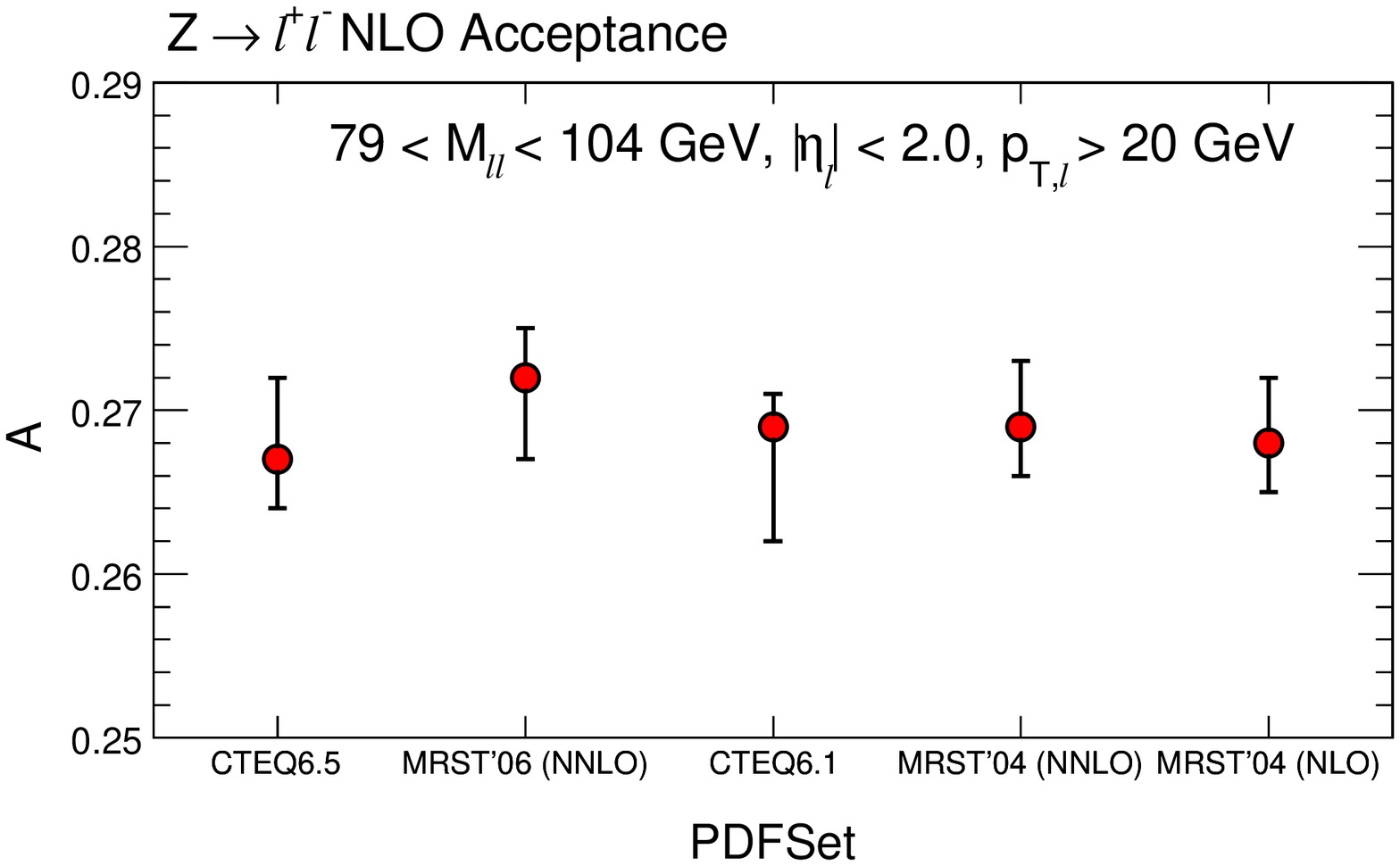,width=7cm} &
\epsfig{file=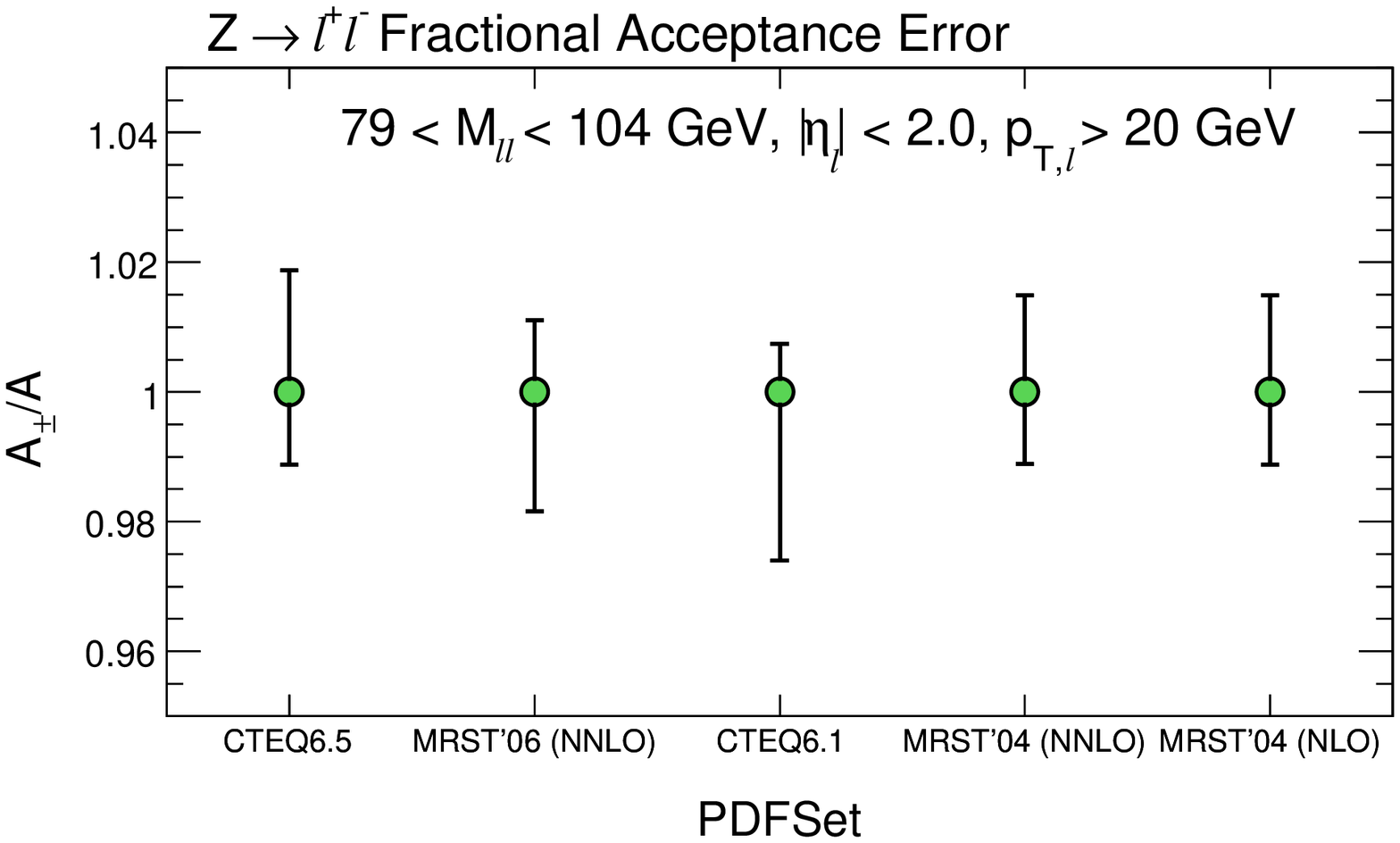,width=7cm}\\
\multicolumn{2}{c}{(c)} \\[3em]
\end{tabular}
\caption{ Comparison of $Z/\gamma^* \to \ell^{+}\ell^{-}$ ($\ell=e$ or $\mu$)
acceptances $A$, with several recent PDF calculations for acceptance regions 
(a) Cut 1, (b) Cut 2, and (c) Cut 3, as
defined in Table~\ref{table:acceptance}. The left-hand plots show the total
acceptance and the right hand plots show the fractional error on the acceptance.}
\label{fig:pdf_acc_mu}
}

\FIGURE[ht]{
\begin{tabular}{cc}
\epsfig{file=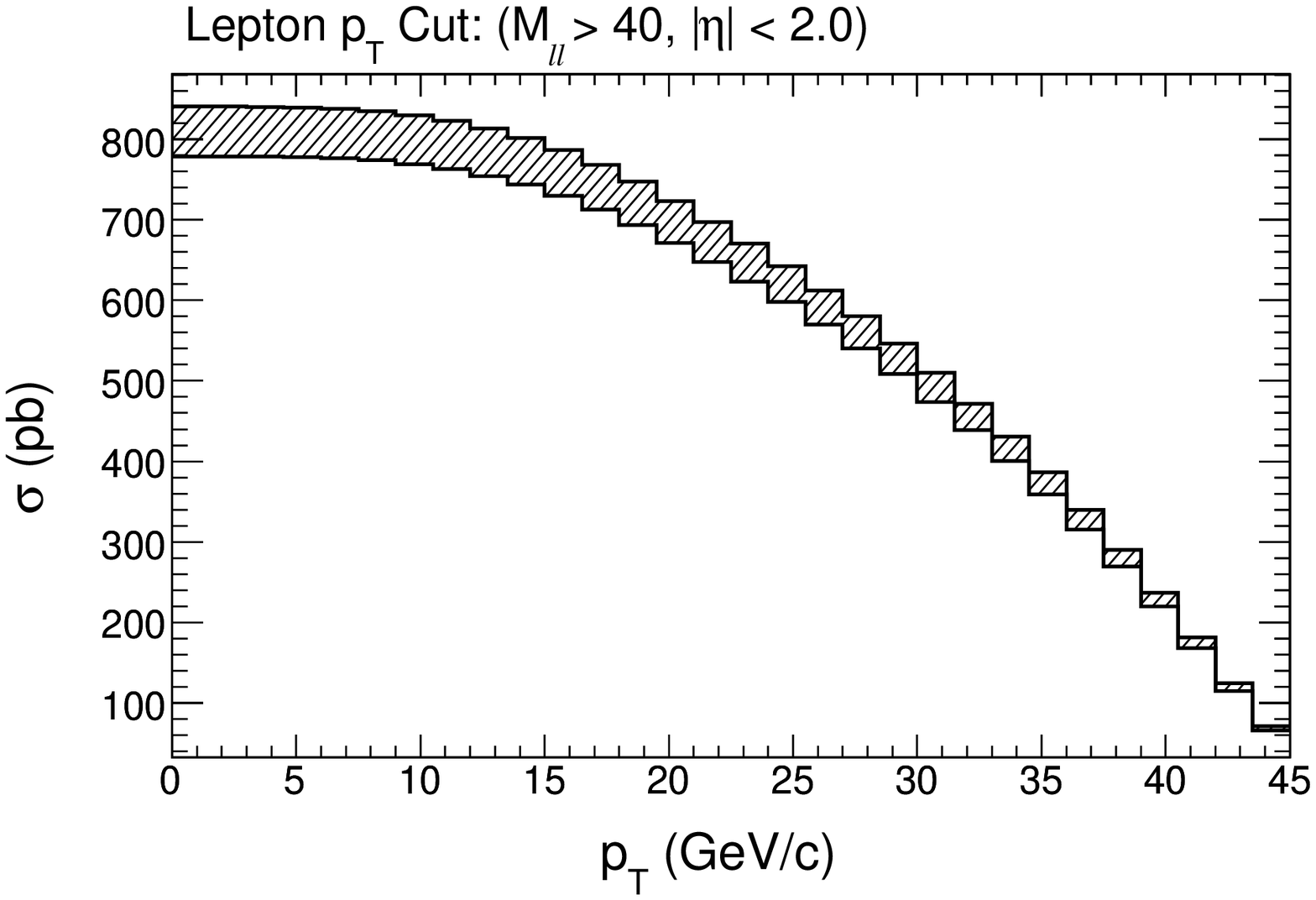,width=7cm} &
\epsfig{file=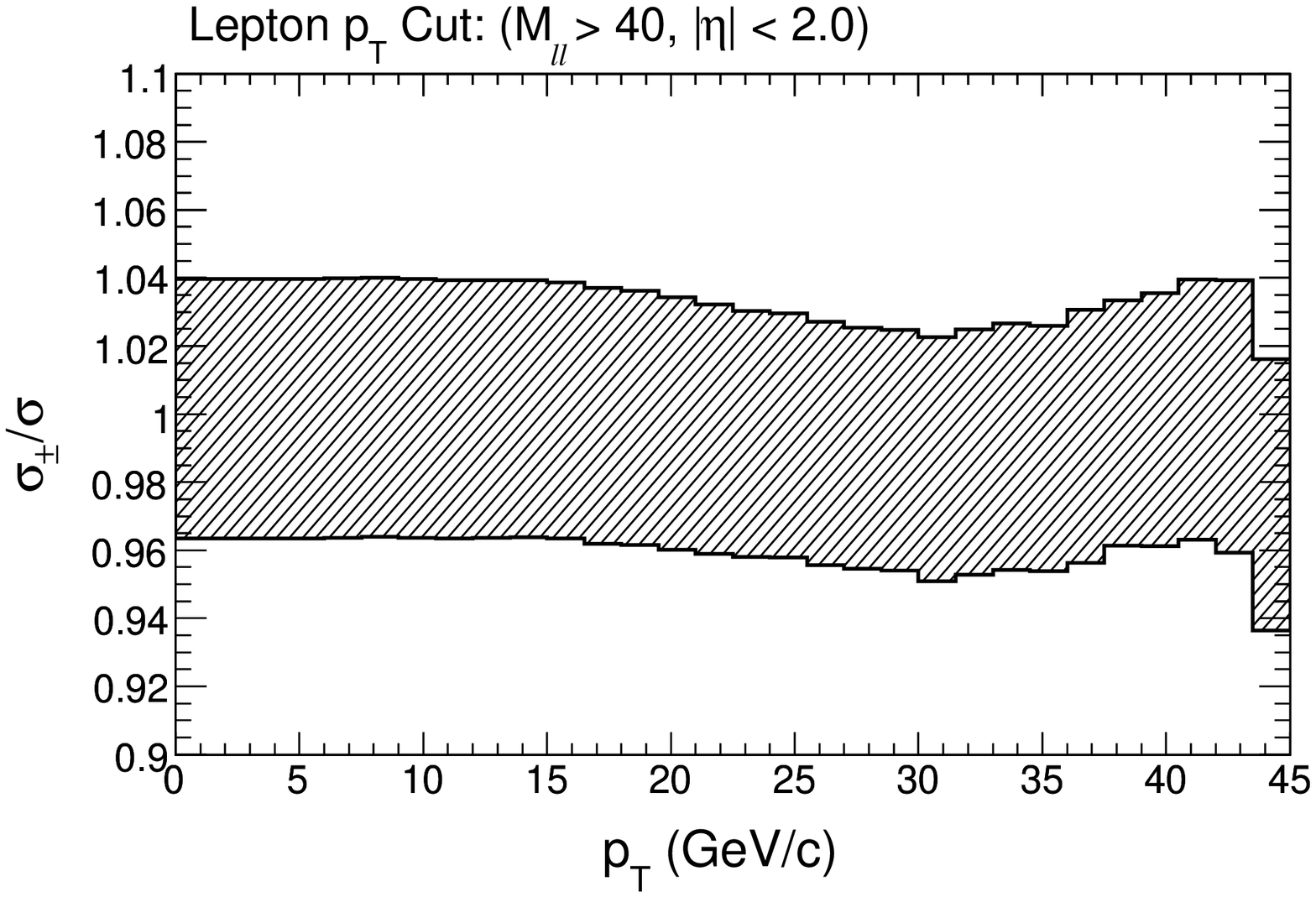,width=7cm}\\
\multicolumn{2}{c}{(a)} \\[3em]
\epsfig{file=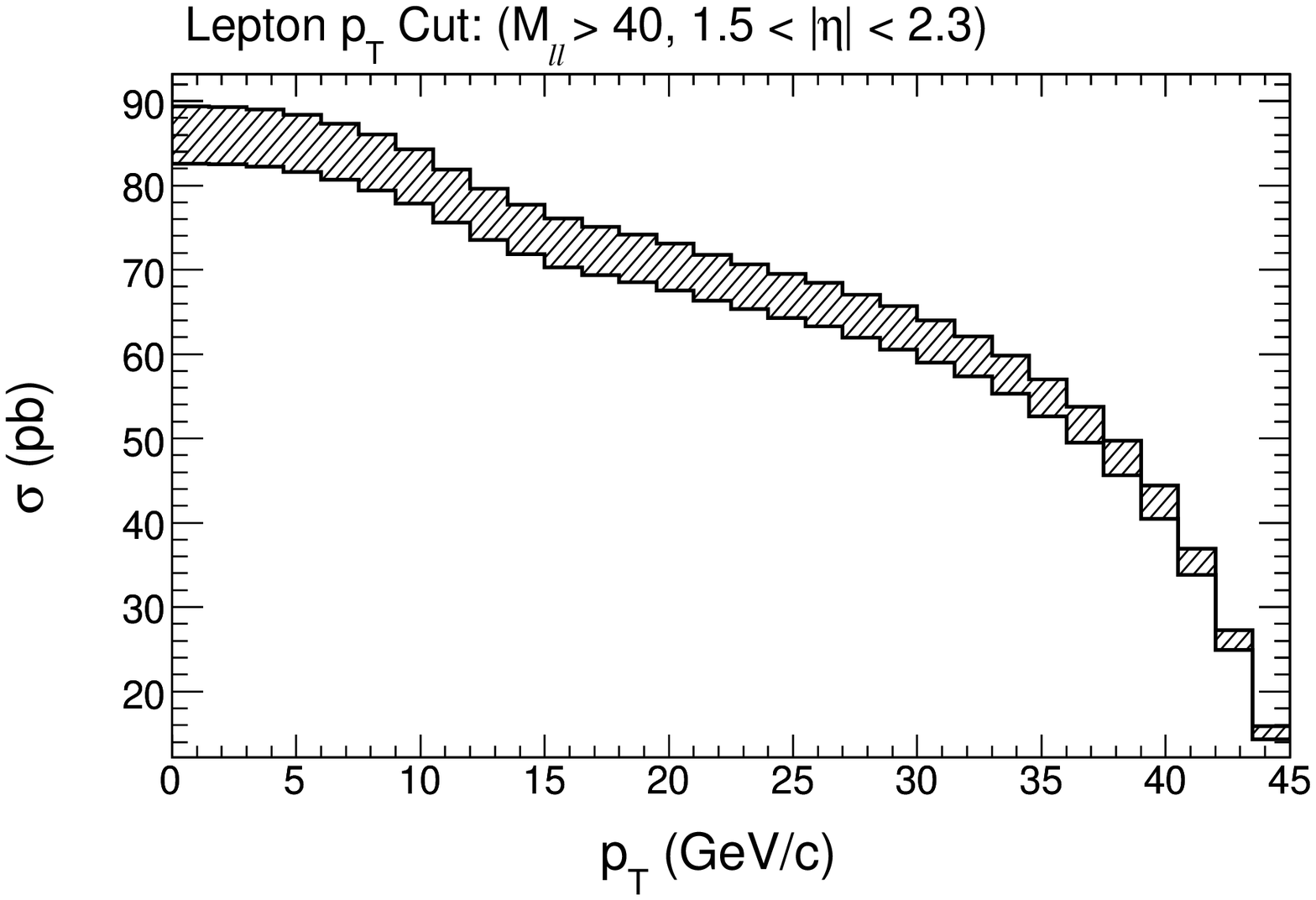,width=7cm} &
\epsfig{file=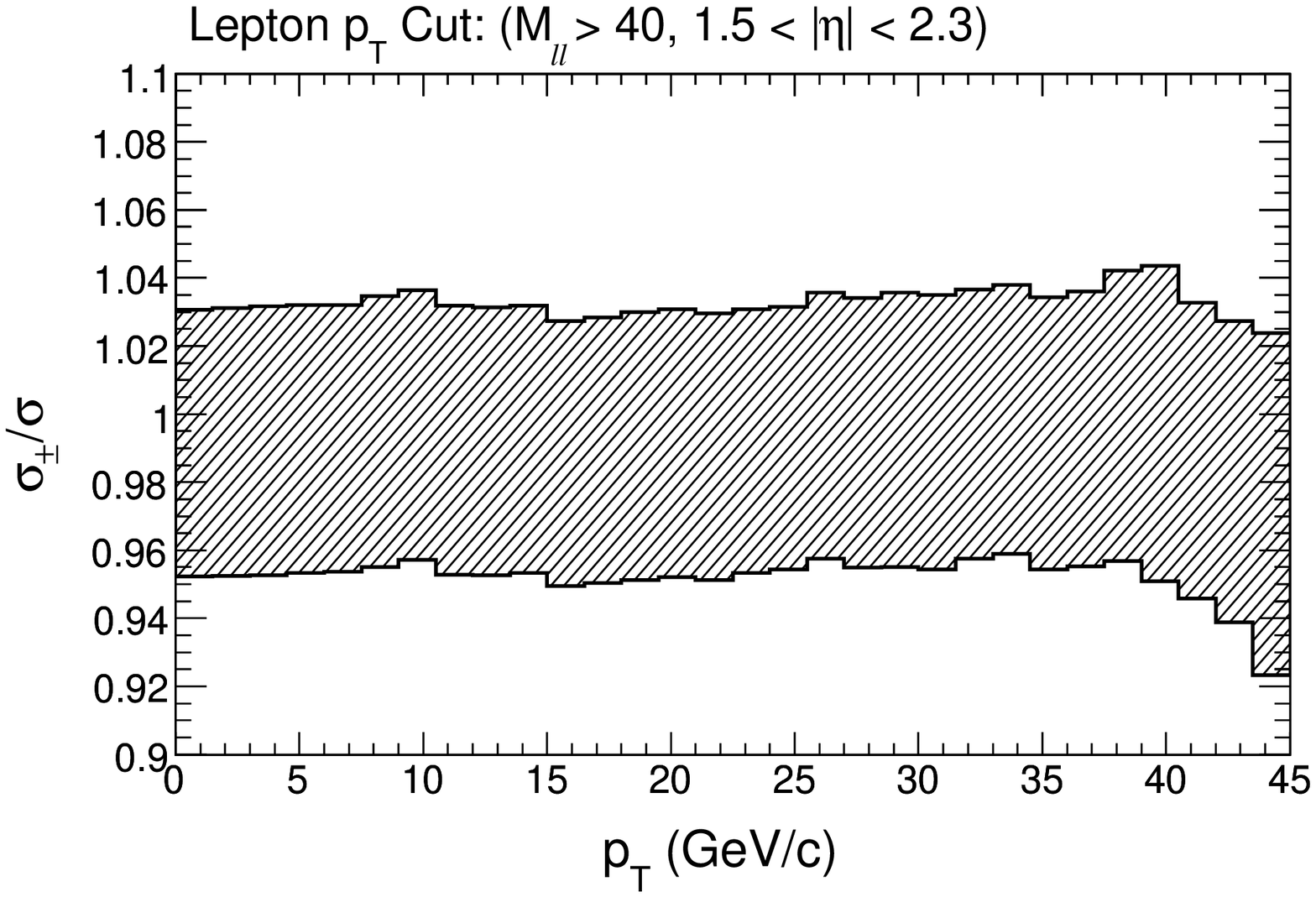,width=7cm} \\
\multicolumn{2}{c}{(b)} \\[3em]
\epsfig{file=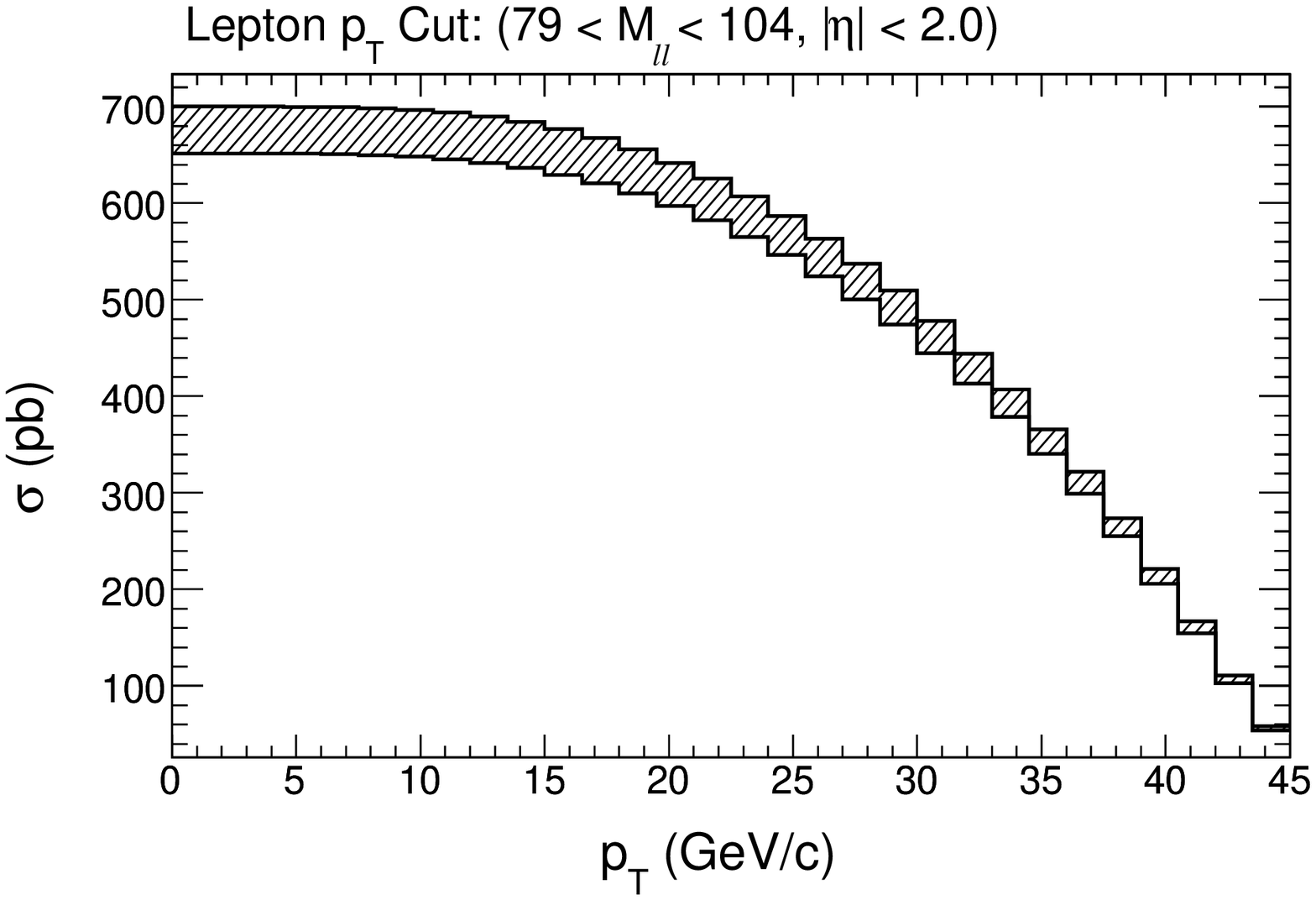,width=7cm} &
\epsfig{file=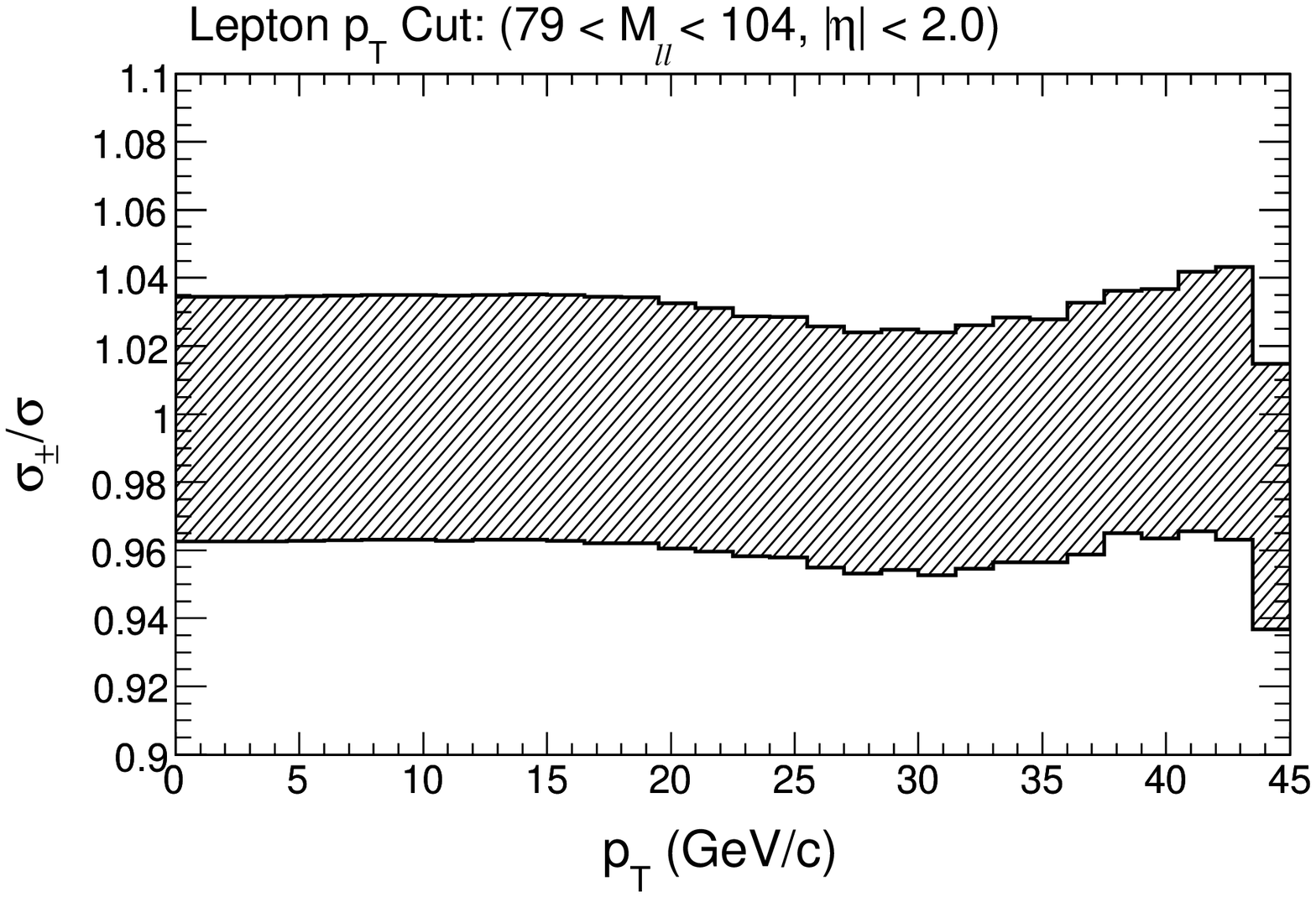,width=7cm} \\
\multicolumn{2}{c}{(c)} \\[3em]
\end{tabular}
\caption{ The $Z/\gamma^* \to \ell^{+}\ell^{-}$ cross-section $\sigma$ ($\ell =
e$ or $\mu$), as a function of the $\pT$ cut for acceptance regions
 (a) Cut 1, (b) Cut 2, and (c) Cut 3, as defined in
  Table~\ref{table:acceptance}. For each acceptance region we fix the
  invariant mass and $|\eta|$ cuts at their specified values, and vary only
  the $\pT$ cut. The figures on the right show the relative errors in the cross sections.}
\label{fig:pdf_xs_vs_pt_mu}
}

\FIGURE[ht]{
\begin{tabular}{cc}
\epsfig{file=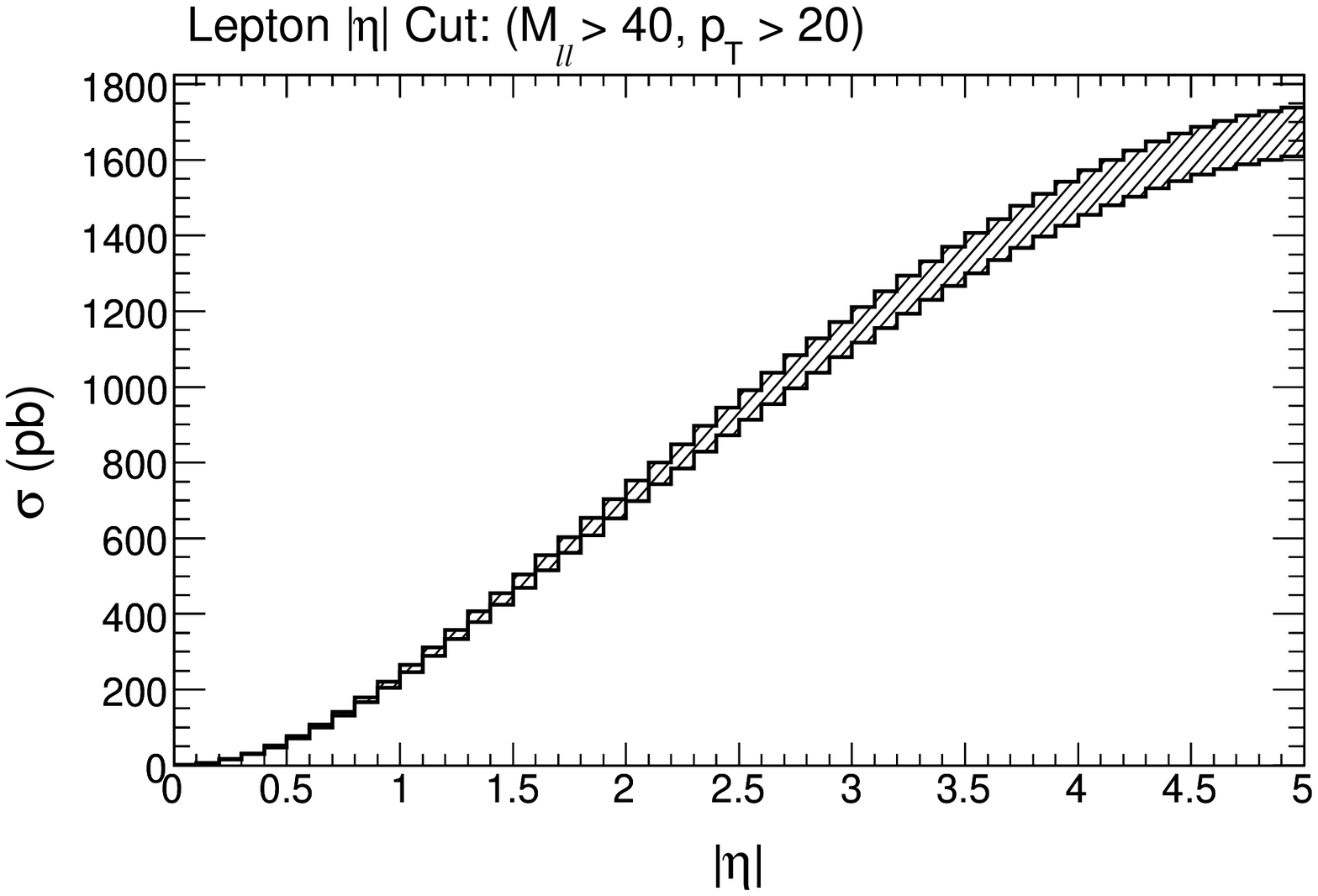,width=7cm}&
\epsfig{file=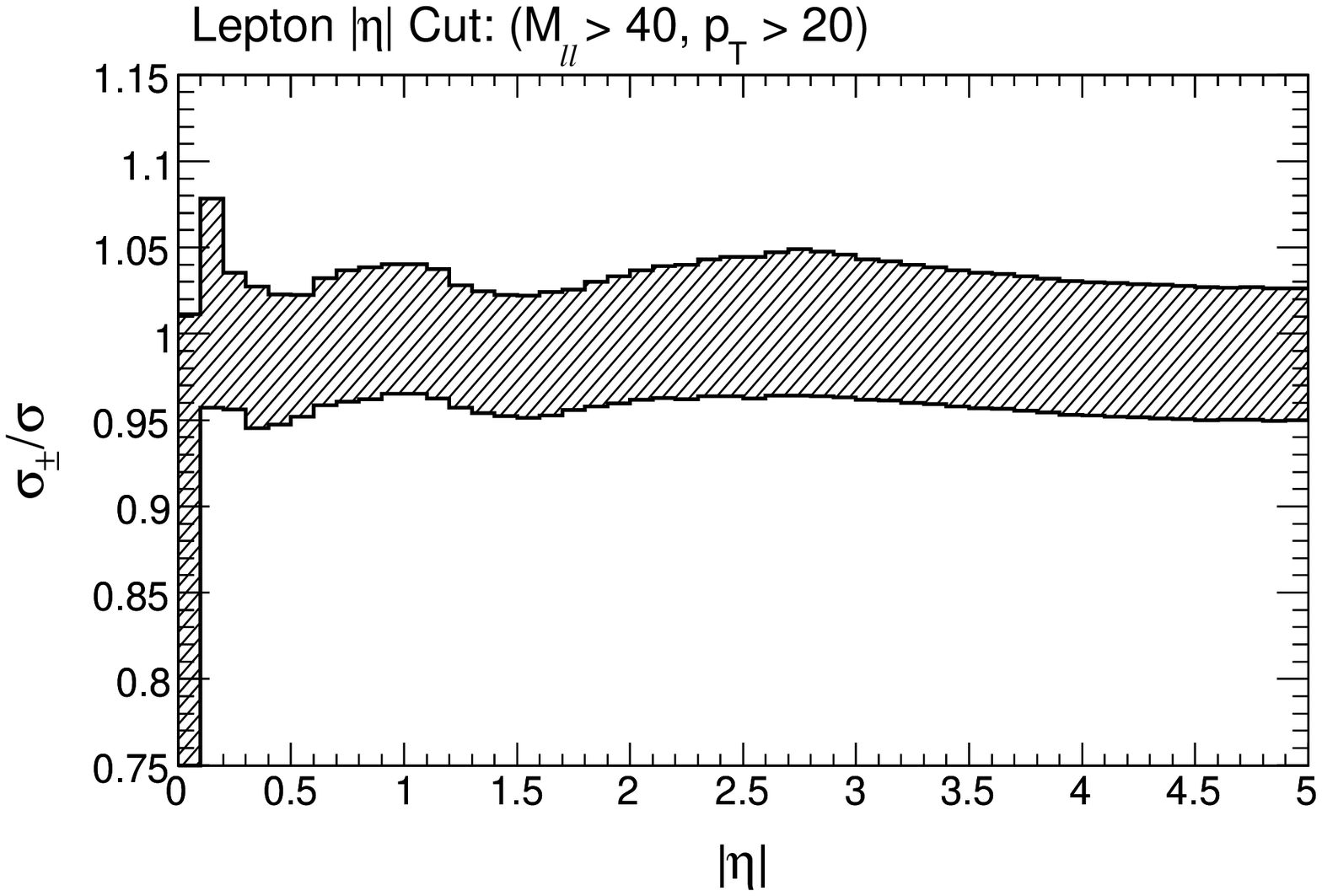,width=7cm}\\
\multicolumn{2}{c}{(a)} \\[3em]
\epsfig{file=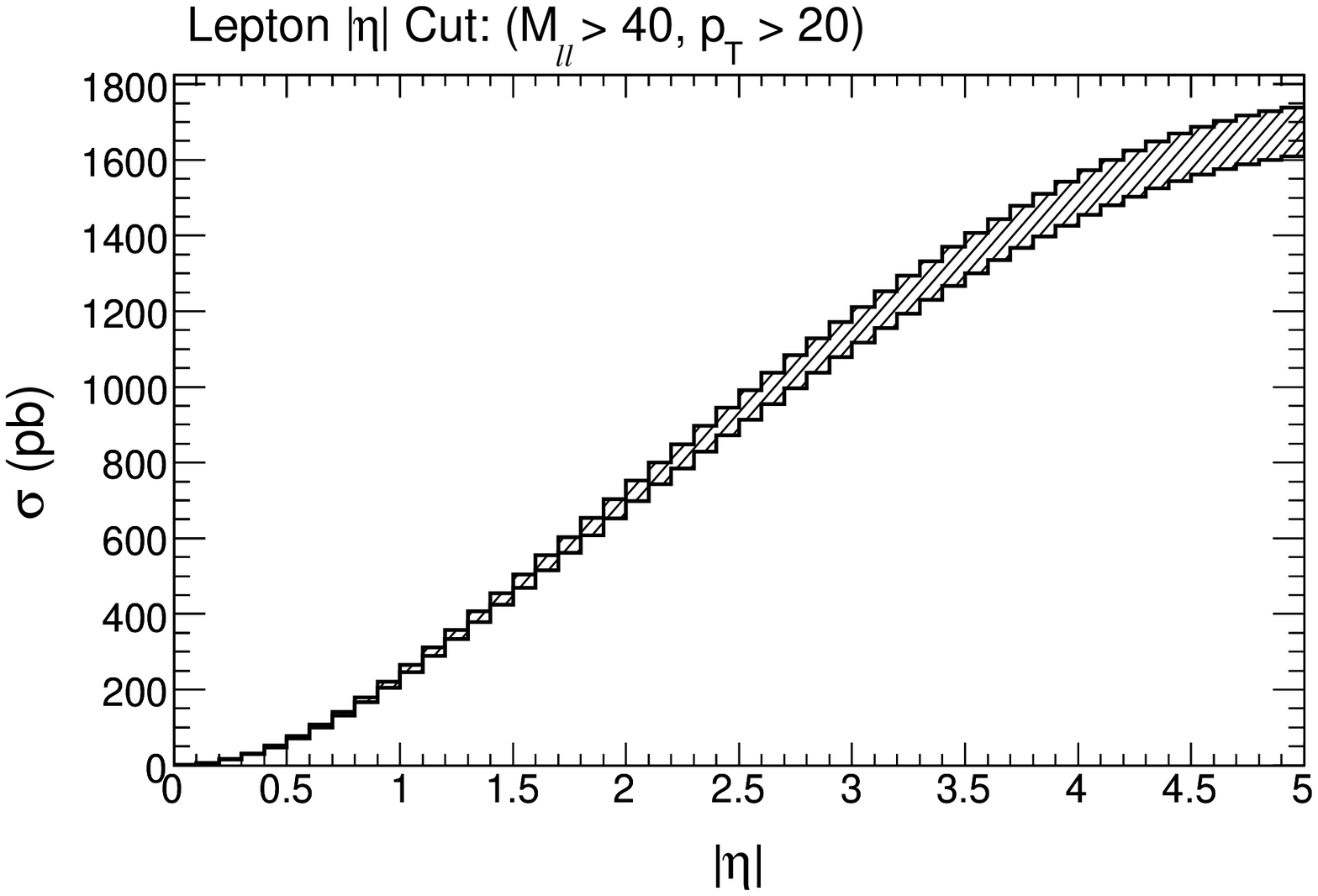,width=7cm} &
\epsfig{file=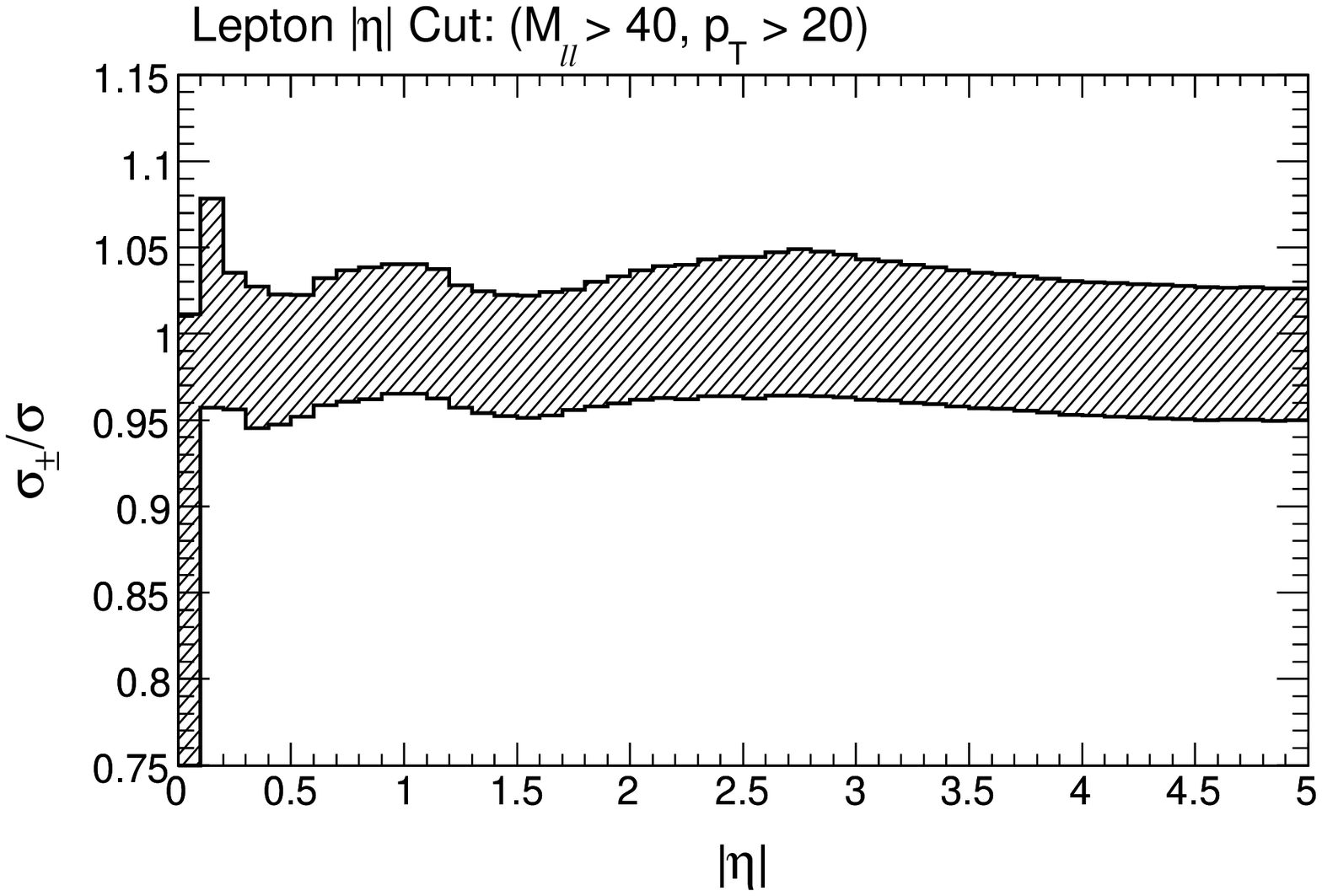,width=7cm} \\
\multicolumn{2}{c}{(b)} \\[3em]
\epsfig{file=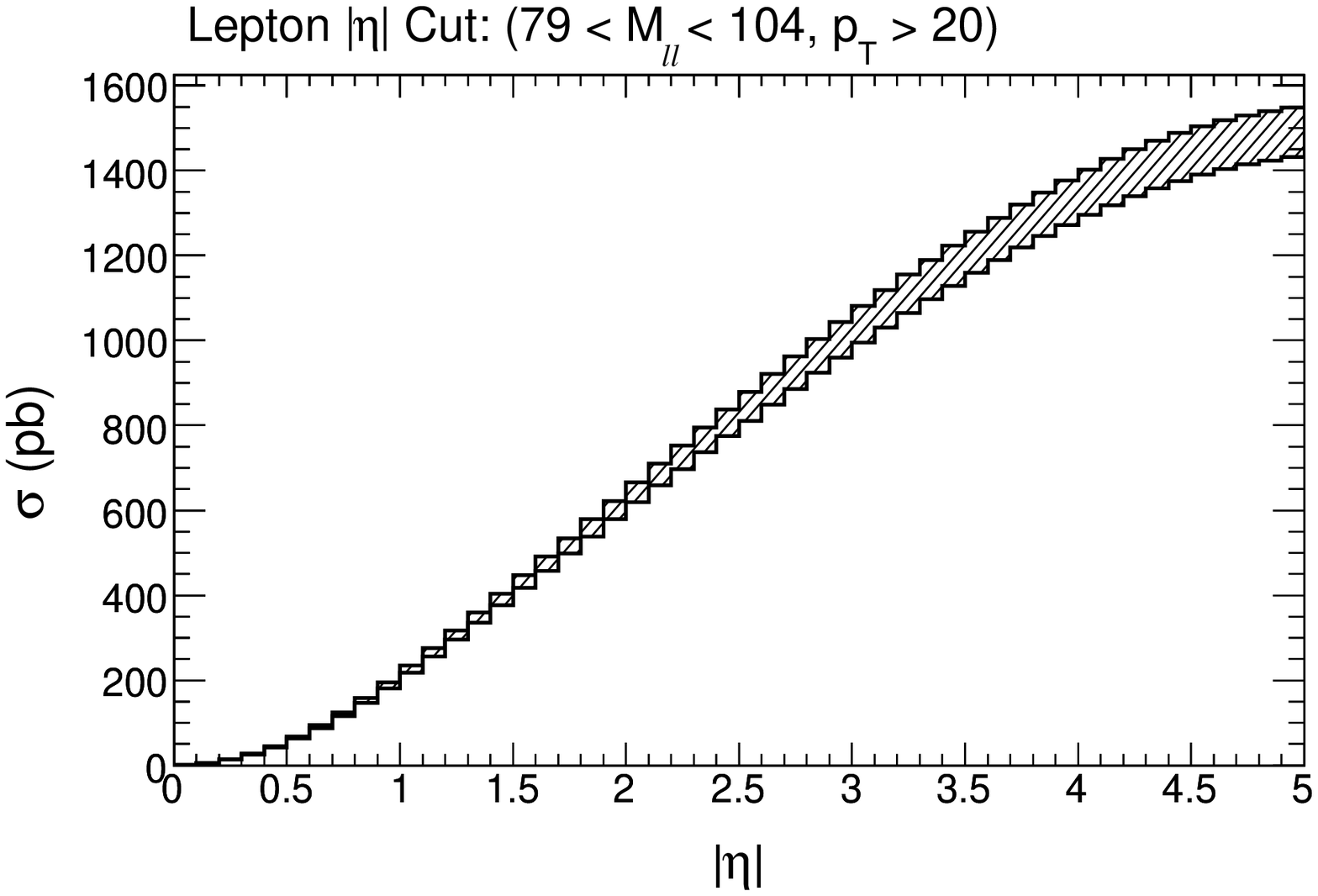,width=7cm} &
\epsfig{file=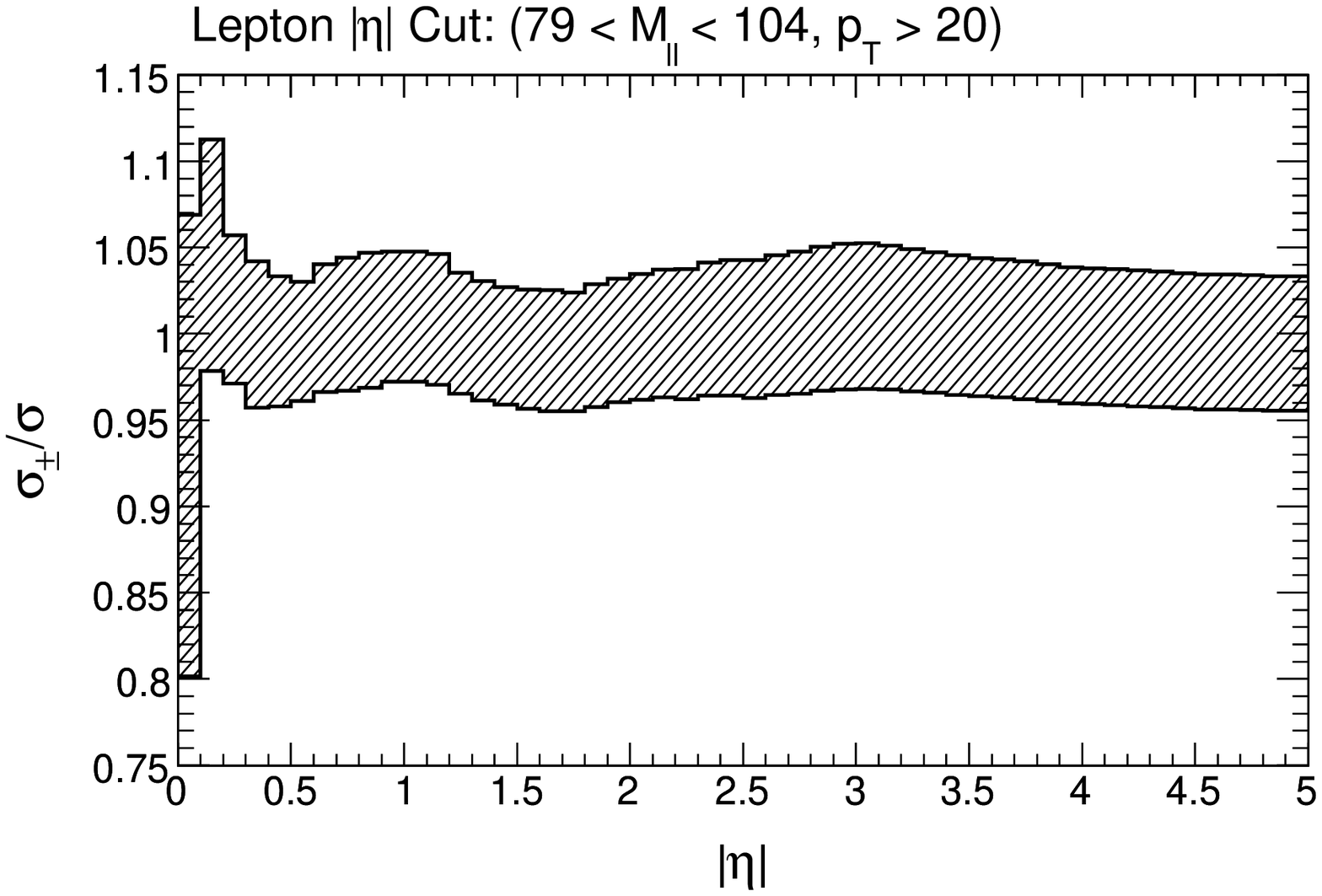,width=7cm} \\
\multicolumn{2}{c}{(c)} \\[3em]
\end{tabular}
\caption{ The $Z/\gamma^* \to \ell^{+}\ell^{-}$ cross-section $\sigma$
($\ell=e$ or $\mu$), as a function of the $|\eta|$ cut for acceptance regions
 (a) Cut 1, (b) Cut 2, and (c) Cut 3, as defined in
  Table~\ref{table:acceptance}. For each acceptance region we fix the
  invariant mass and \pT cuts at their specified values, and vary only
  the $|\eta|$ cut. The figures on the right show the relative errors in the cross sections.}
\label{fig:pdf_xs_vs_eta_mu}
}

\FIGURE[t]{
\begin{tabular}{cc}
\epsfig{file=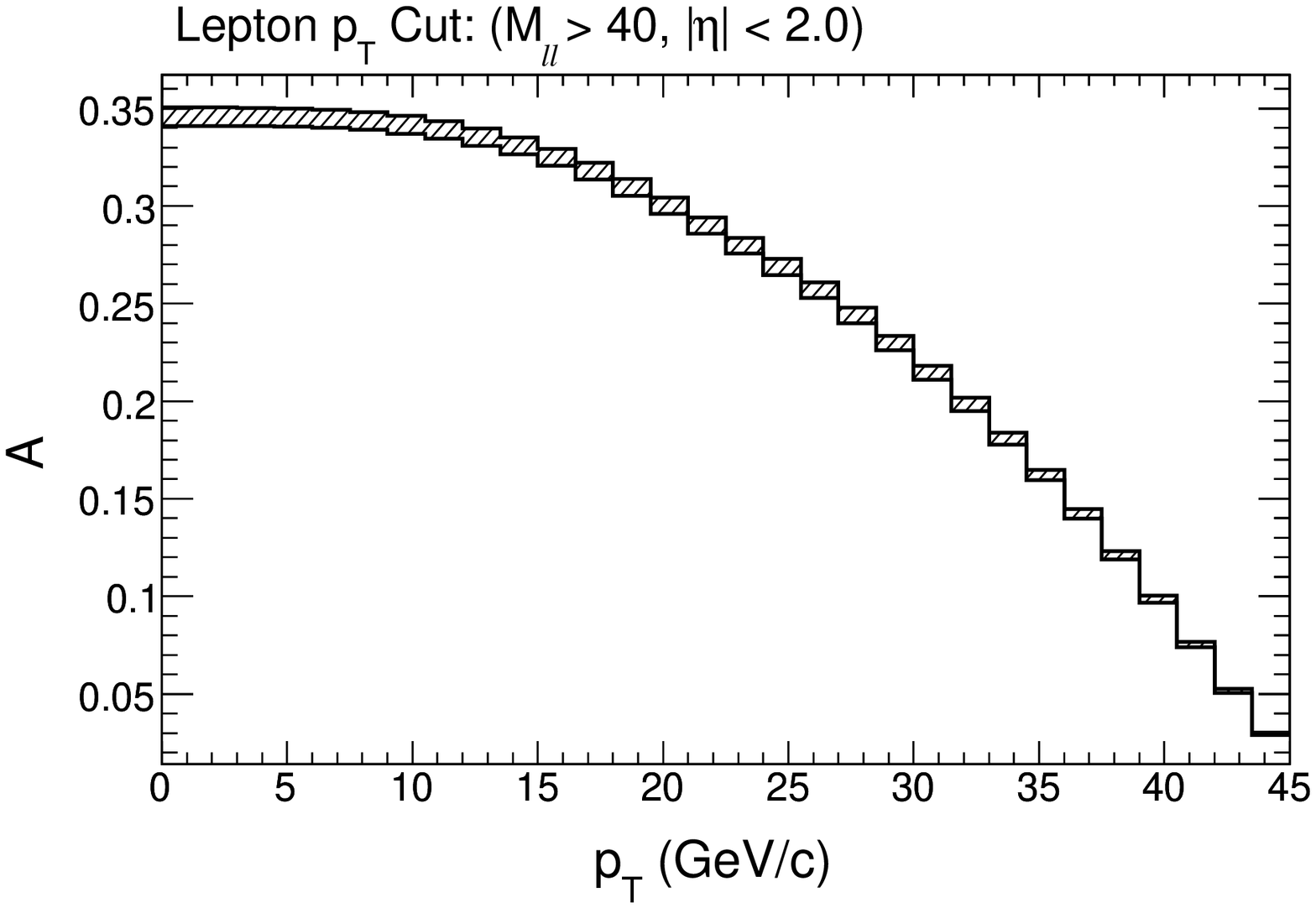,width=7cm}&
\epsfig{file=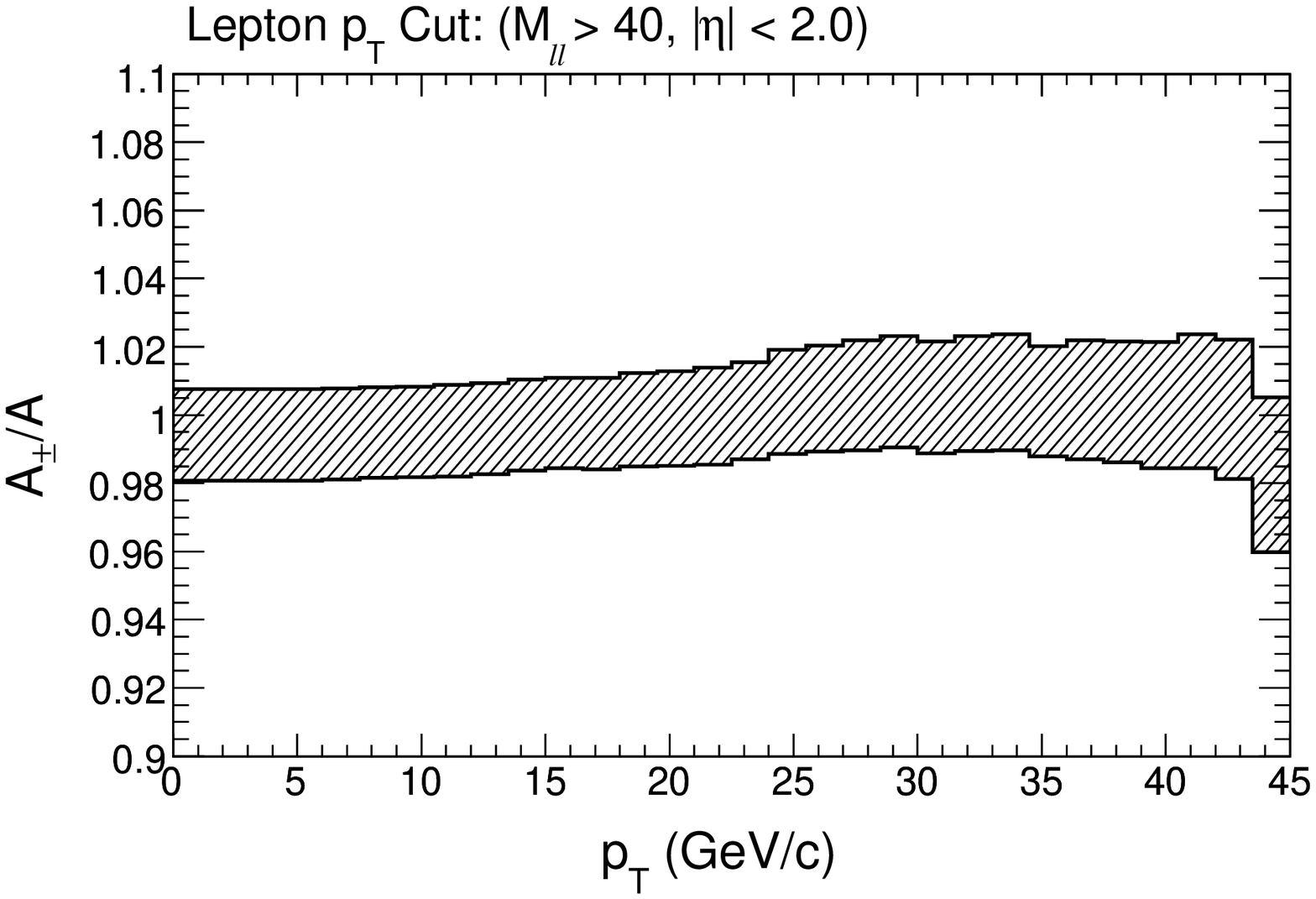,width=7cm}\\
\multicolumn{2}{c}{(a)} \\[3em]
\epsfig{file=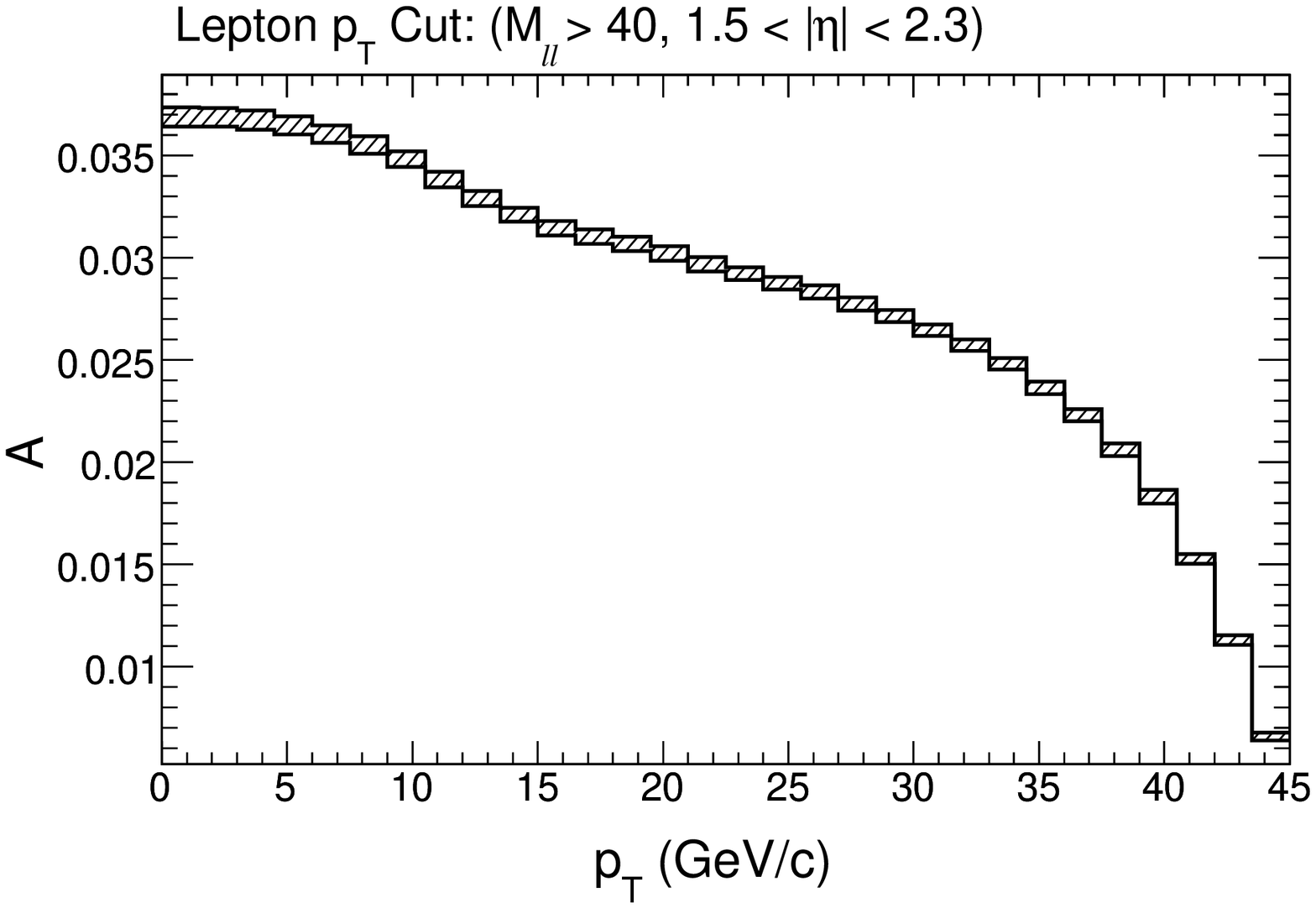,width=7cm} &
\epsfig{file=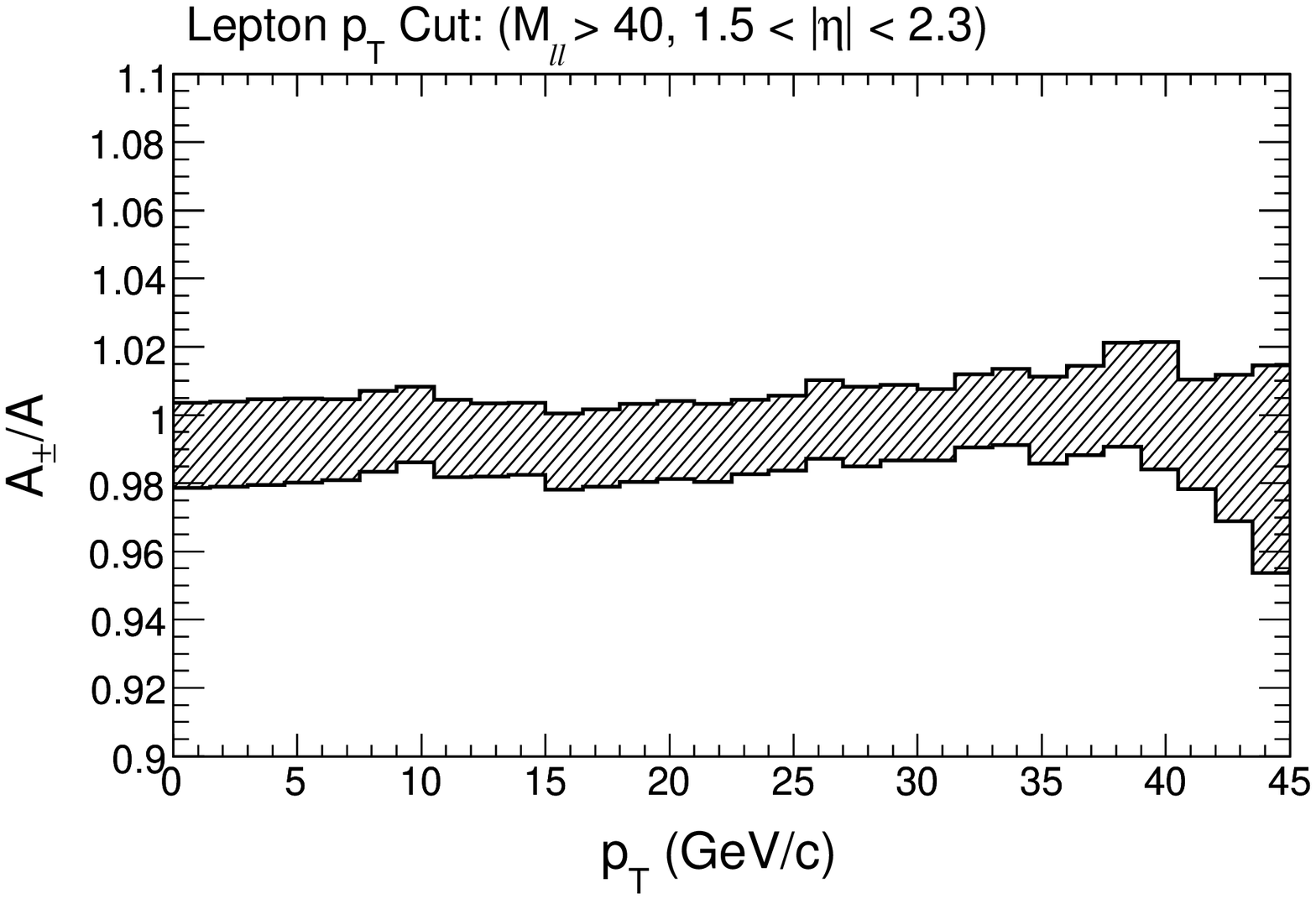,width=7cm} \\
\multicolumn{2}{c}{(b)} \\[3em]
\epsfig{file=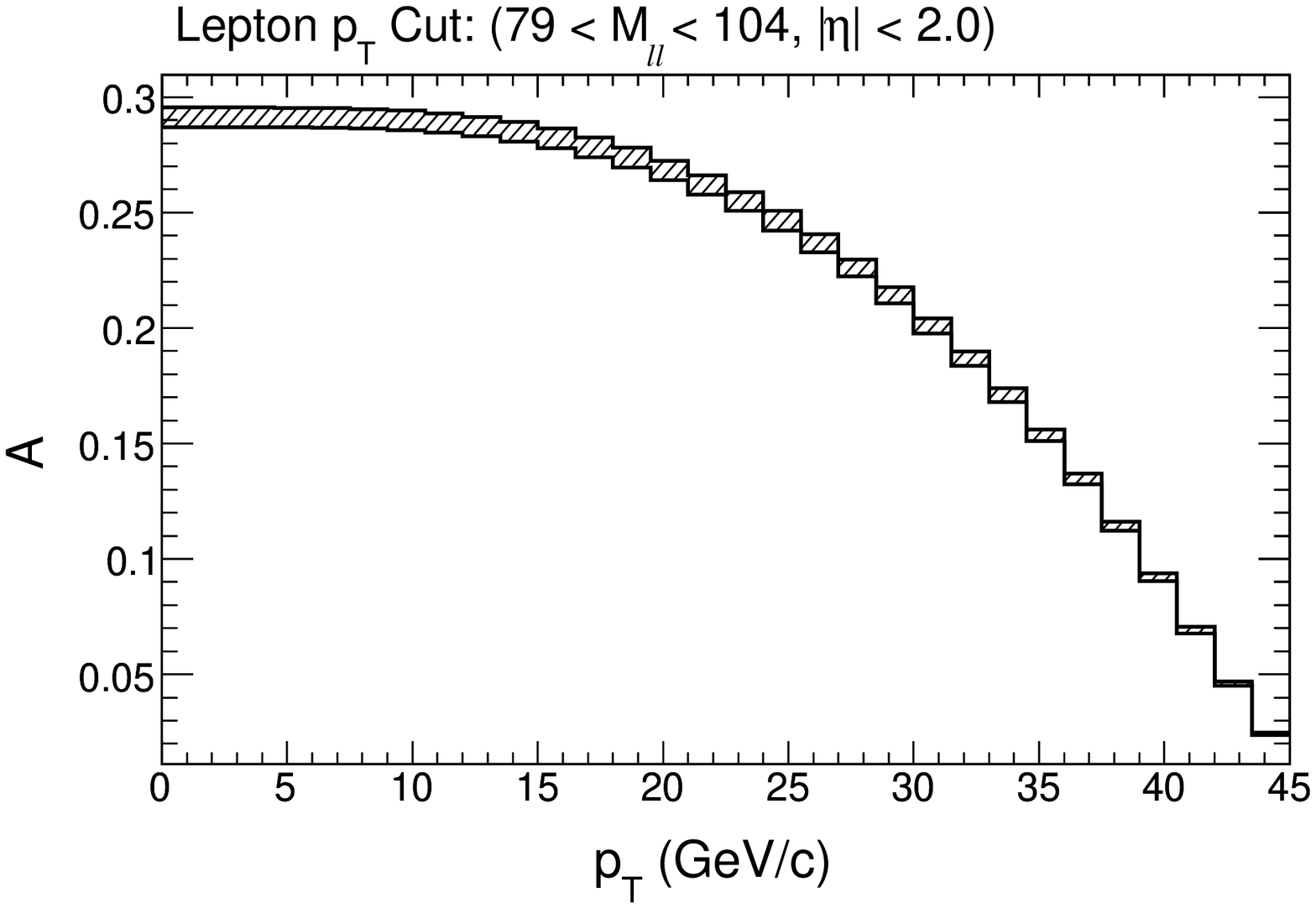,width=7cm} &
\epsfig{file=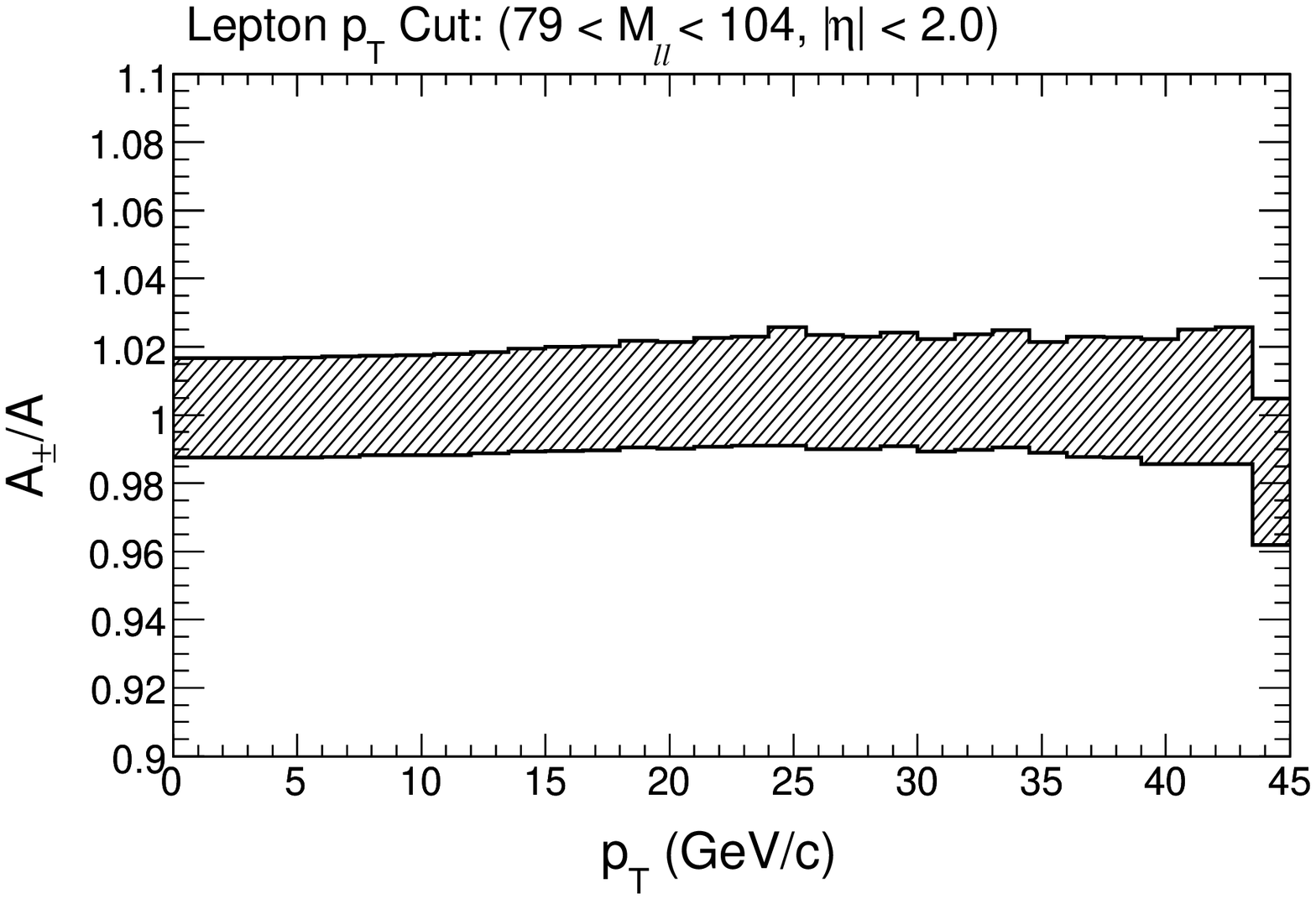,width=7cm} \\
\multicolumn{2}{c}{(c)} \\[3em]
\end{tabular}
\caption{ The $Z/\gamma^* \to \ell^{+}\ell^{-}$ acceptances $A$, as a
  function of the $\pT$ cut for acceptance regions
 (a) Cut 1, (b) Cut 2, and (c) Cut 3, as defined in
  Table~\ref{table:acceptance}. For each acceptance region we fix the
  invariant mass and $|\eta|$ cuts at their specified values, and vary only
  the $\pT$ cut. The figures on the right show the relative errors in the
  acceptances.}
\label{fig:pdf_acc_vs_pt_mu}
}

\FIGURE[ht]{
\begin{tabular}{cc}
\epsfig{file=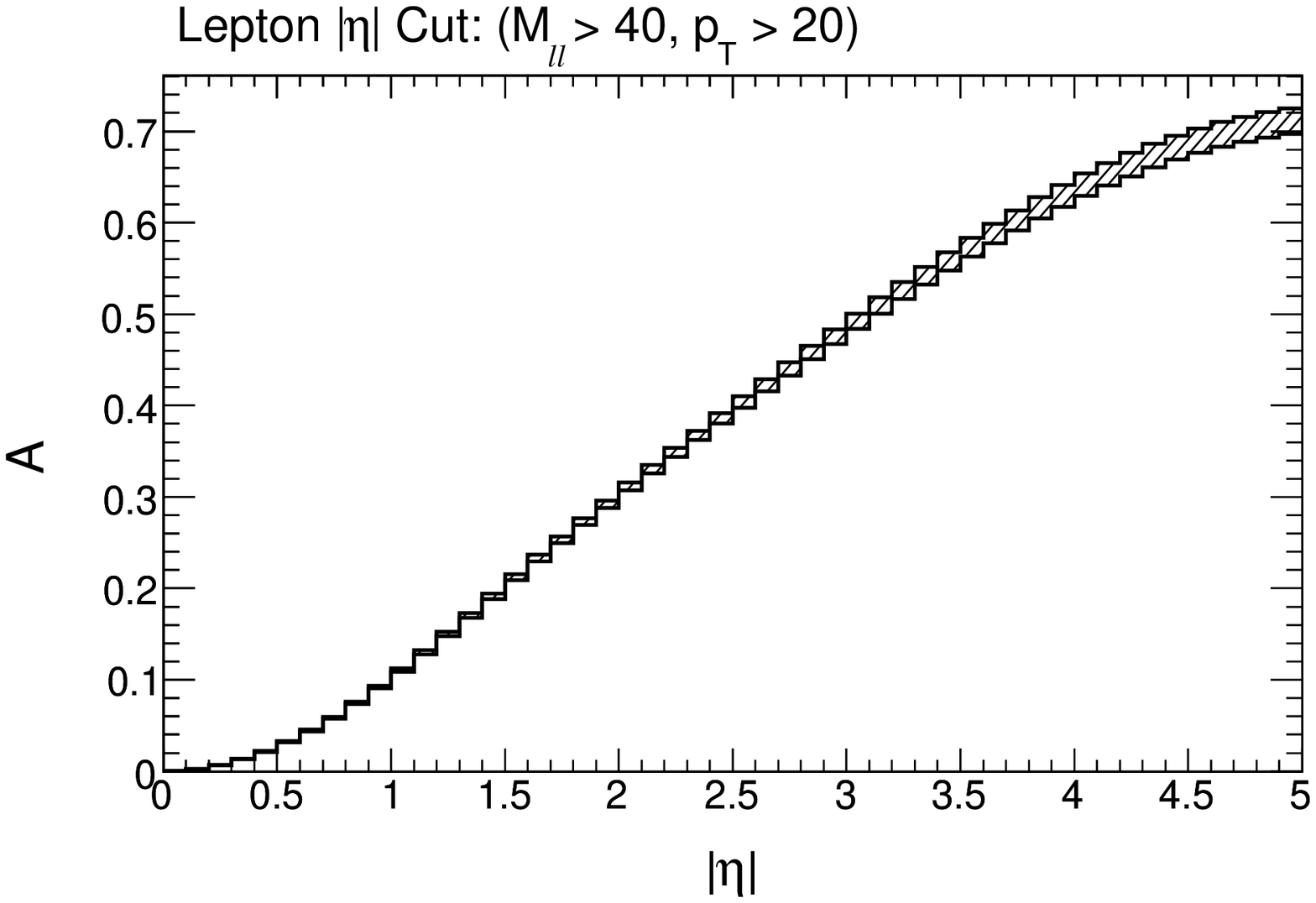,width=7cm}&
\epsfig{file=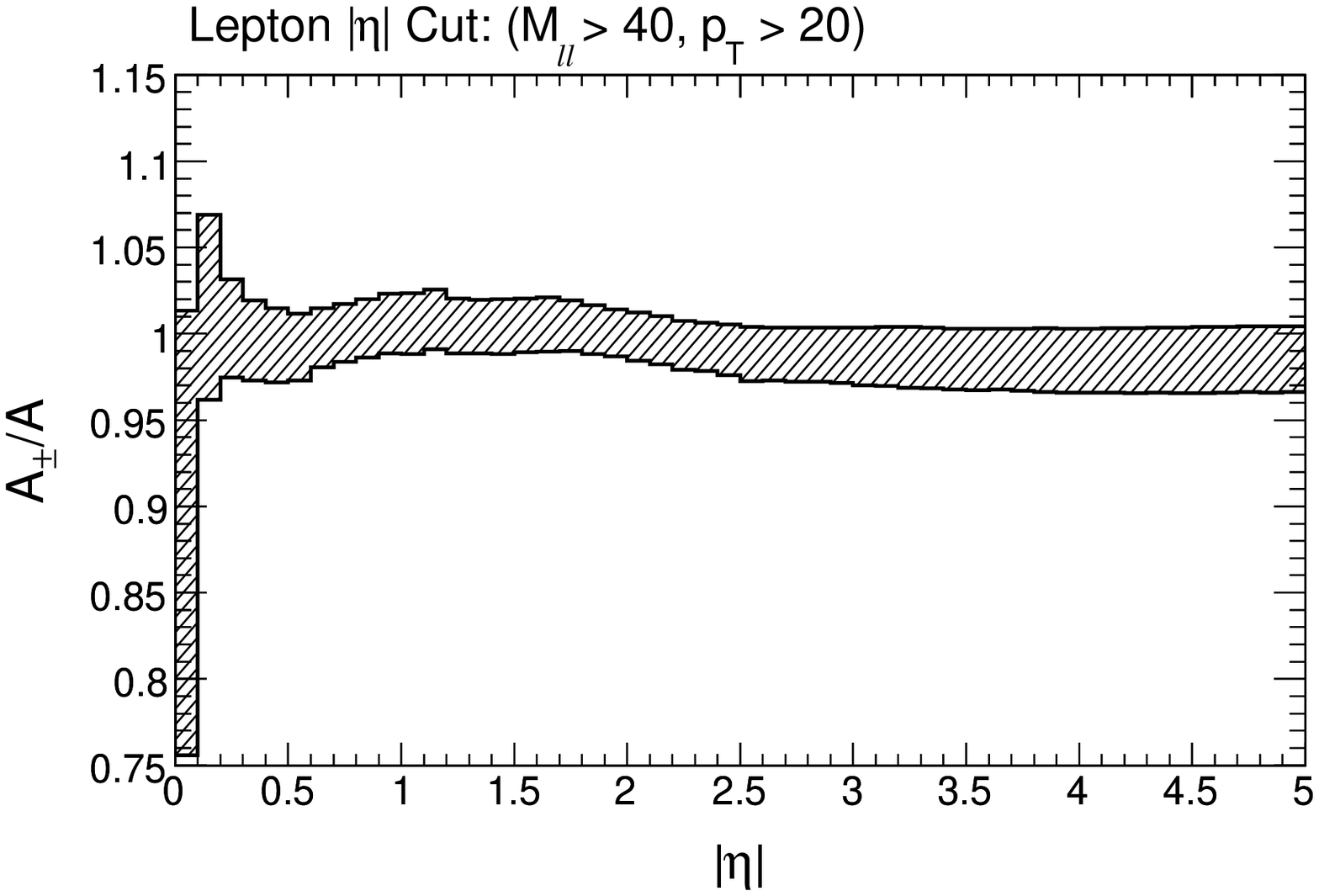,width=7cm}\\
\multicolumn{2}{c}{(a)} \\[3em]
\epsfig{file=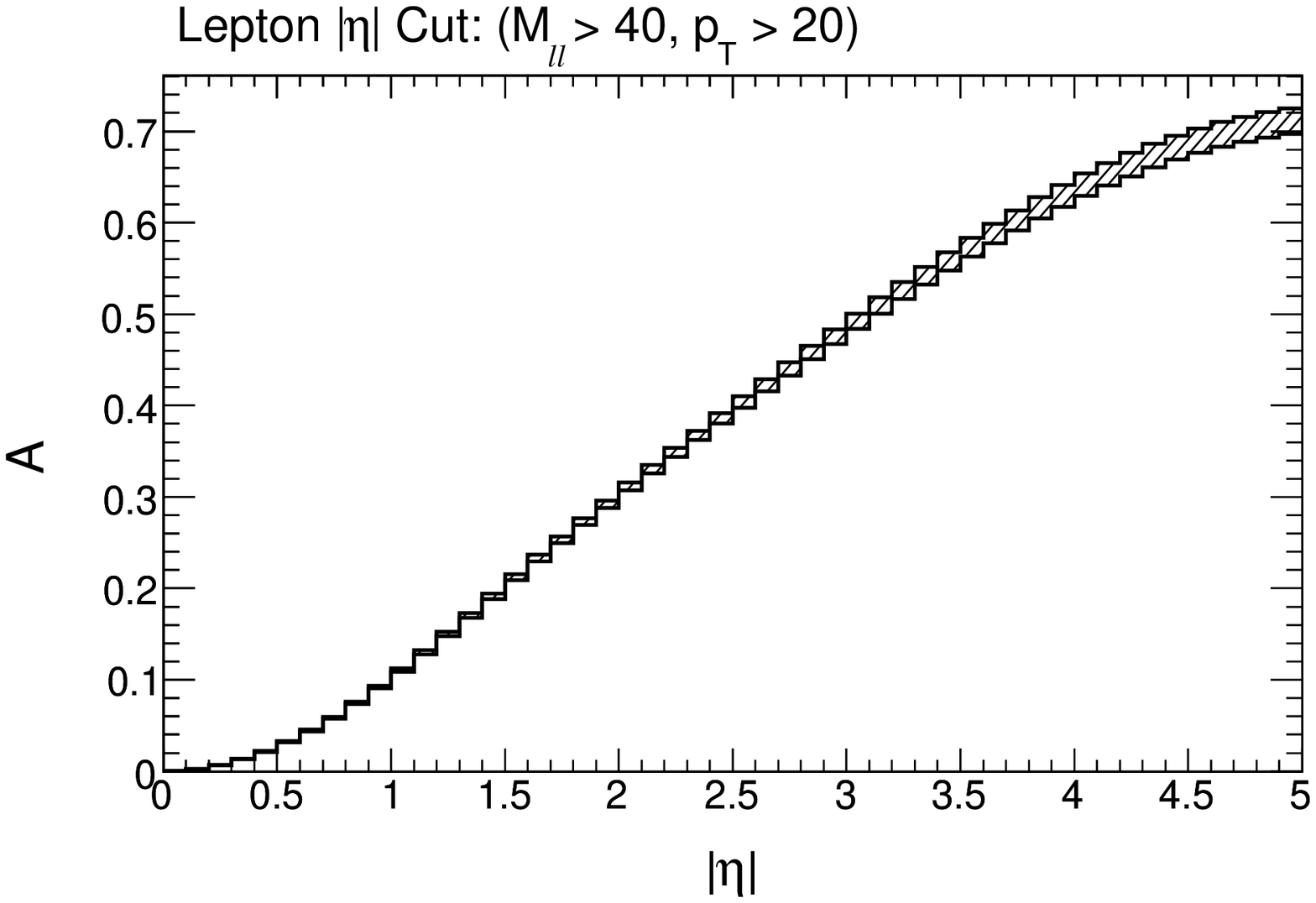,width=7cm} &
\epsfig{file=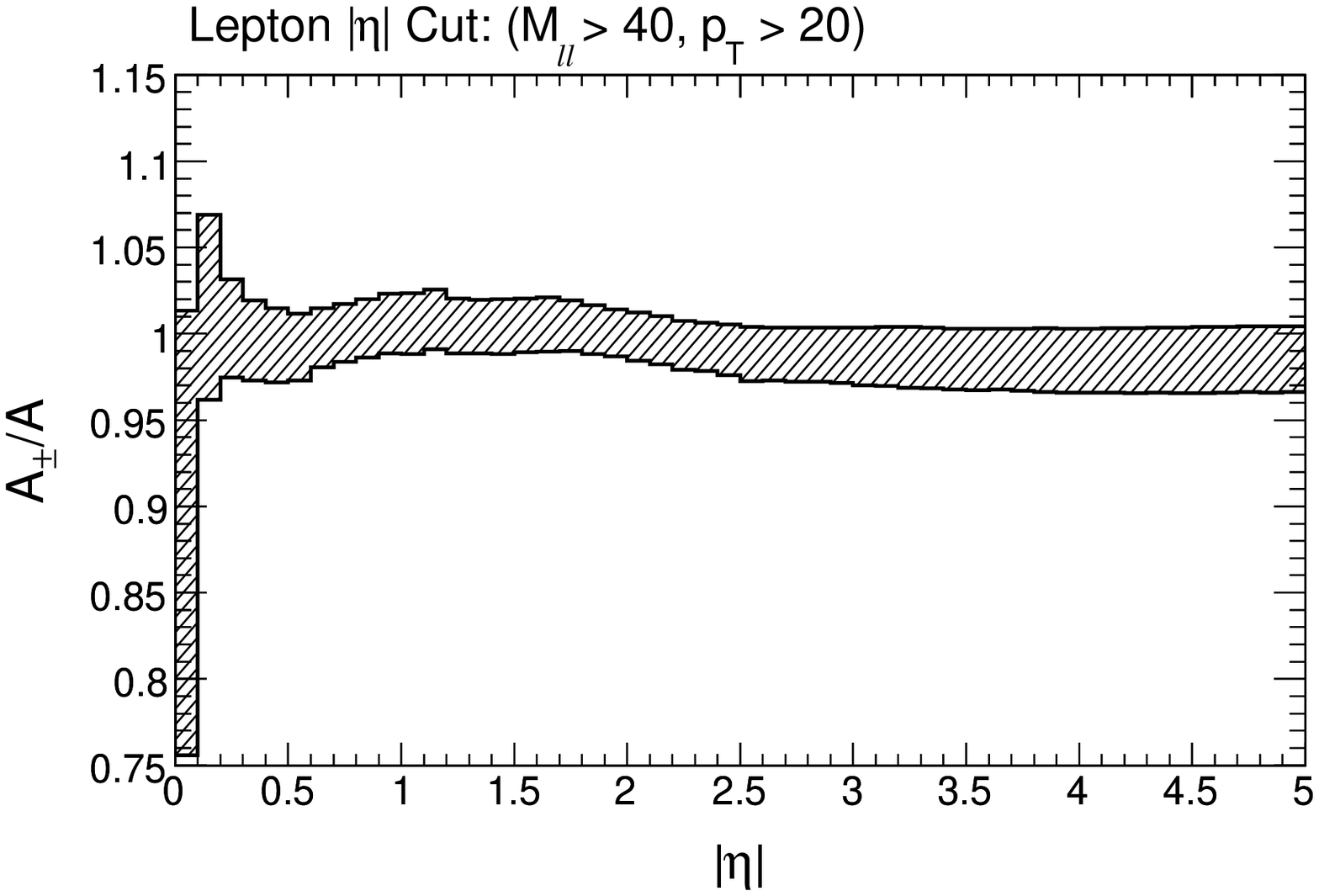,width=7cm} \\
\multicolumn{2}{c}{(b)} \\[3em]
\epsfig{file=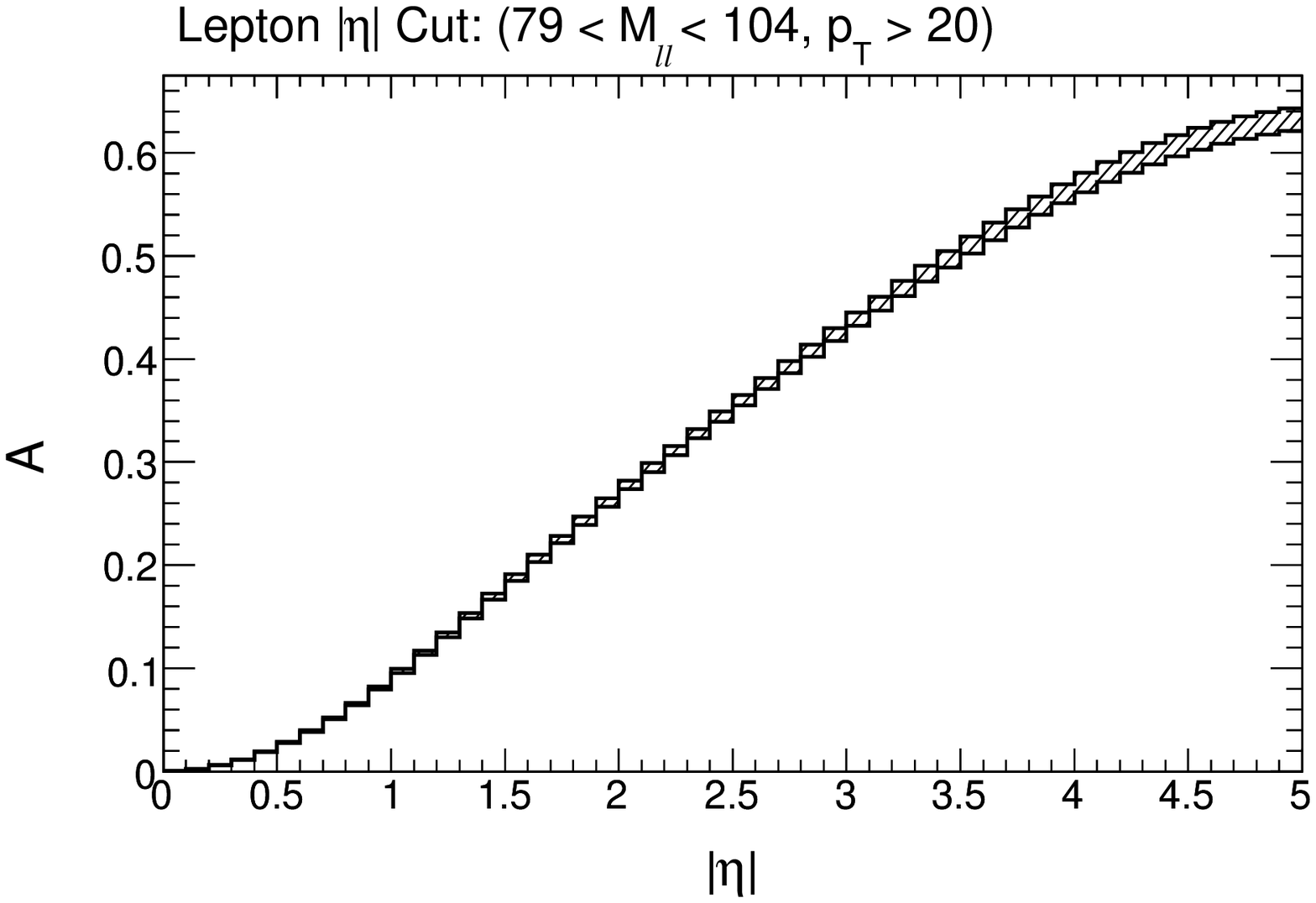,width=7cm} &
\epsfig{file=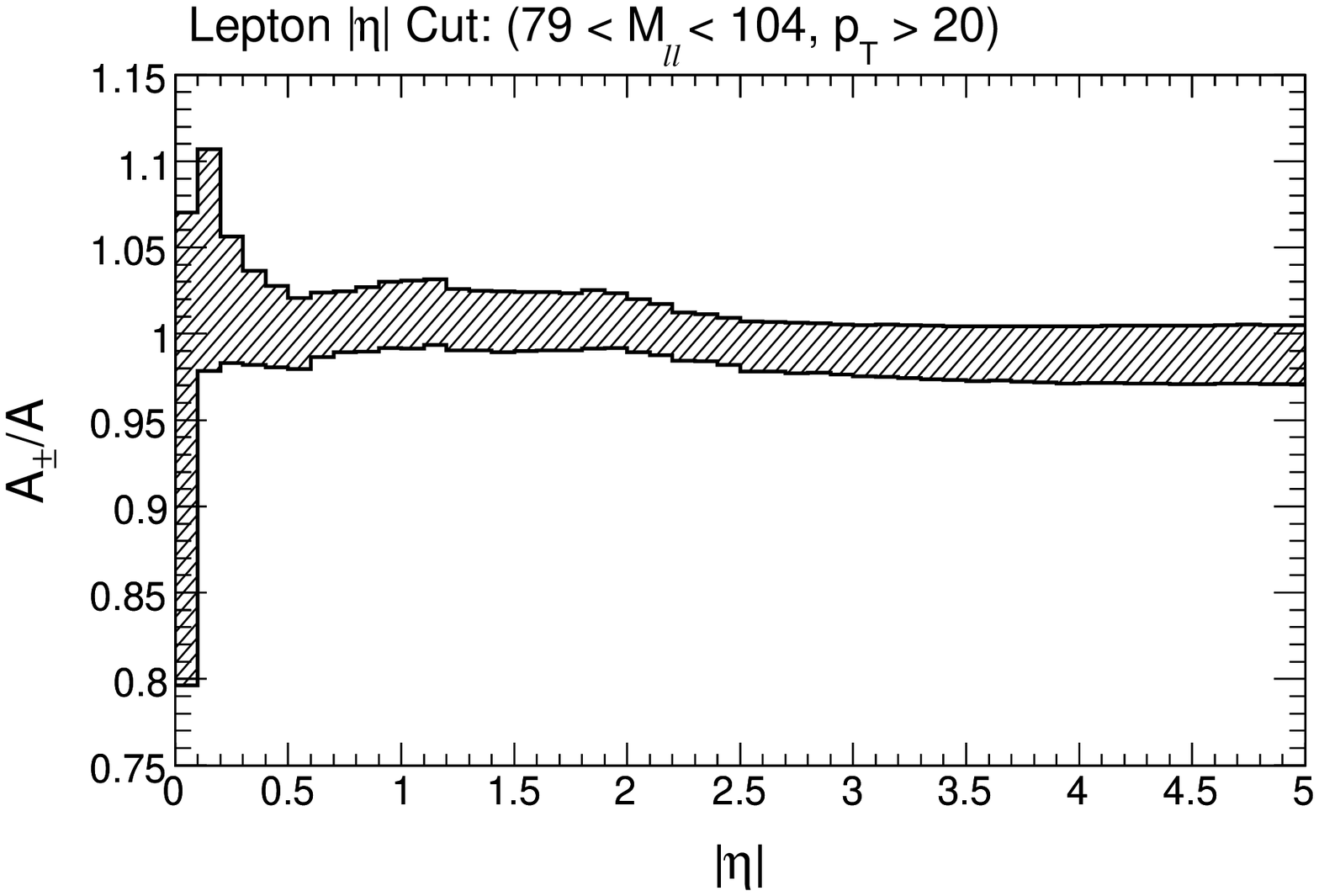,width=7cm} \\
\multicolumn{2}{c}{(c)} \\[3em]
\end{tabular}
\caption{ The $Z/\gamma^* \to \ell^{+}\ell^{-}$ acceptances $A$, as a
  function of the $|\eta|$ cut for acceptance regions
 (a) Cut 1, (b) Cut 2, and (c) Cut 3, as defined in
  Table~\ref{table:acceptance}. For each acceptance region we fix the
  invariant mass and \pT cuts at their specified values, and vary only
  the $|\eta|$ cut. The figures on the right show the relative errors in the
  acceptances.}
\label{fig:pdf_acc_vs_eta_mu}
}

\end{document}